\documentclass[iop,apj]{emulateapj}

\usepackage{natbib}
\usepackage{hyperref}

\begin{document}

\title{Infrared Continuum and Line Evolution of the Equatorial Ring around SN 1987A}
\author{Richard G. Arendt\altaffilmark{1,2}, 
Eli Dwek\altaffilmark{2}, 
Patrice Bouchet\altaffilmark{3}, 
I. John Danziger\altaffilmark{4}, 
Kari A. Frank\altaffilmark{5},\\
Robert D. Gehrz\altaffilmark{6},
Sangwook Park\altaffilmark{7}, 
Charles E. Woodward\altaffilmark{6}
}

\affil{$^1$CRESST/UMBC; richard.g.arendt@nasa.gov,\\
$^2$Observational Cosmology Lab, Code 665, NASA/GSFC, Greenbelt, MD 20771, USA\\
$^3$Laboratoire AIM Paris-Saclay, CEA-IRFU/SAp, CNRS, Universit\'e Paris Diderot, F-91191 Gif-sur-Yvette, France\\
$^4$INAF-Osservatorio Astronomico di Trieste, via G.B. Tiepolo 11, 34143 Trieste, Italy\\
$^5$Department of Astronomy and Astrophysics, Pennsylvania State University, University Park, PA 16802, USA\\
$^6$Minnesota Institute for Astrophysics, School of Physics and Astronomy,
University of Minnesota,\\
116 Church Street, SE, Minneapolis, MN 55455, USA\\
$^7$Department of Physics, University of Texas at Arlington, Arlington, TX 76019, USA\\
}

\begin{abstract}
\hspace{0pt}{\it Spitzer} observations of SN 1987A have now spanned more than a decade.
Since day $\sim4,000$, mid-infrared (mid-IR) emission has been dominated by that 
from shock-heated dust in the equatorial ring (ER).
From 6,000 to 8,000 days after the explosion, {\it Spitzer} observations included
broadband photometry at 3.6 -- 24 $\mu$m, and low and moderate resolution
spectroscopy at 5 -- 35 $\mu$m. Here we present later {\it Spitzer} 
observations, through 
day 10,377, which include only the broadband measurements at 3.6 and 4.5 $\mu$m. 
These data show that the 3.6 and 4.5 
$\mu$m brightness has clearly begun to fade after day $\sim8,500$,
and no longer tracks the 
X-ray emission as well as it did at earlier epochs. 
This can be explained by the destruction 
of the dust in the ER on time scales 
shorter than the cooling 
time for the shocked gas. We find that the evolution of the 
late time IR emission is 
also similar to the now fading optical emission. We provide
the complete record of the IR emission lines,
as seen by {\it Spitzer} prior to day 8,000.
The past evolution of the gas as seen by the IR emission lines seems 
largely consistent with the 
optical emission, although the IR [\ion{Fe}{2}] and 
[\ion{Si}{2}] lines show different, peculiar velocity structures.
\end{abstract}


\section{Introduction}

SN 1987A presents, thus far, a unique opportunity for the  
detailed study of the evolution of a supernova into a supernova remnant
\citep[see reviews by][]{mccray:1993,mccray:2007}.
Its proximity allows the imaging of small scale structure of both the SN ejecta 
and the circumstellar environment. It also is sufficiently bright that 
at most wavelengths high S/N observations can be obtained very quickly, even 
at high spectral resolution.
SN 1987A was not a prototypical Type II supernova, 
exhibiting a subluminous light curve with a late maximum,
and having a known blue supergiant progenitor \citep[e.g.][]{arnett:1989}.
However, a number of other blue supergiant stars have been found to have 
similar circumstellar structures \citep{brandner:1997b,brandner:1997a,
smith:2007,smith:2013,muratore:2015,gvaramadze:2015}.

Shortly after its launch, 
the {\it Spitzer Space Telescope} \citep{werner:2004, gehrz:2007} 
began a long-term campaign of imaging and 
spectroscopy of SN 1987A at wavelengths from 3.6 -- 160 $\mu$m. At the near- and mid-IR 
wavelengths {\it Spitzer} has been able to monitor thermal emission from warm dust 
in the SN. Despite insufficient angular resolution to resolve detailed structure, 
this emission has been identified \citep{bouchet:2006} as arising from dust in the equatorial ring (ER) 
that was formed by winds during mass loss episodes of the SN progenitor star 
\citep[or binary system, e.g.][]{podsiadlowski:1989}.

Dust is also known to be present in SN 1987A's ejecta. 
Photometric and spectroscopic evidence of dust formation in the ejecta
began to appear after day $\sim450$ \citep{danziger:1989,suntzeff:1990}.
From days 615 -- 1,144, mid-IR spectrophotometry by \cite{moseley:1989a,dwek:1992a}
and \cite{wooden:1993} monitored the emission of the dust directly.
However at much later times, the ejecta has cooled and faded, and 
has been too faint to be detected by {\it Spitzer} at mid- or far-IR wavelengths.
At present times, the ejecta dust has been detected 
in the far-IR with larger, more sensitive instruments such as {\it Herschel} and 
ALMA \citep{matsuura:2011,Lakicevic:2012,kamenetzky:2013,Indebetouw:2014,zanardo:2014,matsuura:2015,dwek:2015}.
The $\sim0.4-0.7$~$M_\sun$ of cold ejecta dust seen by Herschel \citep{matsuura:2011}
is greatly in excess of the $\sim2\times 10^{-6}$~$M_\sun$ of warmer circumstellar
dust located in the ER \citep{bouchet:2006}.

The SN ejecta were first beginning to impact the ER creating hotspots visible 
in the high resolution {\it HST} images at 
days $\sim$3,000 -- 4,400 \citep[and references therein]{lawrence:2000}. 
{\it Spitzer} observations commenced near day 6,100 \citep{bouchet:2006} while 
the number and brightness of hotspots around the ER were increasing
\citep{groningsson:2008a,fransson:2015}. 
This epoch is also when the soft X-ray light curve turned up
\citep{park:2005,park:2006}, indicating that
the shock started interacting with dense protrusions from
the ER. 
Thus {\it Spitzer} data are well suited for assessing what kind of
dust exists in the circumstellar environment and whether this dust can survive
in the hot shocked environment downstream of the SN blast wave. 

\cite{bouchet:2004}, using the T-ReCS instrument \citep{telesco:1998,debuizer:2005}
on the Gemini South 8m telescope, detected 10 $\mu$m emission from SN 1987A
at Day 6,067 and resolved it as being dominated by the ER (i.e. pre-existing circumstellar dust), 
with only a small contribution from the ejecta (i.e. newly-formed SN dust).
\cite{bouchet:2006} presented later T-ReCS observations (Day 6,526) 
showing the ejecta to be fading and the ER brightening. 
They also analyzed {\it Spitzer} IRS spectroscopy at day 6,190, 
revealing that the dust possesses strong silicate features 
at 10 and 20~$\mu$m, which are typical of silicate grains in the ISM. 
The derived dust temperature of $\sim180$~K (with a surprising lack of evidence for variations
in dust temperature) is far warmer than typical ISM dust temperatures ($\sim20$~K)
and provides a clear indication of strong heating of the dust.
The observed dust temperature was determined to be consistent with collisional
heating of dust in the shocked X-ray emitting gas. Radiative heating
was estimated to produce dust temperatures of only $\sim125$~K. 
While radiative heating was not ruled out, this does suggest it is not the 
dominant dust heating mechanism.

About 1,000 days later, \cite{dwek:2008} reported that further {\it Spitzer} observations
had shown the dust was unchanged in temperature and increasing in brightness, but 
not increasing as rapidly as the X-ray emission. The decreasing ratio of IR to X-ray 
emission (IRX) was interpreted as destruction of dust behind the blast wave on timescales
shorter than that for the cooling of the shocked, X-ray emitting gas. However, this 
conclusion was based on only two epochs of data.

\cite{dwek:2010} presented the full set of {\it Spitzer}'s cryogenic imaging 
and spectroscopy (continuum results only), through day~7,983. Over this longer baseline 
it became apparent that, although there appeared to be a marginally significant 
decrease in the IRX after the first {\it Spitzer} observations, the IRX ratio was constant 
for all further observations from days 6,500 to 8,000. The favored interpretation 
was that the grain destruction timescale was sufficiently long that no major destruction
of the dust had yet taken place. They predicted that grain destruction 
may become appreciable after day $\sim9,200$. This work also drew attention to the short wavelength 
continuum emission, which was present at $\lesssim 8$ $\mu$m, 
in excess of the $\sim180$ K silicate dust. This indicates a hotter dust component, but
lacking any spectral features, the composition and temperature of this hot component were indeterminate.

In this paper, we examine continued {\it Spitzer} observations of SN 1987A. 
This paper is an observational update to, and relies upon, previous
analysis of the {\it Spitzer} photometry and spectroscopy
presented in the works of \cite{bouchet:2006}, \cite{dwek:2008}, and \cite{dwek:2010}.
In the {\it Spitzer} post-cryo era, only the 3.6 and 4.5 $\mu$m 
imaging capability remains, but continued observations reveal
the evolution of the emission of the hot dust component (the component 
emitting at $\lambda<8$~$\mu$m), which may provide clues
to the nature of this dust. Section 2 summarizes the characteristics of 
the {\it Spitzer} instruments and data used in this research. 
In Section 3, we present the latest 
{\it Spitzer} light curves of SN 1987A, and characterize the IR evolution 
with respect to the X-ray emission, and the optical emission. In Section 4, 
we present the evolution of the IR emission lines during {\it Spitzer}'s cryogenic 
mission (days 6,000 - 8,000).  The results are discussed in Section 5 and 
summarized in Section 6.

\section{Spitzer Instruments and Data}

The {\it Spitzer Space Telescope} carries three instruments, 
all of which were used for observation of SN 1987A. 

The Multiband Imaging Photometer for {\it Spitzer} \citep[MIPS;][]{rieke:2004} provides broadband imaging at 24, 70, and 160 $\mu$m. It also 
has an SED mode to provide very low resolution spectra over 52 - 97 $\mu$m.
Over the course of the cryogenic mission, the SN was only detected in 
the 24 $\mu$m band. No lines or continuum were detected in the SED mode.

The Infrared Spectrograph \citep[IRS;][]{houck:2004} provides low resolution 
($R\sim100$) in two modules (short-low = SL and long-low = LL) covering 5.2-38 $\mu$m.
it also provides high resolution ($R\sim600$) spectra in two modules 
(short-high = SH and long-high = LH) covering 9.9-37.2 $\mu$m. The low resolution 
spectra are best for study of the continuum, and the high resolution spectra 
are best for the line emission.

The Infrared Array Camera \citep[IRAC;][]{fazio:2004} provides broadband imaging
at 3.6, 4.5, 5.8, and 8 $\mu$m. The IRS low resolution spectra provide overlap with 
the two longer bands, but do not cover the shorter two bands. These shorter 
two bands are the only channels that remain operational on {\it Spitzer} since 
the telescope warmed up after the depletion of cryogen on day 8,117. 

Figure 1 shows composite IRAC images of SN 1987A at 3.6 and 4.5 $\mu$m. 
Combining all epochs of targeted observations after day 7,000 provides sufficient depth, dithering, and 
rotation of the PSF, such that mapping at $0\farcs4$ (here) reveals more 
detail than mapping at the detector pixel scale of $1\farcs2$ \citep{bouchet:2006}.
In Figure 1, we can see the SN marginally resolved from Stars 2 and 3.

\begin{figure*}[t] 
   \centering
   \includegraphics[width=7in]{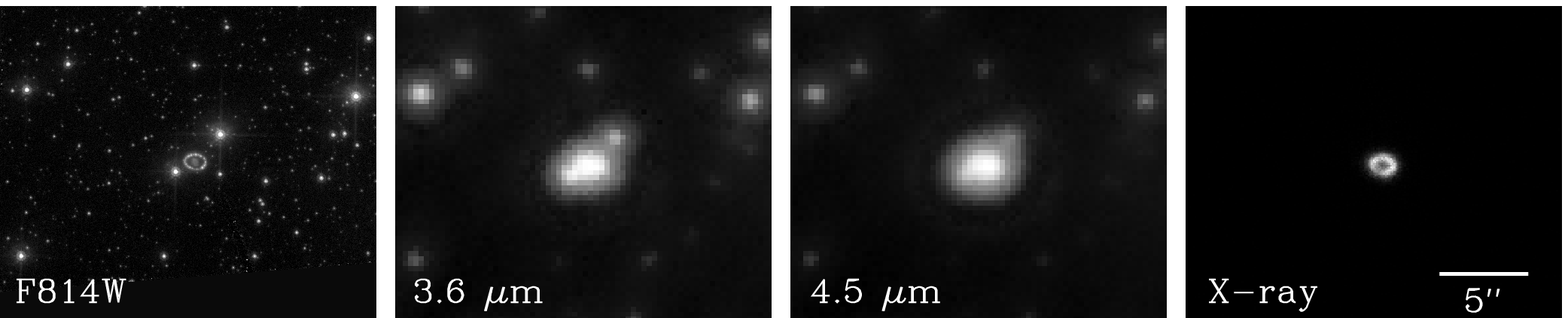}
   \includegraphics[width=7in]{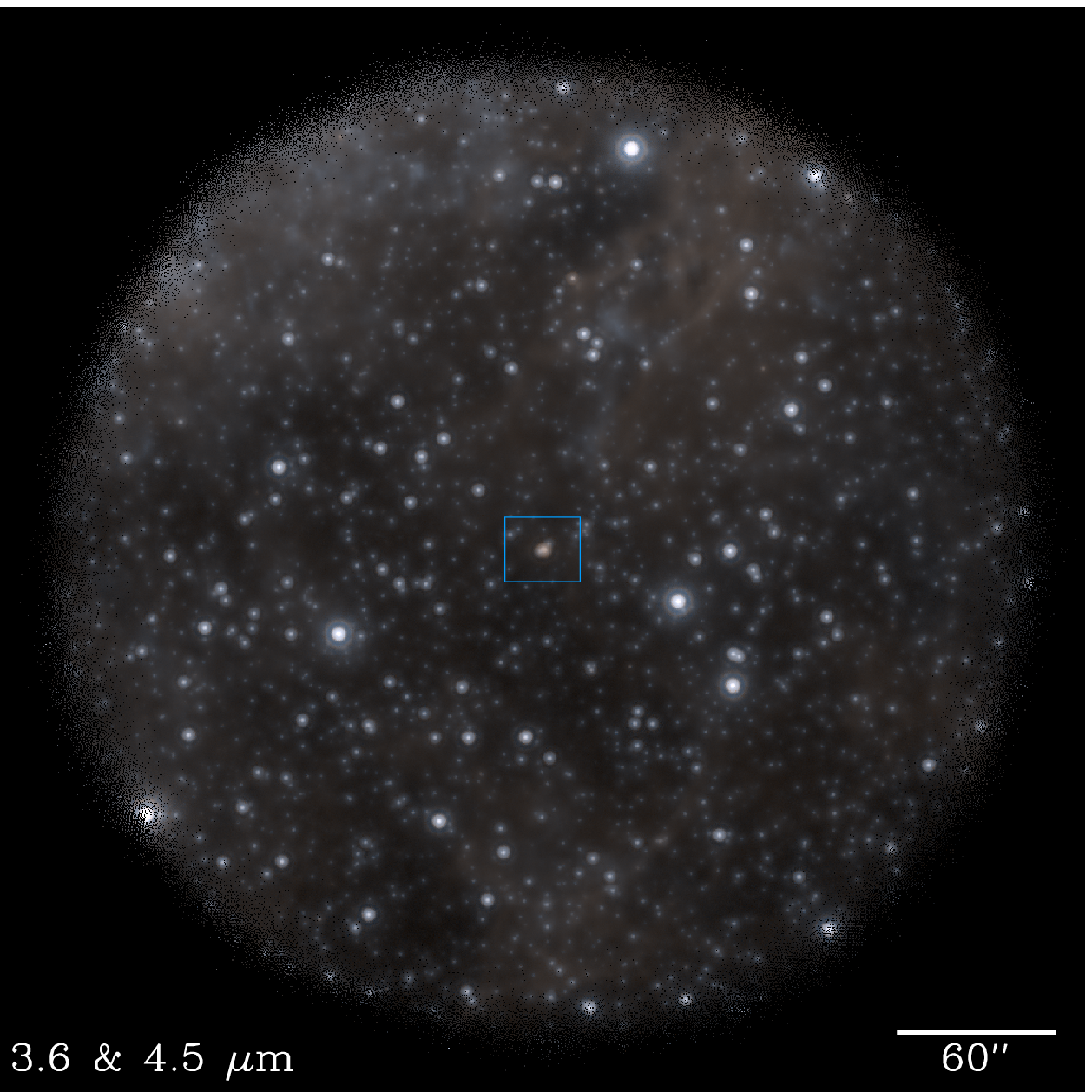} 
   \caption{(top) SN 1987A as seen by {\it Hubble}/ACS (F814W, left), 
   {\it Spitzer}/IRAC (3.6 and 4.5 $\mu$m, middle), and {\it Chandra}/HRC (right).
   In the IRAC images the angular resolution is $\gtrsim 12$ times 
   worse than the ACS image. Stars 2 and 3 to the NW and SE are only 
   barley resolved from the SN. The ER is elongated, but not resolved.
   (bottom) The full IRAC 3.6 and 4.5 $\mu$m 2-color image, 
   with SN 1987A centered in the blue box, reveals that the 
   ER is redder than most stars in the field and that it does not show the 
   symmetric PSF that stellar sources exhibit. The optical and IR images are 
   logarithmically scaled; the X-ray image is scaled linearly.}
   \label{fig:image}
\end{figure*}

\section{Evolution of the IR Continuum}

In sections 3.1 and 3.2 we consider the IR evolution in comparison 
to the X-ray emission. The IR and X-ray emission should be directly
related for collisionally heated dust. In Section 3.3 we look at
the IR evolution in terms of the dust mass and temperature, 
without regard for the heating mechanism. Section 3.4 examines
the relation between the IR and the optical emission, as 
potentially related if radiative heating of the dust is important.
The ``models'' in this section are empirical parametric descriptions 
of the evolution of the IR emission. They can be used as simplified 
descriptions of the data (e.g. for extrapolation or interpolation)
and to identify the behavior of correlations between the IR, and
X-ray and optical emission.
In some cases the model parameters are related to one or more of 
the underlying physical parameters of the system.

\subsection{Infrared and X-ray Brightness Evolution}

\begin{figure}[b] 
   \centering
   \includegraphics[width=3.5in]{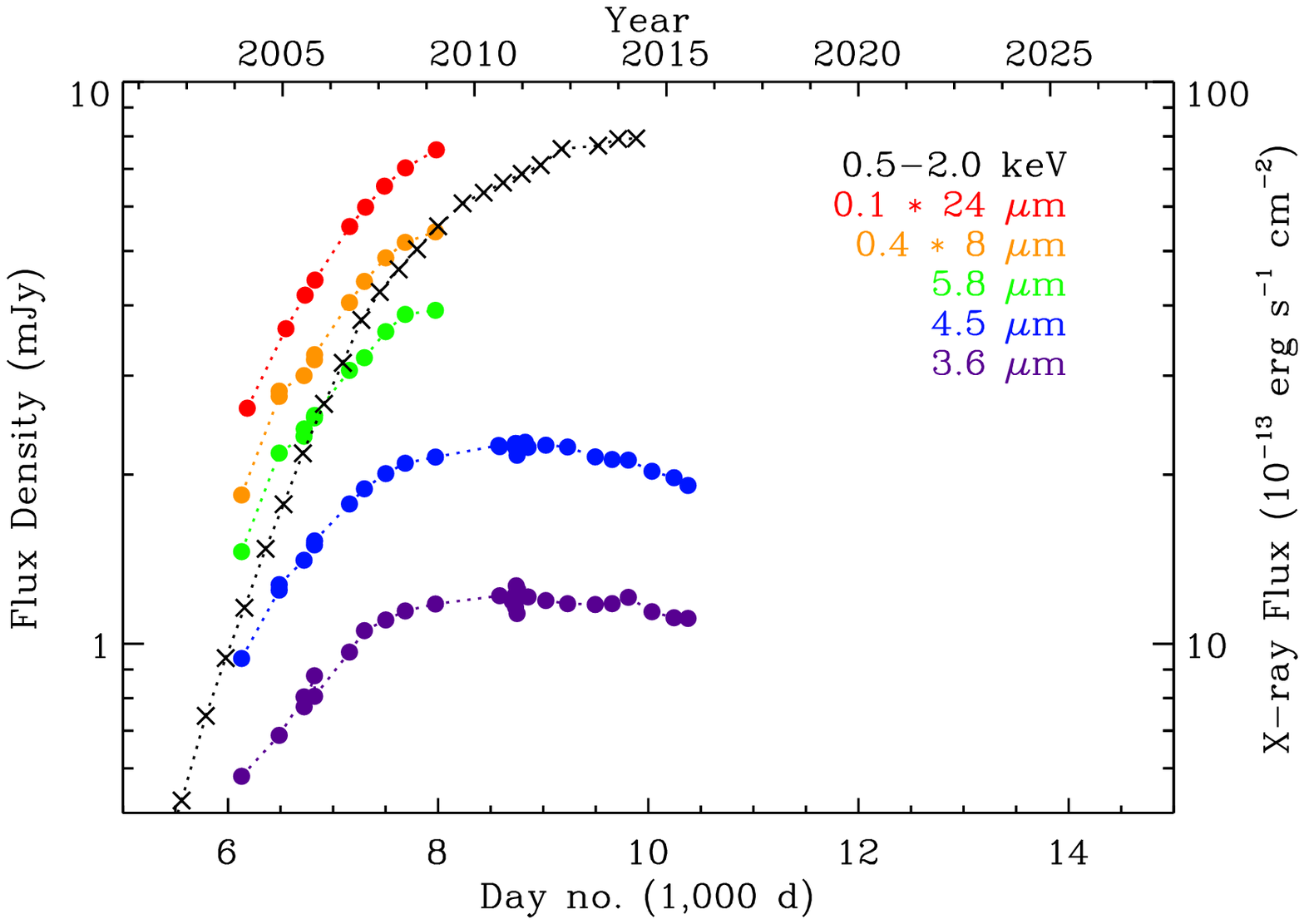} 
   \caption{Evolution of SN 1987A as seen in 
   the 3.6 -- 24 $\mu$m flux densities observed with {\it Spitzer}, 
   and the soft X-ray fluxes detected with {\it Chandra} \citep{helder:2013}. 
   Until 2009, the IR emission
   had been nearly proportional to the soft X-ray emission. 
   {\it Spitzer}/IRAC warm mission data since 2009, have shown that the 
   3.6 and 4.5 $\mu$m flux densities have deviated from the X-ray trend, 
   and have begun to decline slowly.}
   \label{fig:evolution}
\end{figure}

Examination of the cryogenic data had indicated that the ratio of 24 $\mu$m IR to 
X-ray brightness (IRX) was approximately constant between days 6,000 and 8,000. 
However, continued observations at 3.6 and 4.5 $\mu$m have shown a trend 
that clearly diverges from the X-ray emission. 
Figure~\ref{fig:evolution} depicts the observed light curves, as listed
in Table \ref{tab:fluxes}.

The soft 0.5-2 keV in-band X-ray flux is used throughout for comparison with 
the IR emission because it is more representative than 
the harder X-ray emission of the 
forward shock that propagates through the ER
\citep{dwek:1987}. 
The X-ray data used here are those measured with the {\it Chandra} ACIS and 
published by \cite{helder:2013}, to which we add 3 additional epochs of 
similarly observed and reduced {\it Chandra} observations.
A more physically motivated comparison
might use the integrated X-ray emission of the softer component of
2-component fits to the total X-ray spectrum. The temperature
of the soft component varies from 0.2-0.4 keV, and is sufficiently low
that the fraction of emission at energies $>2$ keV is negligible.
The soft component accounts for $\sim33\%$ of the flux in the 0.5-2 keV
band at day 5,000, rising to $\sim41\%$ at day 10,000. However, given 
uncertainties in the corrections of systematic effects in the
X-ray data, and the relatively low signal-to-noise of the data at some epochs,
both of which affect the spectral fitting, the light curve of the integrated
soft X-ray component is much noisier than that of the simple 0.5-2 keV in-band
flux.

\begin{figure*}[ht] 
   \centering
   \includegraphics[width=3.in]{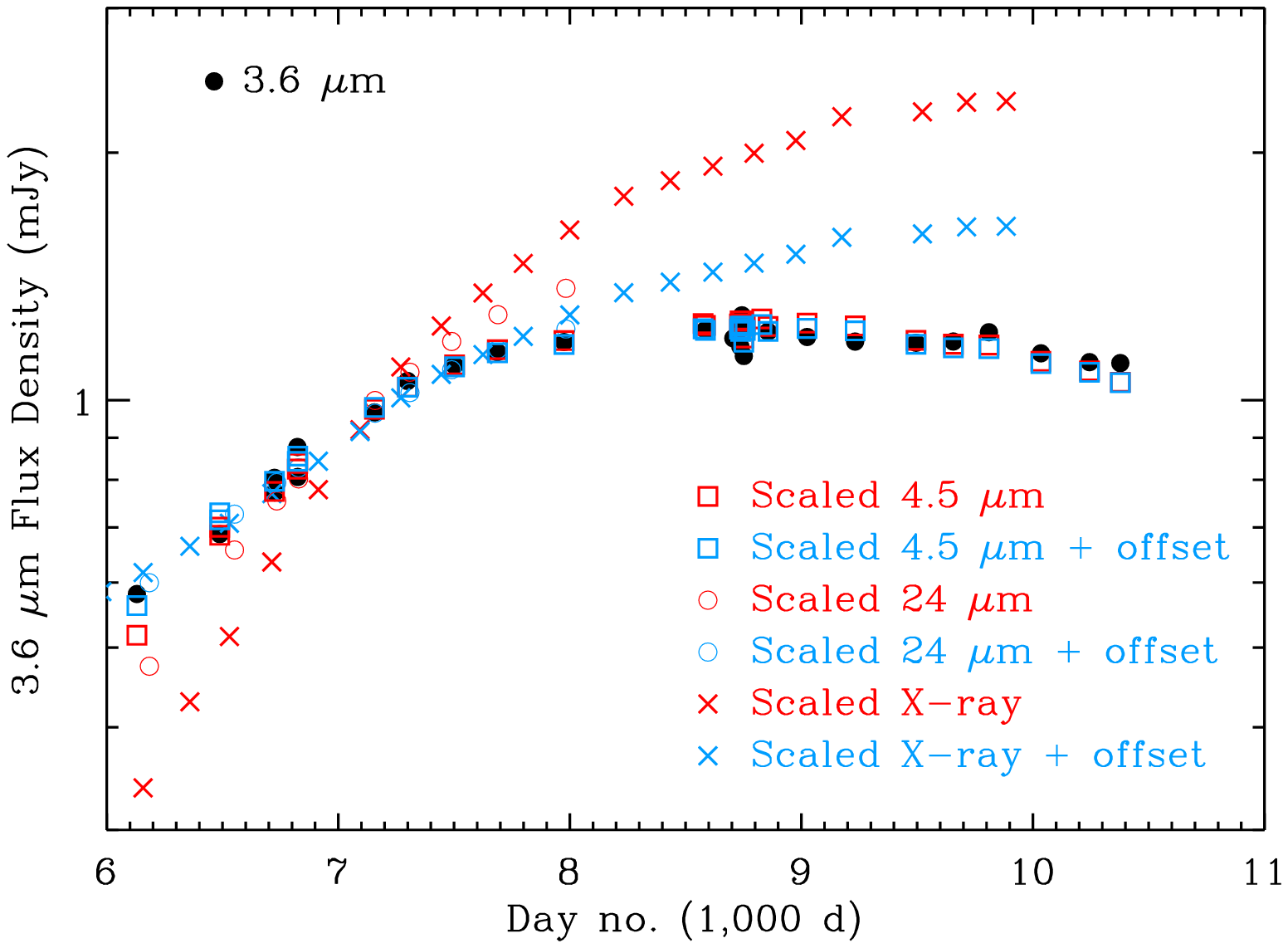} 
   \includegraphics[width=3.in]{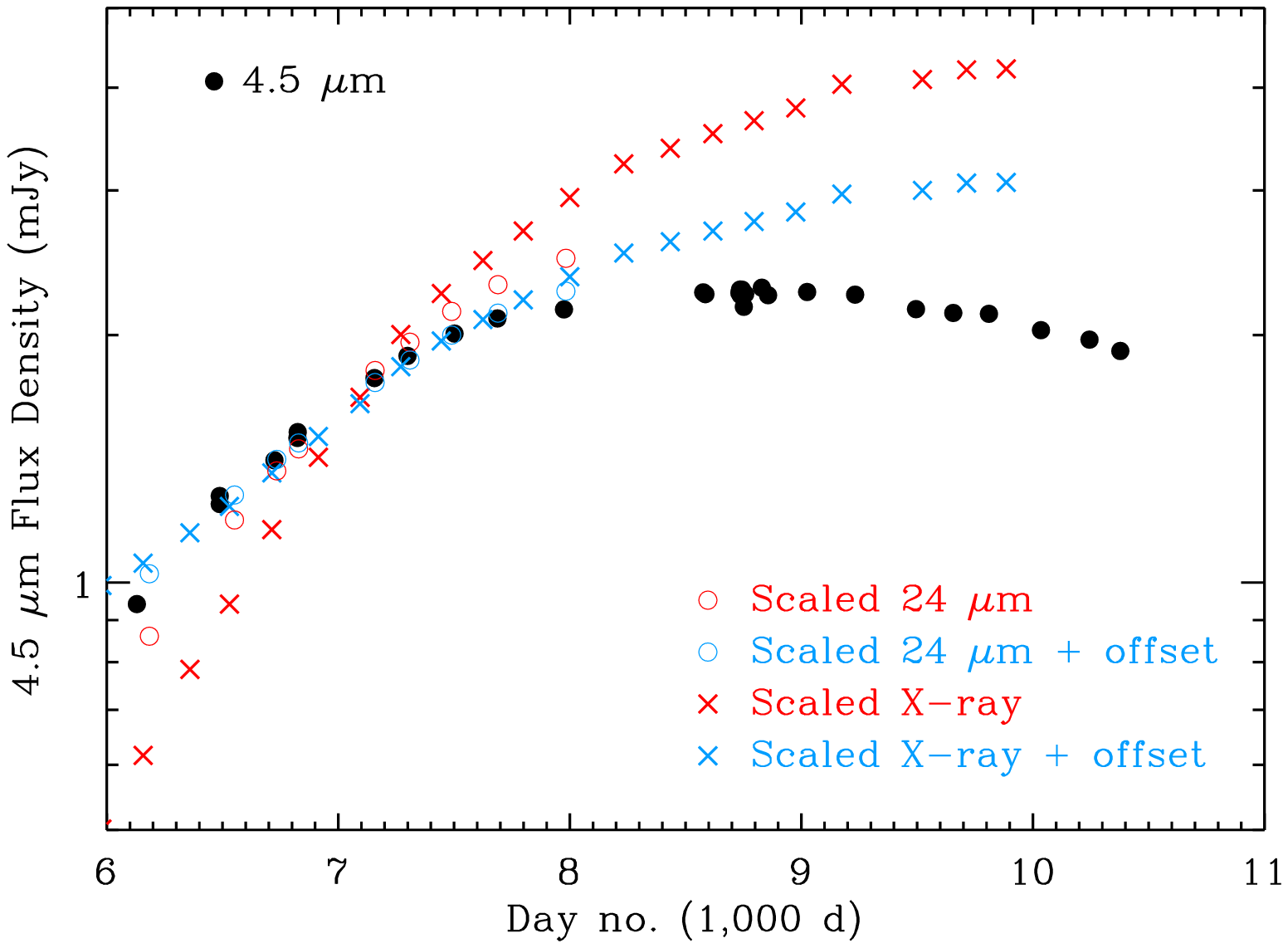}\\ 
   \includegraphics[width=3.in]{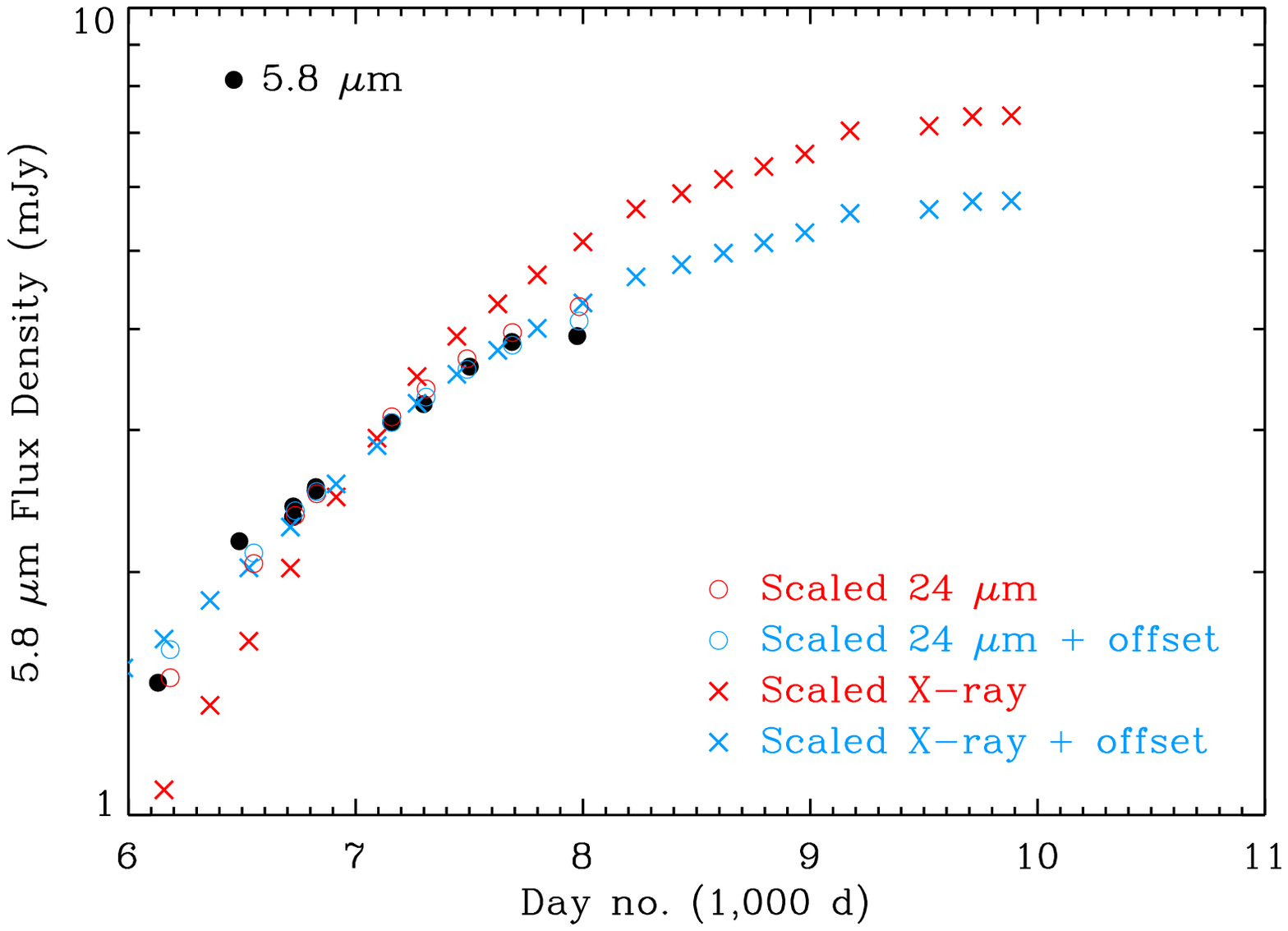} 
   \includegraphics[width=3.in]{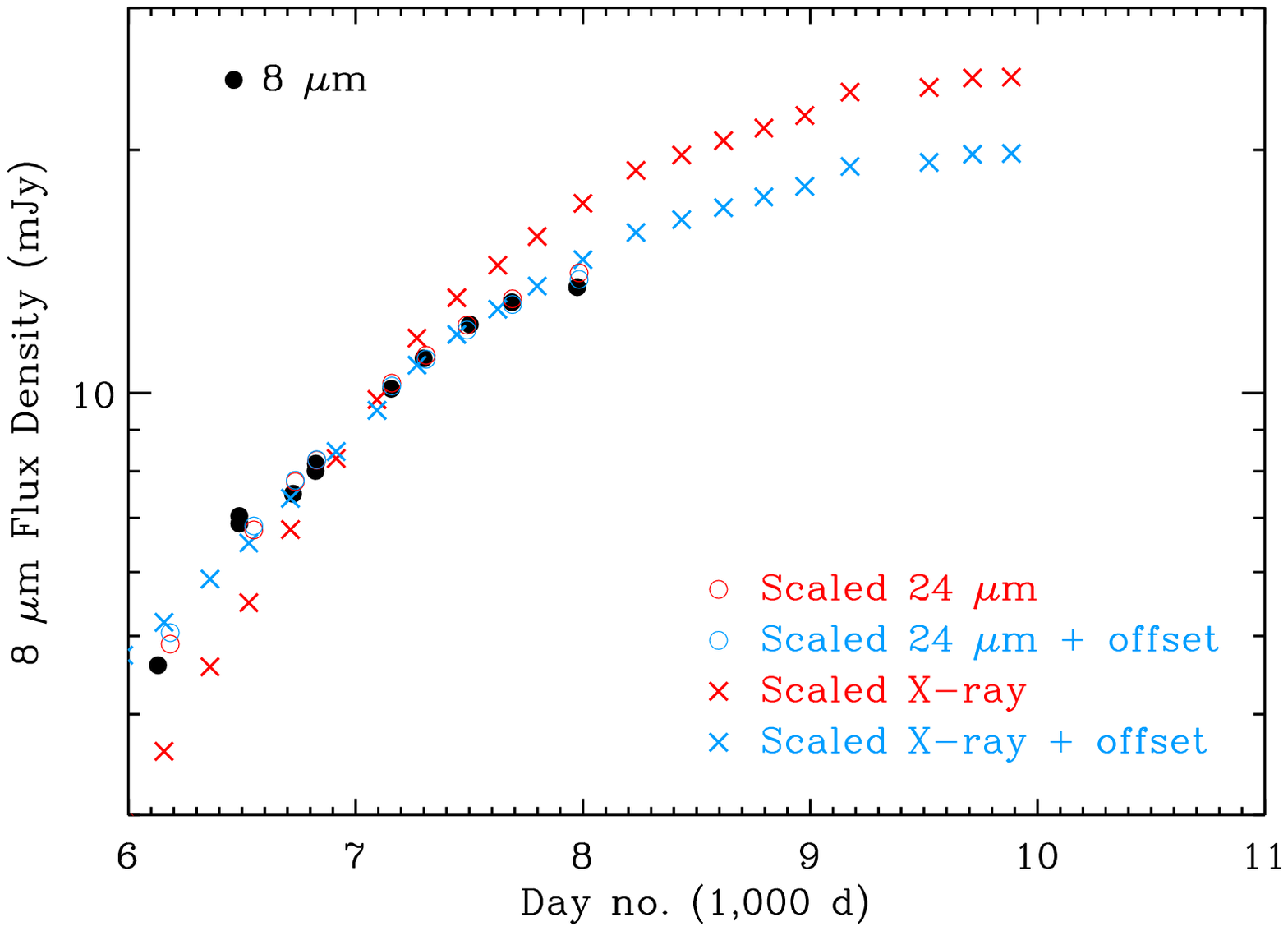}\\ 
   \includegraphics[width=3.in]{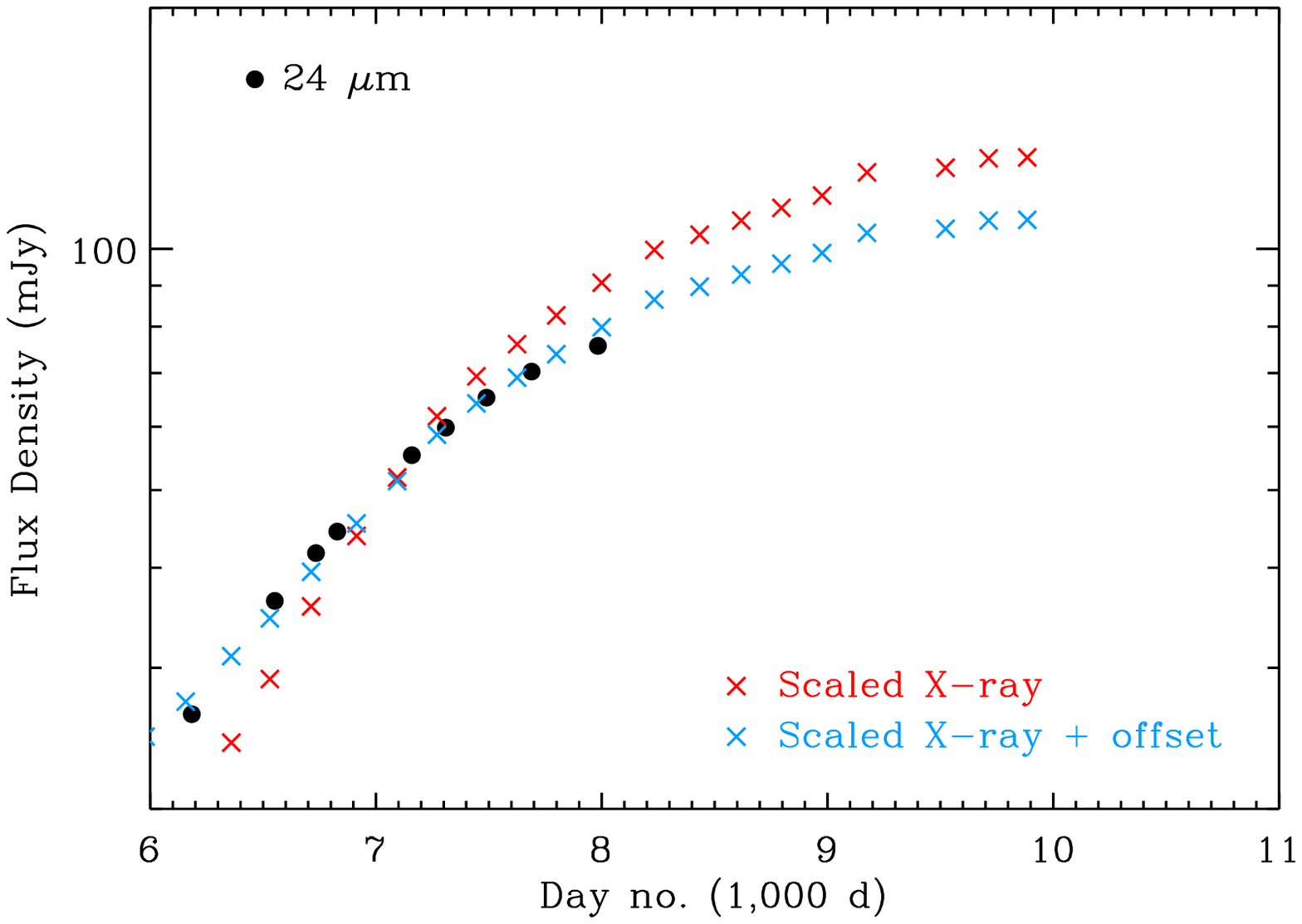} 
   \caption{The evolution of the flux density at 3.6 -- 8 $\mu$m 
   (solid black circles) is 
   compared to that at 24 $\mu$m (open circles) and in the soft X-rays 
   ($\times$). In the bottom panel, the 24 $\mu$m 
   evolution (solid black circles) is similarly compared to 
   the soft X-ray evolution. The scaling of the reference emission 
   is determined both directly:
   $S(\lambda,t) = a(\lambda)\ S(X,t)$ (red symbols), and with the presence of 
   a constant offset: 
   $S(\lambda,t) = a(\lambda)\ S(X,t) + b(\lambda)$ (blue symbols).
   In all cases, only the data from the 
   cryogenic mission (earlier than day 8,000) are used to 
   determine the scale parameters.
   Prior to day 8,000, the IR emission can follow the X-ray emission 
   closely {\it if an additional constant term is present}. 
   However, at short wavelengths (3.6 and 4.5~$\mu$m) the IR emission 
   deviates after day 8,000.}
   \label{fig:scaled_evolution}
\end{figure*}

A strong correlation between the evolution of the X-ray and IR luminosity would 
be an indication that the mid-IR emission of SN 1987A is from 
collisionally heated dust in the ER, as both the IR and X-ray emission 
should be proportional to the mass of shocked material.
Figure~\ref{fig:scaled_evolution} shows the evolution of flux density,
$S(\lambda)$,
at each IR wavelength, $\lambda$,
compared to scaled versions of the soft X-ray emission, $S(X)$:
\begin{equation}
S(\lambda, t) = a(\lambda) S(X, t).
\label{eq:ireqax}
\end{equation}
The empirical fitting parameter $a$ corresponds to a 
constant IRX ratio \citep{dwek:1987,dwek:1992}: 
\begin{equation}
IRX = n_{dust}\Lambda_{dust}(T_{gas})/n_H\Lambda_{gas}(T_{gas}),
\label{eq:irxdef}
\end{equation} 
where $T_{gas}$ is the gas temperature, $\Lambda_{dust}(T_{gas})$ and 
$\Lambda_{gas}(T_{gas})$ are the in-band cooling 
functions of the dust and gas,
$n_{dust}$ is the number density of dust grains, and  
$n_H$ is the number density of hydrogen. This is for collisional 
heating of the dust. For radiative heating, the dust cooling function will 
depend on the radiation field rather than the gas temperature.
The relative importance of radiative vs. collisional heating
of dust in supernova remnants has been discussed by \cite{arendt:1999}
and \cite{andersen:2011}, and in more general circumstances by 
\cite{bocchio:2013}.
In all cases we only use data from days 6,000 to 8,000 to determine 
the parameter $a$. The fit is generally not very good, 
indicating that the simplest model, given by Equation (\ref{eq:ireqax}), fails. 
However, the fit is significantly improved if the model
allows a constant flux density offset, $b$:
\begin{equation}
S(\lambda, t) = a(\lambda) S(X, t) + b(\lambda)
\label{eq:ireqaxb}
\end{equation}
as shown in Figure~\ref{fig:scaled_evolution}.
A non-zero value for the parameter $b$ could indicate 
a systematic error in the measurement of the flux densities
of SN 1987A, e.g. error in the subtraction of the contribution 
of nearby Stars 2 and 3. The parameter $b$ would also 
account for a component of the SN's IR or X-ray emission that 
doesn't exhibit the same correlated temporal variation as the bulk
of the emission.

Using the 24 $\mu$m emission as the reference instead of the X-ray emission
provides a check on whether the IR emission is evolving
in a consistent manner across all wavelengths. 
This comparison is also shown in Figure~\ref{fig:scaled_evolution}.
As with the X-ray emission, the shorter wavelengths are consistent 
with the 24 $\mu$m emission only if a constant emission component 
is present at each IR wavelength. The 
fit parameters ($a$, $b$) required here are consistent with those derived from 
the X-ray comparison given by Equation (\ref{eq:ireqaxb}). 

At 3.6 and 4.5 $\mu$m, the comparison to X-ray emission 
can be extended over for the longer 
time intervals from day 6,000 to day 11,000. In these cases, 
the required constant term is increased such that it accounts for 
nearly all of the emission at day 6,000, and the ratio of the evolving
IR to X-ray emission drops correspondingly. Consistent 
results are obtained when the 3.6 $\mu$m flux density 
evolution is compared directly to that at 4.5 $\mu$m.


\subsection{IRX Evolution}
The empirical models described in the preceding subsection assume
that the ratio of IR to X-ray 
emission should remain constant as the brightness evolves. 
However, a changing value of IRX could eliminate the 
need for a strong non-evolving IR emission component 
in the comparison of IR and X-ray emission. The IRX would 
be expected to change if the physical conditions (temperature,
density, or dust-to-gas mass ratio) in the shocked gas change over time.
The ratio of hard to soft X-ray emission has remained nearly
constant since day $\sim6,000$ \citep[see Table 1 of][]{helder:2013},
therefore a decreasing IRX may indicate ongoing grain destruction
and evolution of the dust to gas mass ratio $n_{dust}/n_{gas}$.
Apart from the explicit dependence of the IRX on the dust to 
gas mass ratio and the gas temperature, see Equation (\ref{eq:irxdef}), 
its numerical value will also depend on the dust grain size distribution,
and on the abundances and ionization state of the gas.

A simple model that would allow divergent but related evolution of
the IR and X-ray emission can be written as:
\begin{equation}
S(\lambda, t) = \alpha\ S(X,t) (t-t_0)^\beta
\label{eq:ireqaxtb}
\end{equation}
where $t_0$ represents the time at which the forward shock began
interaction with the ER, $\alpha$ is the nominal IRX at that time,
and $\beta$ characterizes the evolution of IRX over time. 
If $\beta = 0$ we recover the constant IRX assumption of the model 
in Equation~(\ref{eq:ireqax}).
Assuming $t_0$, we can solve for the free parameters $\alpha$ and $\beta$
by performing a linear fit to: 
\begin{equation}
\log{[S(\lambda, t)/S(X,t)]} = \beta \log{(t-t_0)} + \alpha.
\label{eq:irxeqbta}
\end{equation}
For the choice of $t_0 = 4,600$,
we derive $\beta = -1.0$ for both 3.6 and 4.5 $\mu$m data. However, at this
same $t_0$, $\beta$ increases with wavelength, reaching $-0.6$ for the 
24~$\mu$m data. Because there is a strong correlation between $t_0$ and $\beta$ 
in this model, we can force $\beta = -1.0$ for each wavelength by 
making $t_0(\lambda)$ a function of wavelength, e.g. $t_0(24\ \micron) = 3,270$.

The value of $\beta = -1$ is of particular interest. If the 
volume of the X-ray emitting gas is accumulating as a 
power law function of $t-t_0$, integrated from $t_0$ to $t_1$, but the 
volume of the IR emitting region is restricted to a layer representing 
a finite duration behind the shock front, i.e an integral from 
$t_1-\Delta t$ to $t_1$, then the IRX should evolve as $(t-t_0)^{-1}$
for $\Delta t \ll (t-t_0)$
\citep[see Fig.~7 of ][]{dwek:2010}.
The limited thickness of the IR-emitting region would arise if the dust 
destruction time scale is short compared to the cooling time of the gas.

\begin{figure*}[t] 
   \centering
   \includegraphics[width=3.in]{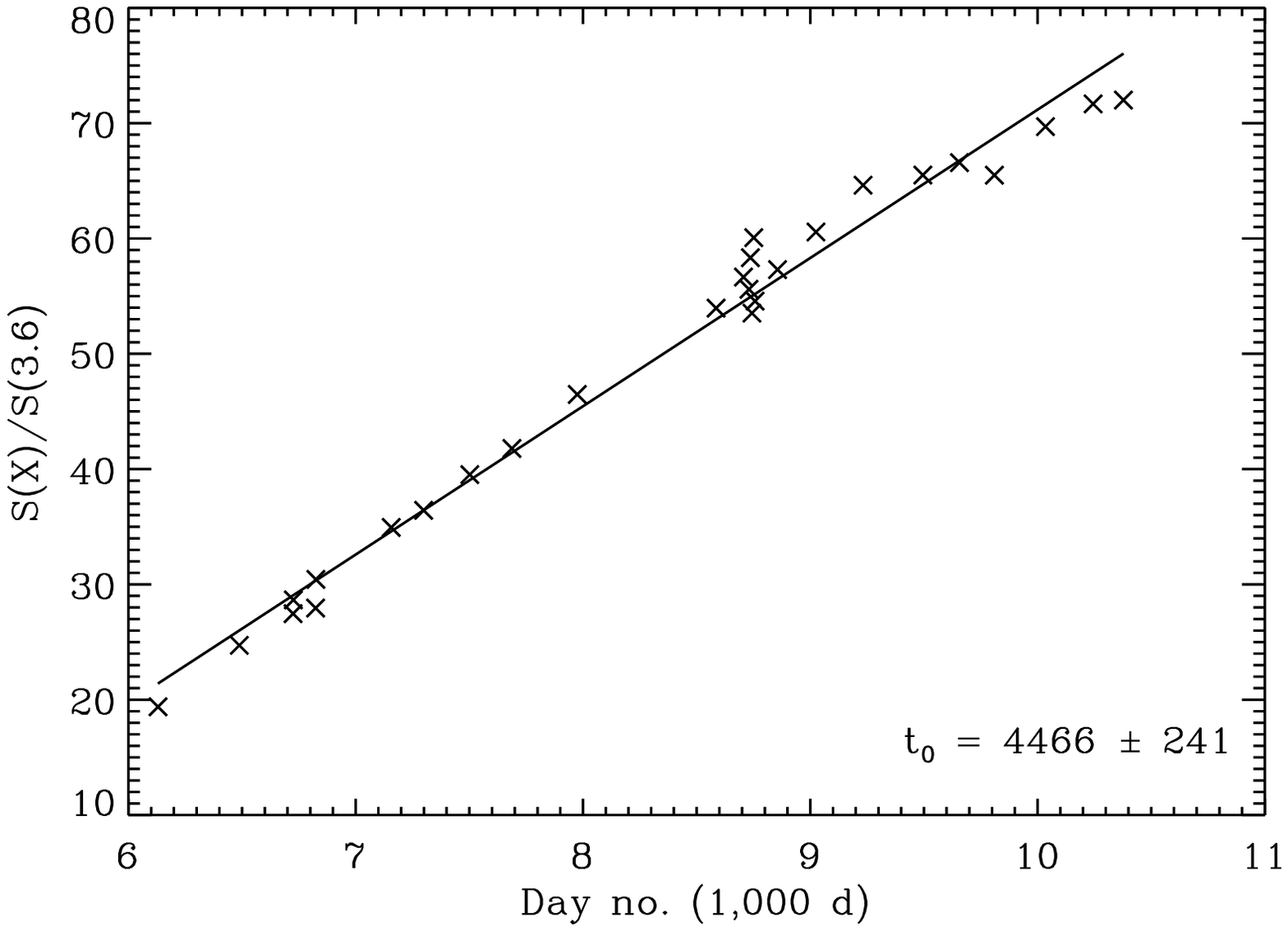}
   \includegraphics[width=3.in]{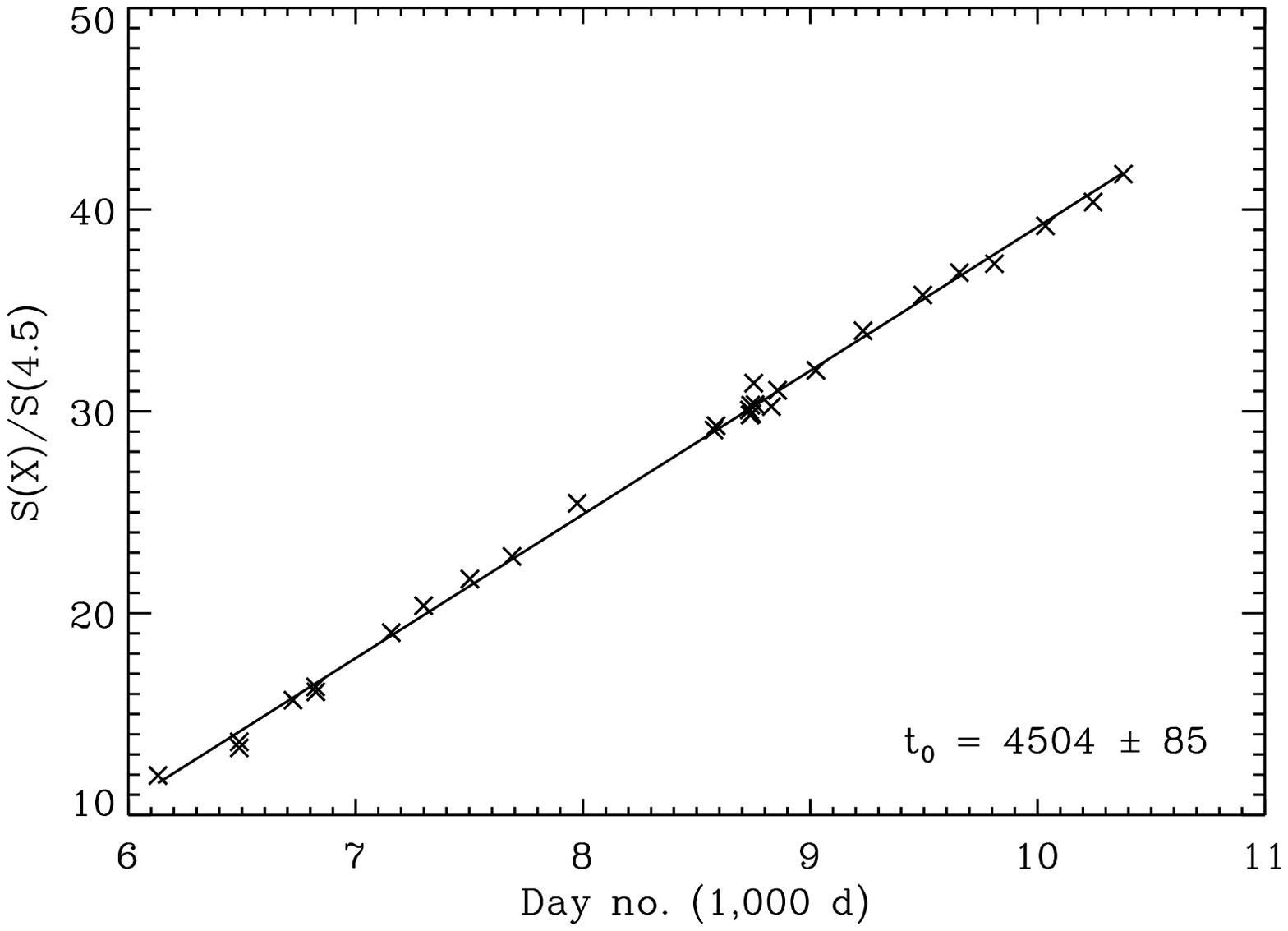}\\
   \includegraphics[width=3.in]{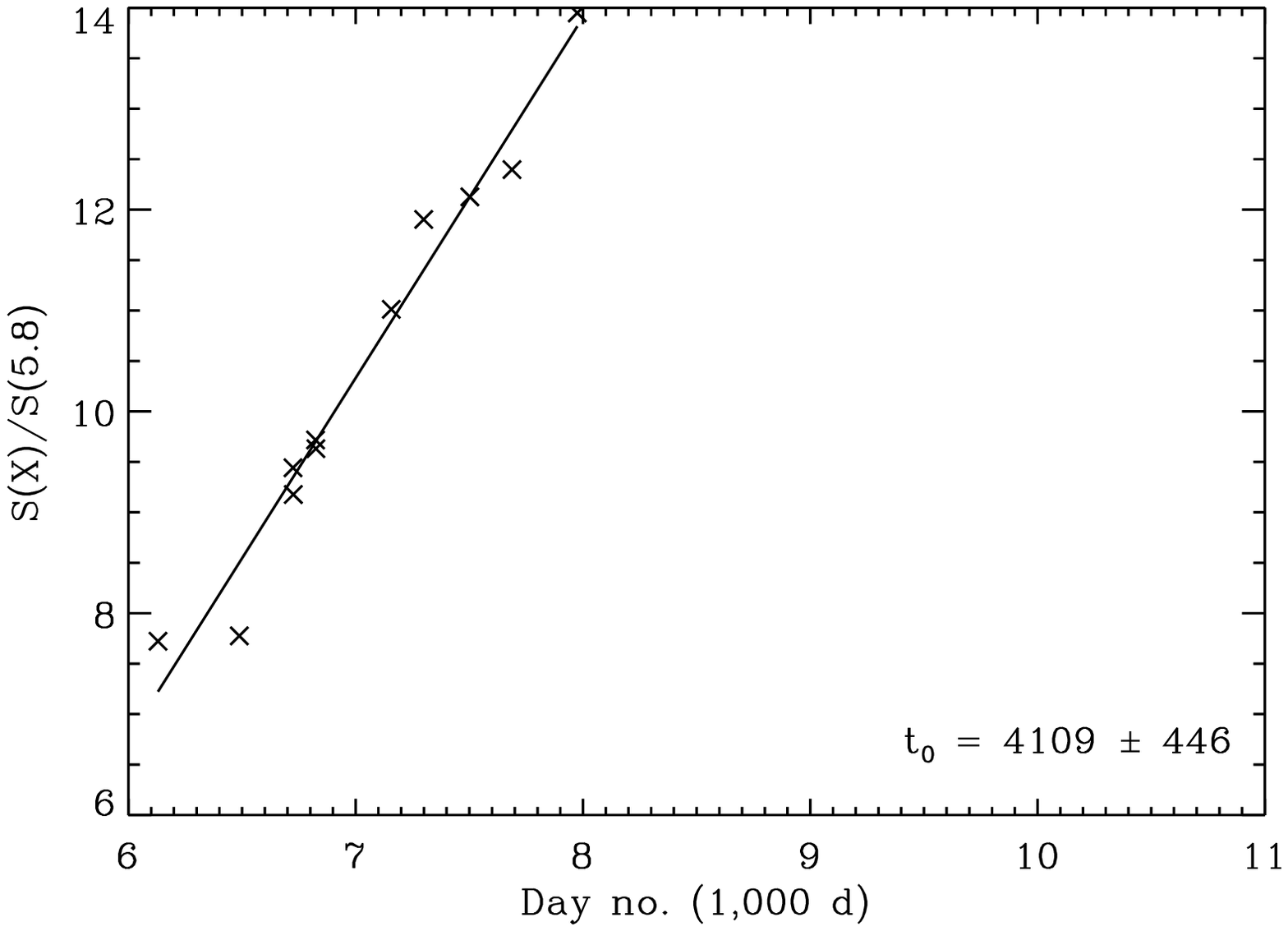}
   \includegraphics[width=3.in]{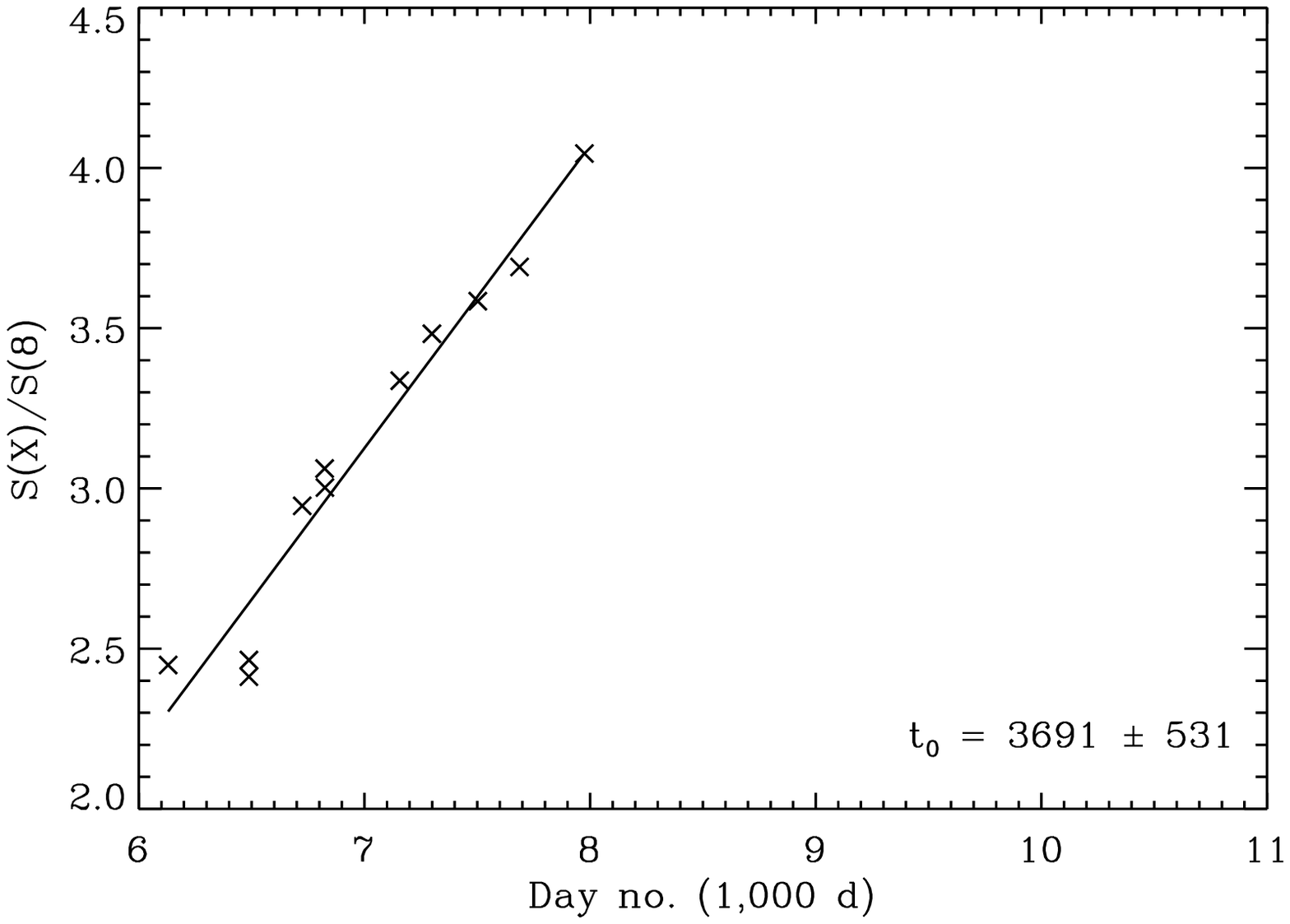}\\
   \includegraphics[width=3.in]{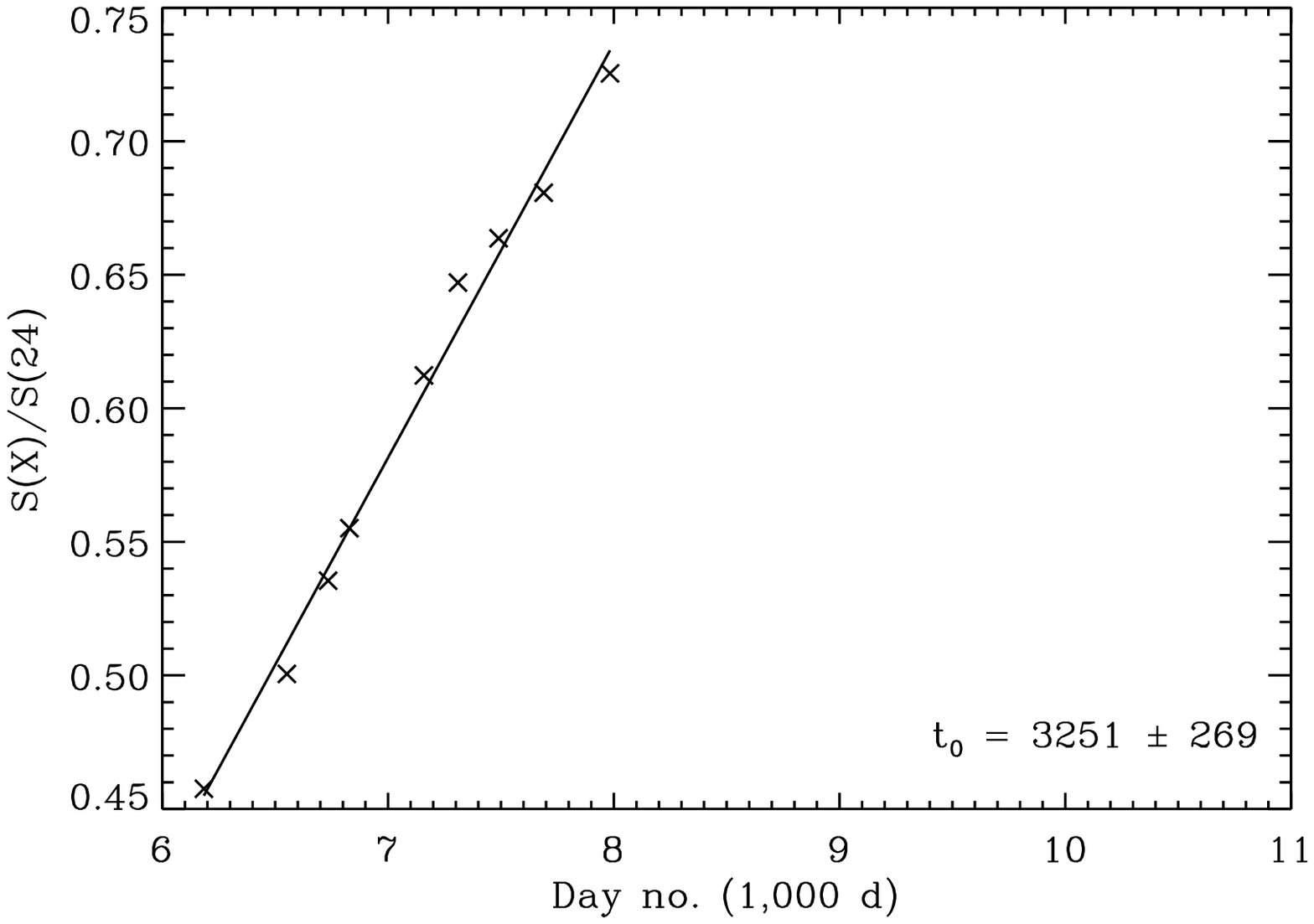}
   \caption{Correlation between $t$ and $S(X,t)/S(\lambda,t)$. The $x$-axis intercepts
   of these linear fits correspond to $t_0$ in Equation ($\ref{eq:t0fit}$).
   Averaged over all 5 wavelengths, $\langle t_0 \rangle = 4,376\pm78$~d.}
   \label{fig:t0fits}
\end{figure*}

\begin{figure}[t] 
   \centering
   \includegraphics[width=3.5in]{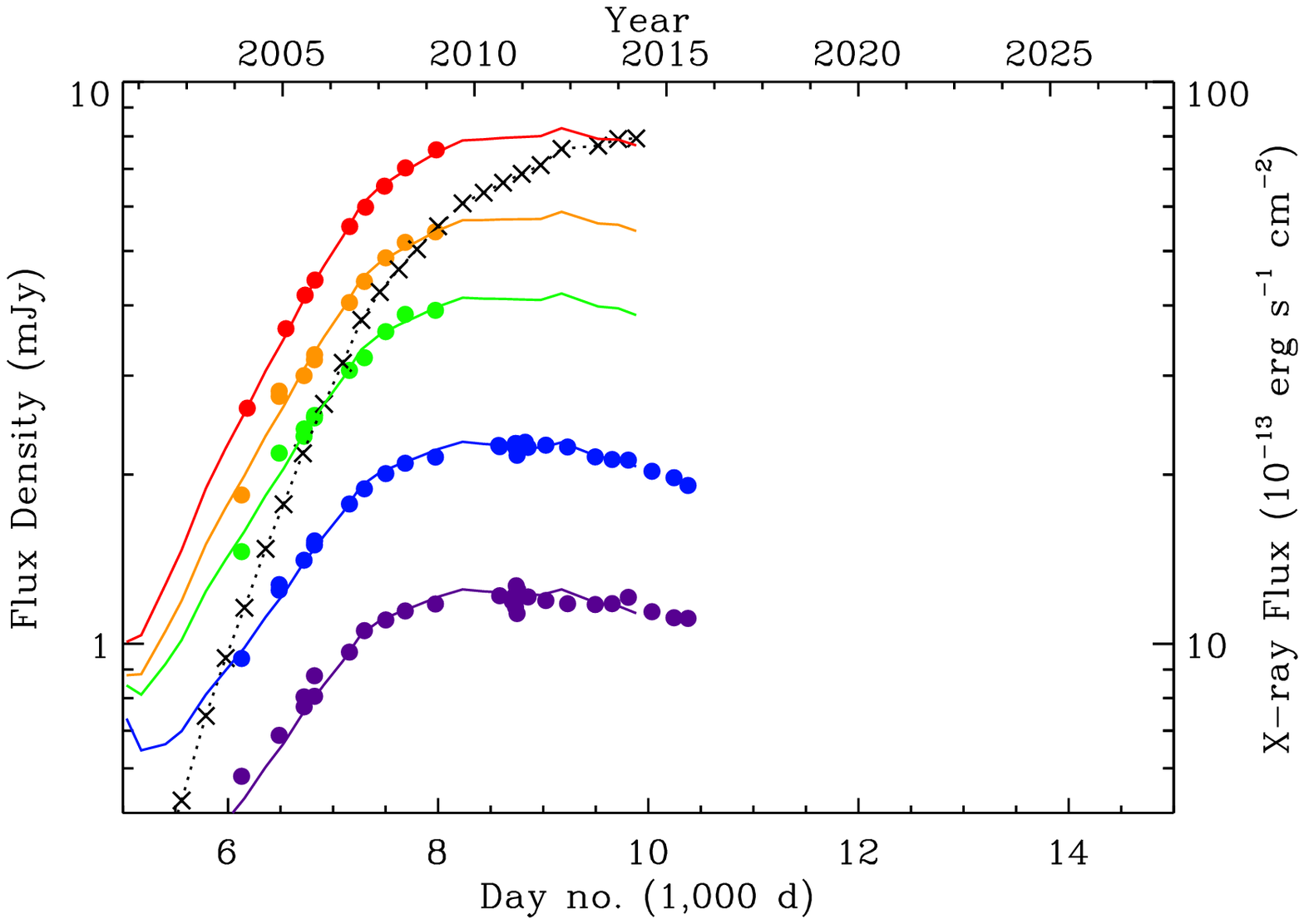}
   \caption{Observed IR flux densities ($\bullet$) can be well fit by 
   models of the IR emission in the form of 
   $S(\lambda,t) = \alpha S(X,t)/[t-t_0(\lambda)]$ 
   (colored lines, coded as in Figure \ref{fig:evolution}). 
   The observed X-ray emission is indicated by the black $\times$ symbols.}
   \label{fig:finalmodel}
\end{figure}

Therefore if we make the assumption that $\beta = -1$, then we can 
 derive $t_0$ and $\alpha$ via a linear fit to:
\begin{equation}
S(X,t)/S(\lambda, t) = \alpha^{-1} (t - t_0).
\label{eq:t0fit}
\end{equation}
Such fits are shown in Figure~\ref{fig:t0fits}. 
If we use this prescription to transform the observed X-ray fluxes
to IR flux densities, then the agreement is very good throughout the 
span of the IR observations (Figure~\ref{fig:finalmodel}). 
Extrapolation back to earlier times will fail if $t-t_0$ is less than the 
dust destruction timescale.
The extrapolation of the 24~$\mu$m emission
to later times may be verified in the future with SOFIA \citep{young:2012}
or {\it JWST} \citep{gardner:2006}.

\subsection{Dust Mass and Temperature Evolution}

Comparison of the IR and X-ray emission in the previous subsections
indicates that the IRX does not remain constant. The changes
may indicate that the decay timescale of the IR emission is 
much shorter than for the X-ray emission. Decreasing IR emission 
can occur if the dust cools or is destroyed. 

The following models specifically examine whether the 
IRAC and MIPS data indicate evolution of dust mass, 
dust temperature, or both.
These models use two dust compositions with independent temperatures to fit
the IRAC and MIPS SED 
of SN 1987A at 9 epochs between days 6,000 and 8,000, when observations 
at all wavelengths were obtained. Guided by the spectral models 
presented by \cite{dwek:2010}, the warm dust component
giving rise to the 8-35~$\mu$m emission
is assumed to be silicate, with a temperature of $\sim 180$~K, 
and the hot dust component 
giving rise to the 3 -- 8~$\mu$m emission is assumed to be 
amorphous carbon, with a temperature of $\sim 460$~K.
The hot dust component could alternately be  
any other composition with a generally featureless spectrum (e.g. iron),
and with temperatures that could be as much as $\sim100$~K lower
\citep{dwek:2010}. These components are derived from fits to the
{\it Spitzer} IRS low resolution spectra (5 -- 30 $\mu$m) obtained
between days 6,000 and 8,000. The warm dust is consistent with 
collisional heating by the X-ray emitting gas, and with even 
small dust grains residing at their equilibrium temperatures 
\citep{bouchet:2006,dwek:2008}. The hotter
dust can also be collisionally heated, if the grain sizes are 
sufficiently small, dependent on the unknown dust composition 
\citep{dwek:2010}. Small carbon grains can reach these high 
temperatures under thermal equilibrium, but other compositions
may require stochastic heating to reach sufficiently high temperatures
\citep{dwek:2010}. However, if transiently heated grains are 
producing the hot component, then they should radiate more strongly
at lower temperatures, producing more emission at $\lambda > 10$~$\mu$m
which would dilute the relative strength of the 10 and 20~$\mu$m silicate 
features.

The fitting is constrained at different epochs
by assuming that the temperature and dust mass evolve as 
functions of time:
\begin{equation}
S(\lambda, t) = \sum_{i=1}^2 \frac{4\kappa_i(\lambda)}{4\pi d^2} M_i(t) \pi B_\nu[T_i(t)]
\label{eq:model1}
\end{equation}
where $\kappa_i(\lambda)$ is the mass absorption coefficient of the dust, $d$ 
is the distance to the SN, $B_\nu(T) = 2h\nu^3/c^2\ 1/(e^{h\nu/kT}-1)$ 
is the Planck function.
The mass absorption coefficient for silicate is taken from \cite{draine:1984},
and for amorphous carbon is from \cite{rouleau:1991}. 
We choose power law functions of time as a simple means of 
characterizing the evolution of the mass and temperature of separate 
warm (silicate) and hot (e.g. carbon) components:
\begin{equation}
M_i(t) = M_{i,0}\ (t-t_1)^{\mu_i}
\end{equation}
and
\begin{equation}
T_i(t) = T_{i,0}\ (t-t_1)^{\theta_i}.
\end{equation}
\cite{dwek:2010} illustrate how the mass power law index, $\mu$, 
changes for idealized shocks expanding into 1-, 2-, or 3-dimensional
media, under the cases of free or Sedov expansion, and with or without
rapid destruction of dust. 
 
If the temporal evolution is measured with respect to the 
SN explosion ($t_1 = 0$) then $t-t_1$ only varies by a factor of 1.33 
from day 6,000 to 8,000, which forces a large value of $\mu_i \sim 4-5$ to match the observed 
change in brightness. It is more realistic to choose $t_1$ as the start of 
the interaction between the blast wave and the ER. Assuming that
$t_1 =$ 5,500 \citep{dwek:2010}, we find 
that $\mu_1 = 0.87$ (silicate) and $\mu_2 = 0.98$ (carbon). 
The value of $\mu_i \lesssim 1$ could indicate free expansion into a 1-dimensional
medium without grain destruction or 2-dimensional medium with grain destruction,
or Sedov expansion into a 2 dimensional medium \citep[see Figure 7 of][]{dwek:2010}. 
For both dust components $\theta_i \sim 0$, indicating very little evolution of the 
temperature.

\begin{figure}[t] 
   \centering
   \includegraphics[width=3.in]{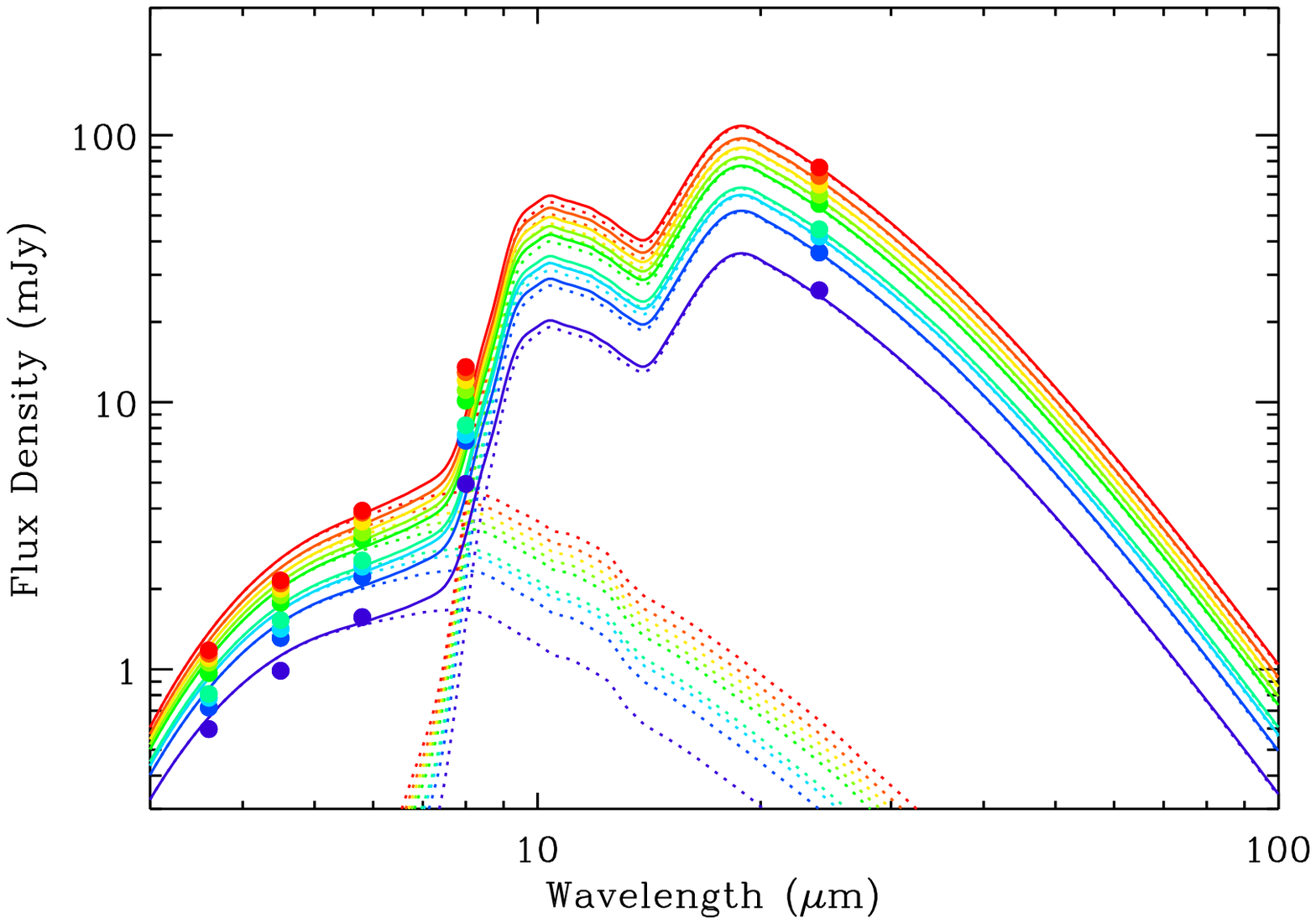} 
   \caption{Fits of a two-temperature, two-composition model to the 
   SED of SN 1987A at 9 different epochs show little apparent
   evolution of the spectrum. The warm (190~K) component is silicate; the 
   hot (525~K) component is assumed to be amorphous carbon. 
   The evolution of the mass and dust temperature
   of each component are constrained to be a power law 
   function of $t-t_1$, with $t_1 = 5,500$.}
   \label{fig:spectral_evolution}
\end{figure}

\begin{figure}[t] 
   \centering
   \includegraphics[width=3.in]{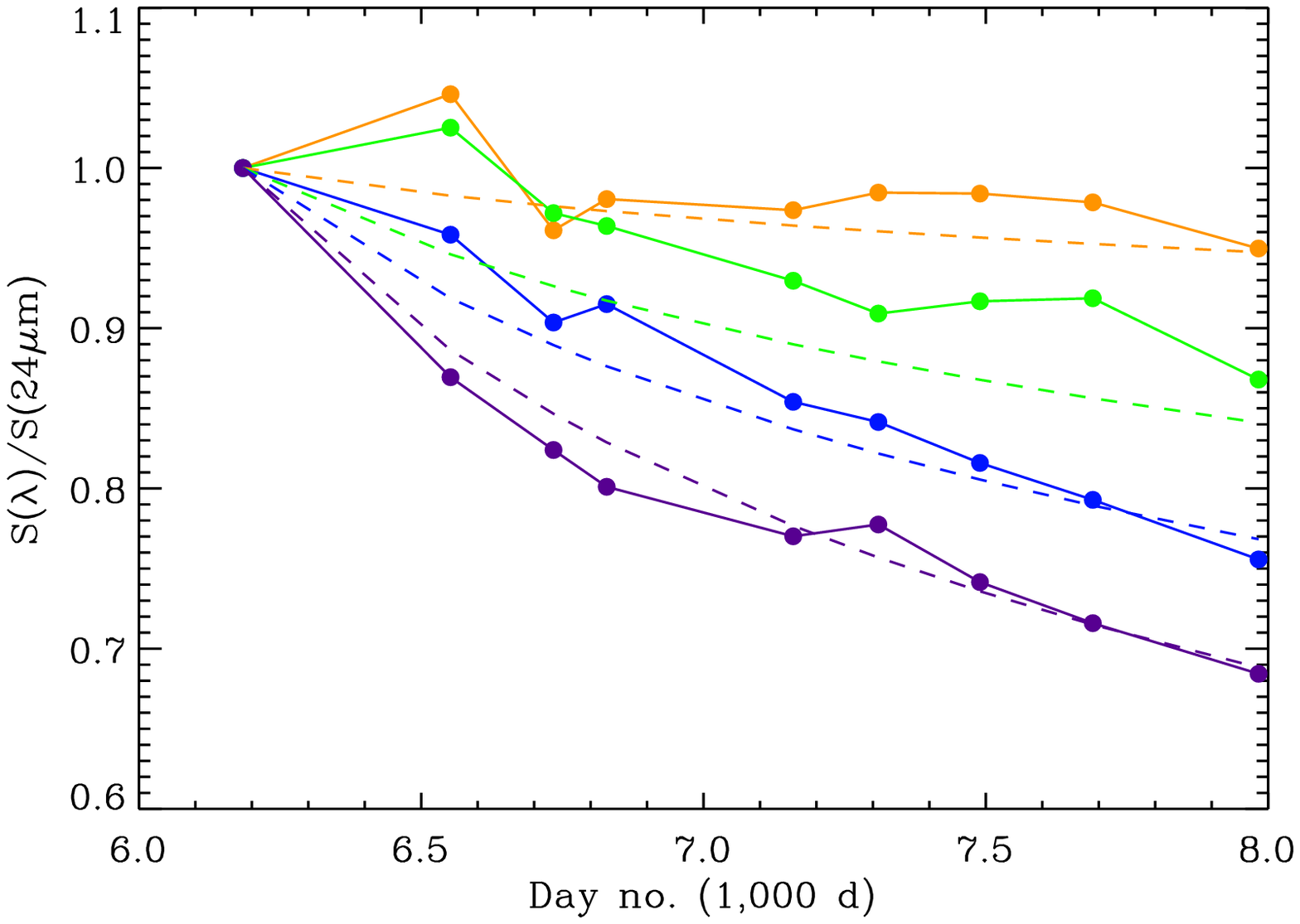} 
   \caption{Evolution of the $\lambda/24\ \micron$ colors for the
   data ($\bullet$) and the model (dashed lines). The colors are normalized with respect 
   to the color at the first epoch. Results at $\lambda = 3.6$, 4.5, 5.8, and 8
   $\mu$m are colored violet, blue, green, and orange, respectively.
   The dashed lines indicate that the model of Equation (\ref{eq:model1})
   provides a reasonable representation of the color evolution for days 
   6,000 - 8,000.}
   \label{fig:color_evolution}
\end{figure}

\begin{figure}[t] 
   \centering
   \includegraphics[width=3.in]{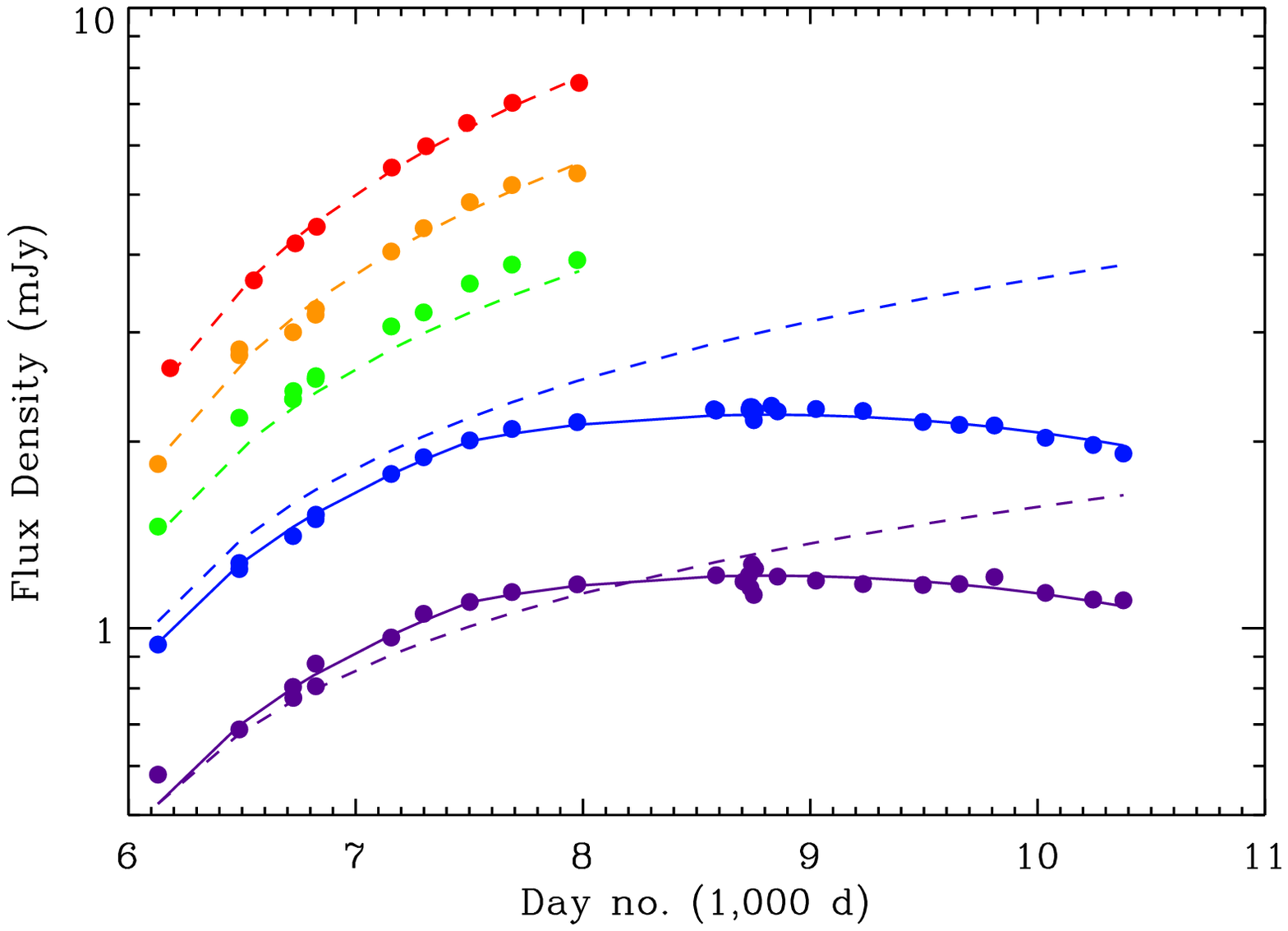} 
   \includegraphics[width=3.in]{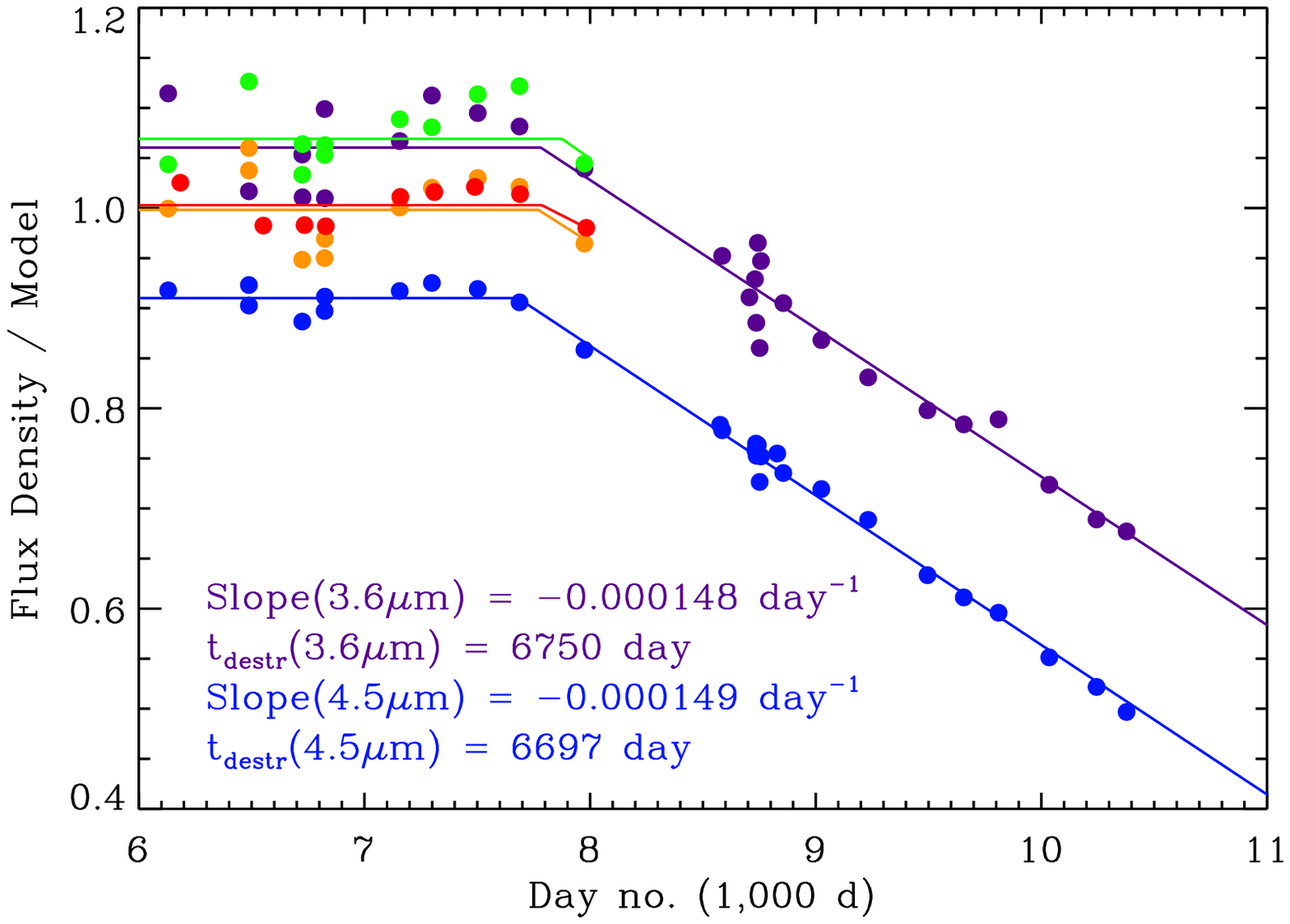} 
   \caption{The top panel shows the evolution of the flux densities for the
   data ($\bullet$) and the model of Equation (\ref{eq:model1}) 
   (dashed lines). Results at $\lambda = 3.6$, 4.5, 5.8, 8, and 24
   $\mu$m are colored violet, blue, green, orange, and red, respectively. 
   At later times the extrapolation of the models is not a good 
   fit to the 3.6 and 4.5 $\mu$m data. The bottom panel shows that the 
   ratio between this model and the data is constant at early times, but 
   appears to fall linearly after day $\sim7,500$. Modifying the model 
   with another factor representing a linear decay
   after day 7,500 [Equation (\ref{eq:model2})], provides a 
   good fit to the 3.6 and 4.5 $\mu$m data (solid lines) in both the
   flux densities (top) and the ratios (bottom).}
   \label{fig:flux_evolution}
\end{figure}

These results are shown in Figure~\ref{fig:spectral_evolution}, 
where the fitted dust temperatures are $T_{1,0} = 190$~K for the warm silicate
component, and $T_{2,0} = 525$~K for the hot amorphous carbon component.
These dust temperatures are warmer, though not significantly, than those 
derived by \cite{dwek:2010}.
Figure~\ref{fig:color_evolution} compares the changes in the $\lambda\ /\ 24$ $\micron$ colors of the
data and the model. These changes are mostly driven by variation of the relative masses
of the warm and hot components, not by changes in the dust temperatures. 
Figure~\ref{fig:flux_evolution} (top)
shows the observed and modeled evolution of
the flux densities. Over the fitted interval of days 6,000 -- 8,000, the model is 
in fair agreement with the data. However, extrapolating the model to later
times at 3.6 and 4.5 $\mu$m shows a clear deviation between the data and the model.

Comparison between the model and observations is shown in 
Figure~\ref{fig:flux_evolution} (bottom). The ratio between the model and data 
is constant until about day 7,500, after which the data fall linearly with 
respect to the model. 
Thus, we have constructed an altered model for the 3.6 and 4.5 $\mu$m data.
Only the hot carbon dust component is used, because the warm silicate dust
provides negligible emission at these wavelengths.
The dust temperature is still a free parameter, but it is held 
constant because the previous models suggested very little evolution in 
temperature. 
Starting at day 7,500, a linear decrease in the mass is applied. The 
slope of this decrease indicates a destruction timescale for the dust, 
$t_{\rm destr}^{-1}$, which is a free parameter to be derived from the data.
The form of the model is thus:
\begin{eqnarray}
\label{eq:model2}
S(\lambda, t) & = &\frac{4\kappa(\lambda)}{4\pi d^2} M_0\ (t-t_0)^{\mu} \ \pi B_\nu(T) \\
 & &  \hspace{2in} {\rm for}\ t < 7,500 \nonumber \\
 & = & \frac{4\kappa(\lambda)}{4\pi d^2} M_0\ (t-t_0)^{\mu} [1-(t-7,500)/t_{\rm destr}] \ \pi B_\nu(T) \nonumber \\
 & &  \hspace{2in} {\rm for}\ t > 7,500 \nonumber 
\end{eqnarray}
The fits for this models are shown in 
Figure~\ref{fig:flux_evolution}. The relative accuracy of this model
is $\sim3\%$ at 3.6 $\mu$m and $\sim2\%$ at 4.5 $\mu$m (shown later in Figure
\ref{fig:IRO}).

\subsection{IR -- Optical Evolution}

\begin{figure*}[t] 
   \centering
   \includegraphics[width=3.in]{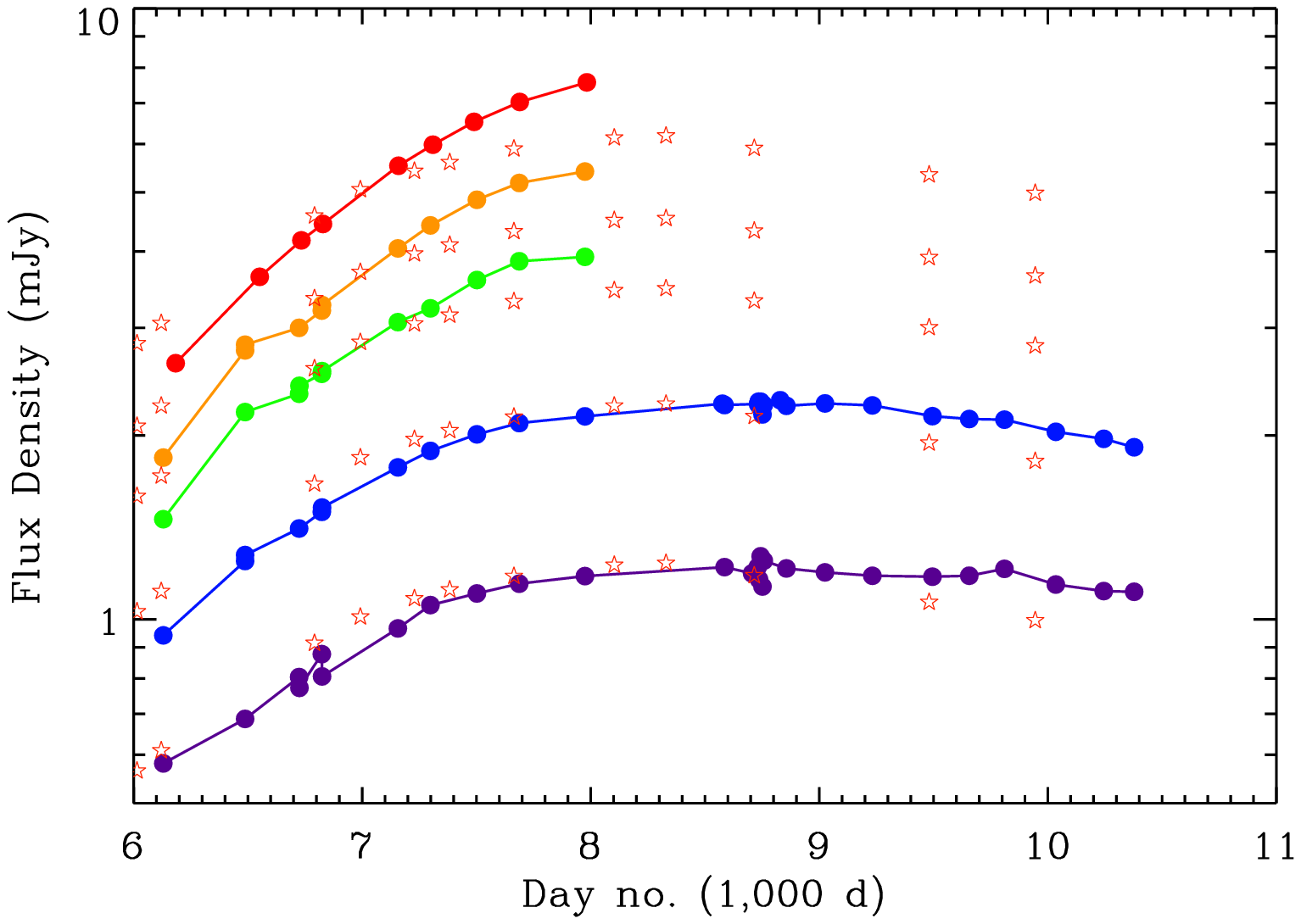}
   \includegraphics[width=3.in]{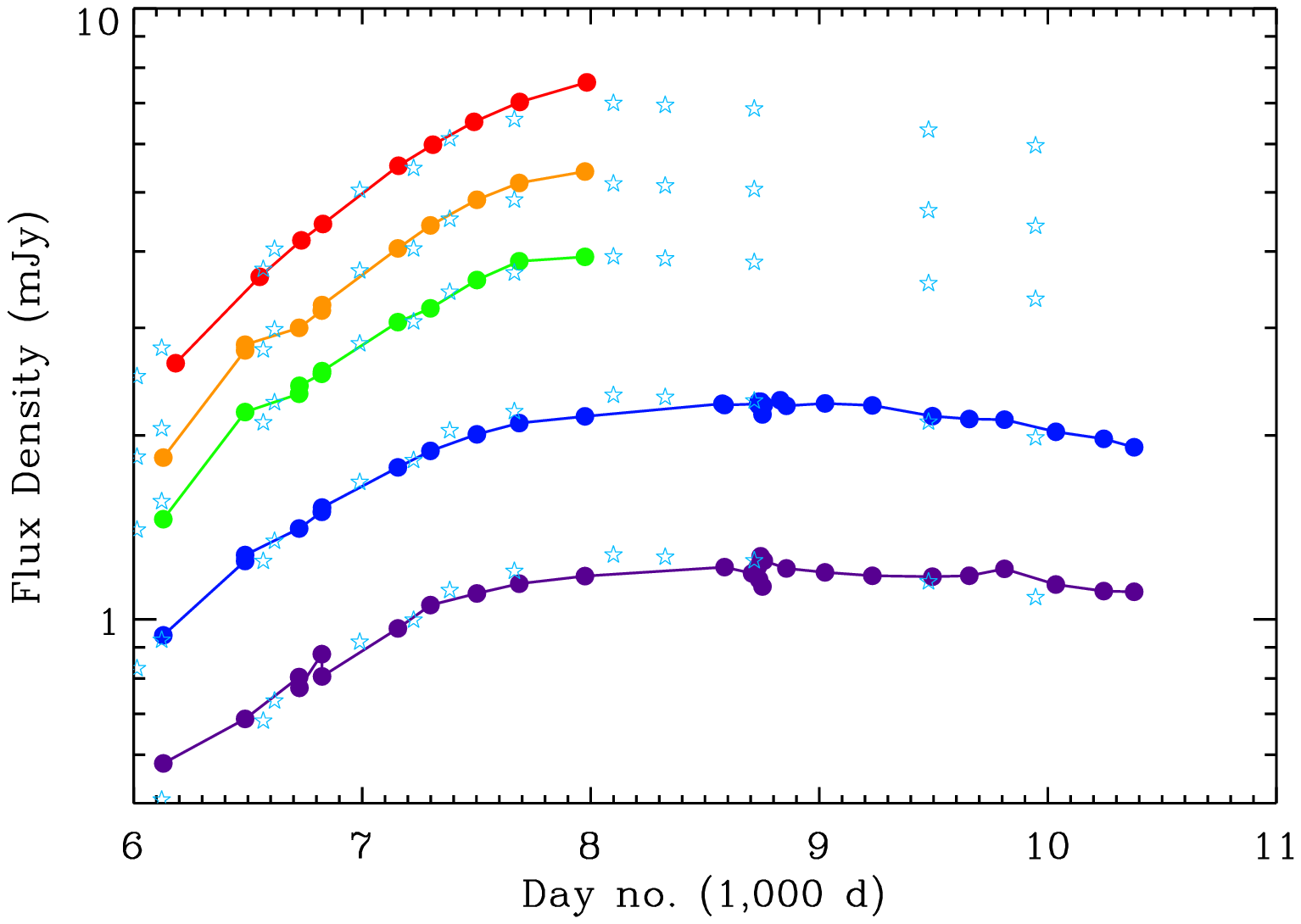} 
   \caption{IR flux density evolution ($\bullet$) in
   the 3.6, 4.5, 5.8, 8, and 24 $\mu$m bands (violet -- red, 
   scaled as in Figure \ref{fig:evolution})
   compared to the scaled optical evolution ($\bigstar$) in 
   the R (left) and B (right) bands \citep{fransson:2015}.
   The evolution of the 3.6 and 4.5 $\micron$ emission is similar to that
   of the optical emission, especially the B band.}
   \label{fig:IR-optical}
\end{figure*}

\begin{figure*}[t] 
   \centering
   \includegraphics[width=3.in]{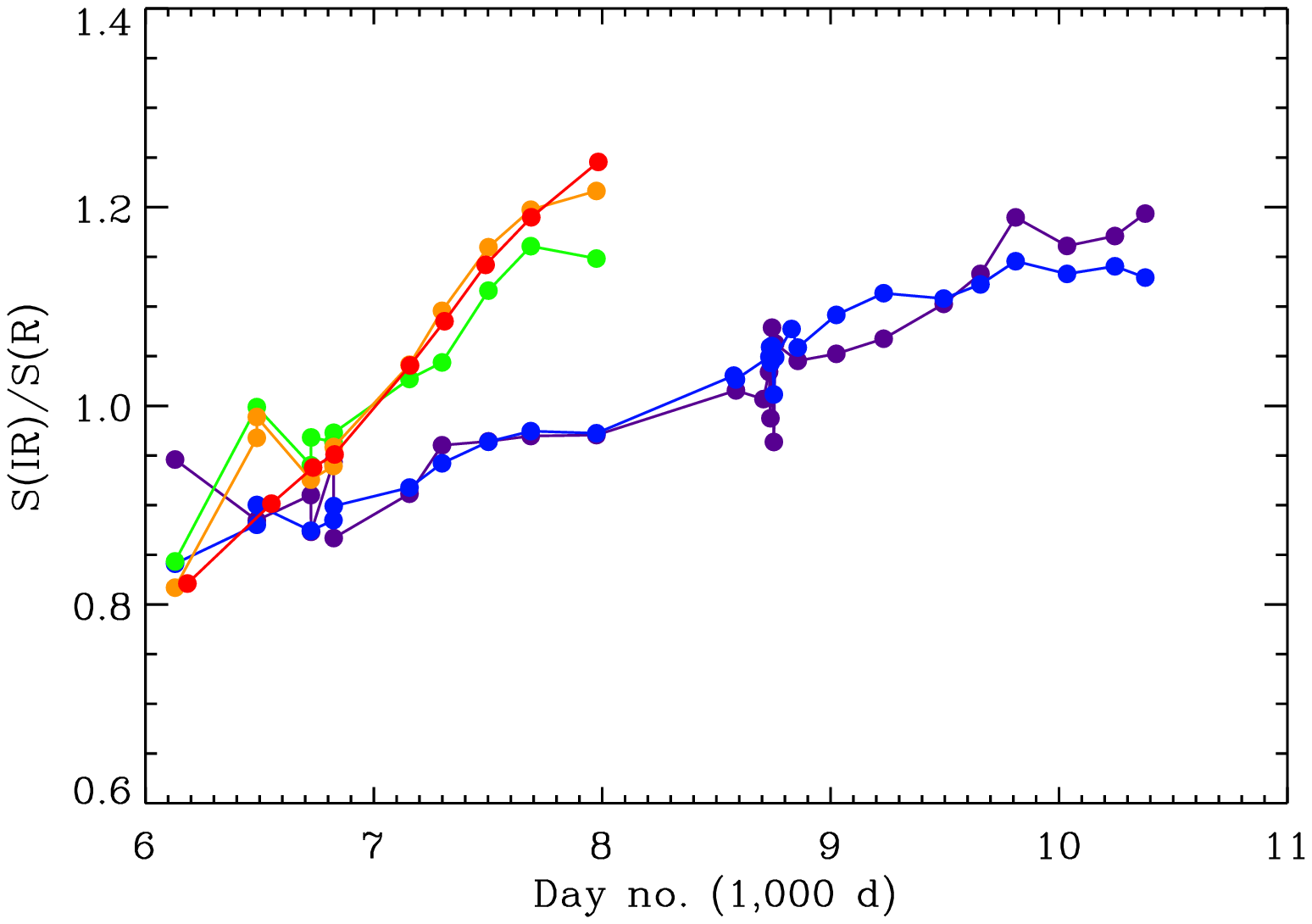}
   \includegraphics[width=3.in]{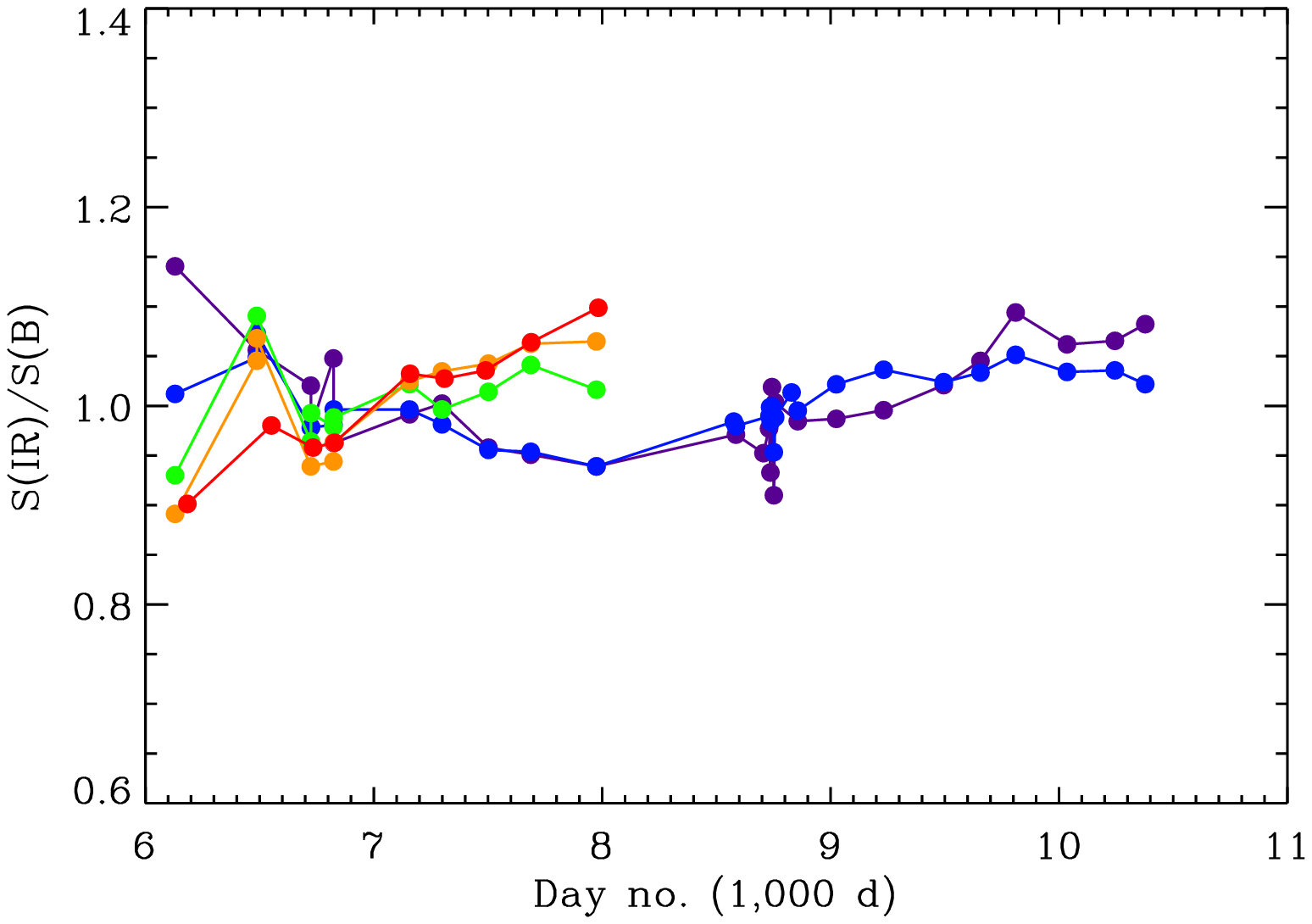}\\
   \includegraphics[width=3.in]{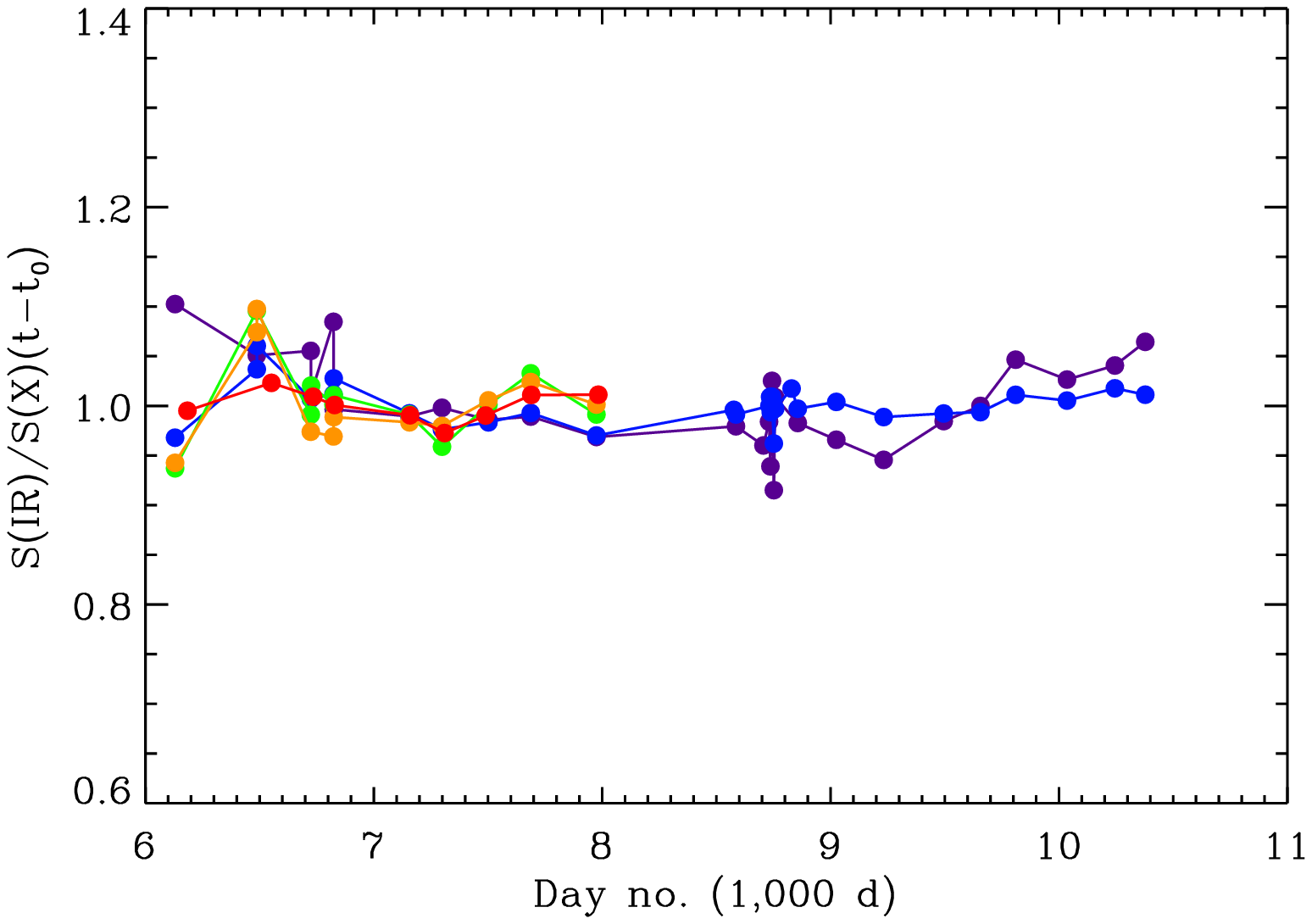} 
   \includegraphics[width=3.in]{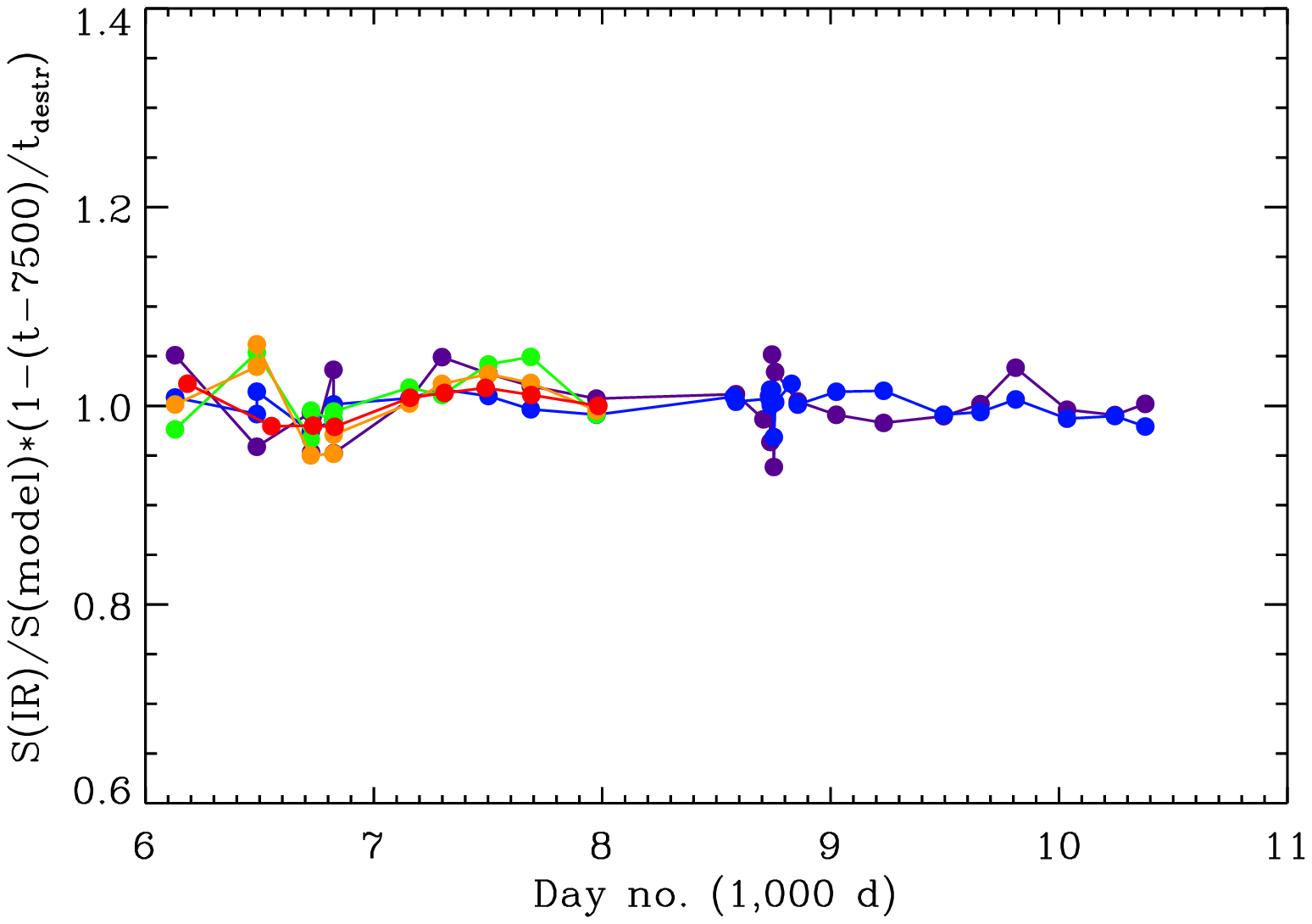}
   \caption{Ratios of the IR to R and B band emission is shown in the top 
   two panels. The ratios are relatively constant with time, especially 
   at 3.6 and 4.5 $\mu$m and with respect to the B band. The bottom left 
   panel shows
   that the IR emission is more accurately modeled as proportional
   to $S(X)/(t-t_0)$. The bottom right panel 
   shows the ratio between the IR emission
   of the model of Equation (10) with 
   a roll off after day 7,500. This model 
   provides the most accurate fit to the IR data.
   }
   \label{fig:IRO}
\end{figure*}

Figure \ref{fig:IR-optical} shows the 
evolution of the IR emission measured by {\it Spitzer} compared to 
the optical emission in the B and R bands measured by {\it HST} 
\citep{fransson:2015}. The correlations are moderately good in both cases, 
but are better for the B band, which is dominated by the emission 
lines of H$\gamma$, H$\delta$, and 
[\ion{S}{2}] 0.4069 $\mu$m \citep{fransson:2015}.
Figure \ref{fig:IRO} (upper panels)
illustrates that the application of Equation (\ref{eq:ireqax}), 
but using the optical instead of the X-ray emission, provides a 
surprisingly effective model of the IR emission, especially when
using the B band optical emission. However, the lower panels 
of the figure show that the IR emission is more closely matched
by the previously discussed functions of the X-ray emission.

These correlations between the optical and the IR emission
may indicate the IR and optical emission originate
in the same, or closely related, regions of the shocked ER. 
However, examination of the radiative luminosity of the SN 
at various wavelengths indicates that
it is unlikely the dust heating can be predominantly 
radiative. Radiation from the ER is dominated 
by UV - X-ray emission with a total 0.01 - 8 keV luminosity 
of $\sim1,460\ L_\sun$ \citep{france:2015}. The ER dust 
has an integrated IR luminosity of $\sim1,450\ L_\sun$.
Radiative heating would require that half of the ER emission
is absorbed by dust within the ring. 
The required optical depth
is inconsistent with gas phase abundances indicating
little depletion into dust \citep[e.g.][]{mattila:2010} and 
spectral analysis which requires 
only Galactic + LMC line of sight extinction corrections
\citep[e.g.][]{pun:2002,groningsson:2008b}.

\section{Evolution of the IR Emission Lines}

\subsection{Line Identification}
\begin{figure*}[ht] 
\begin{center}
  \includegraphics[width=5.5in]{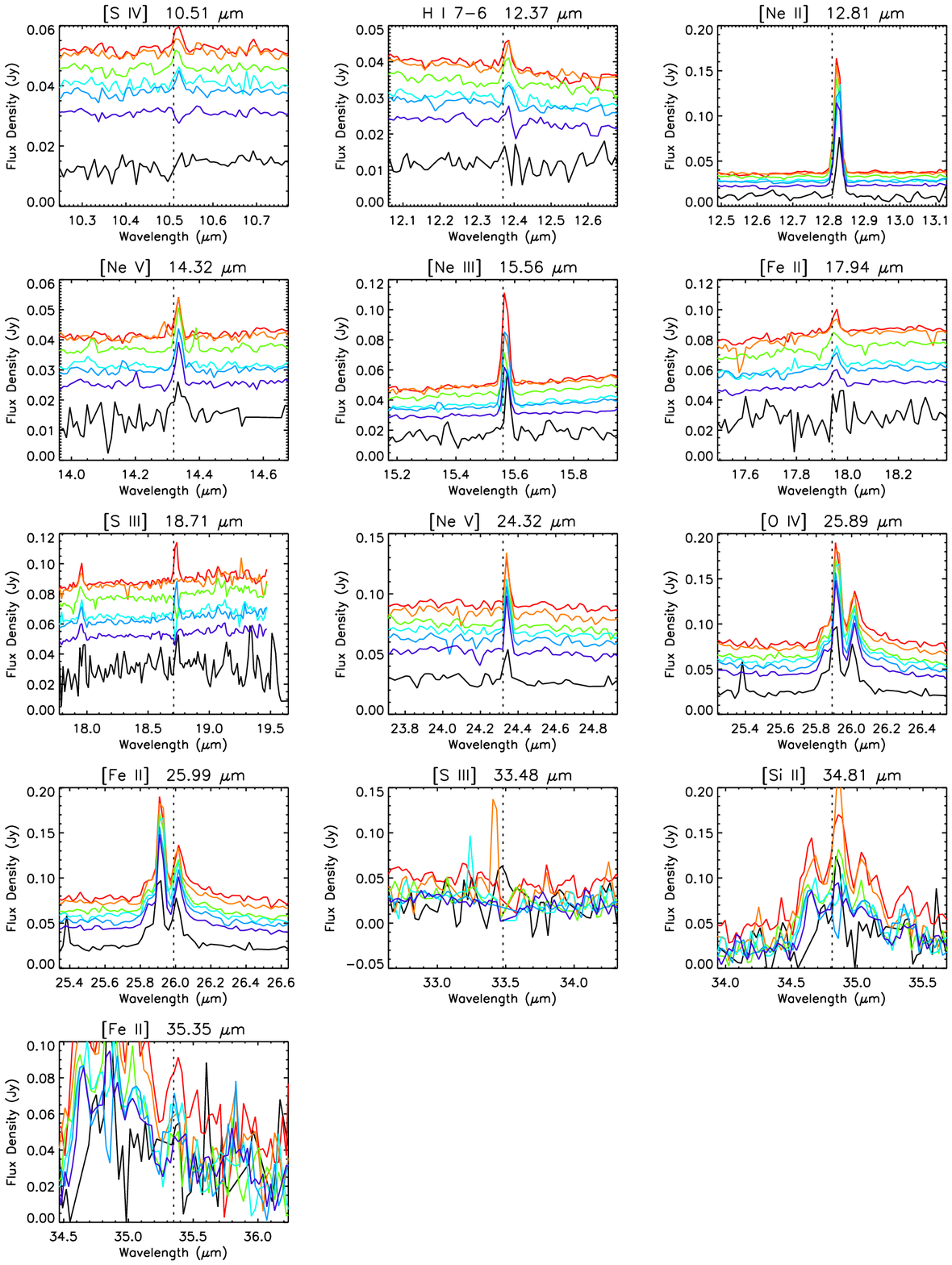}
\end{center}
  \caption{Spectral lines of SN 1987A as seen in the high spectral 
  resolution SH+LH {\it Spitzer} IRS data.
  The rising continuum levels correlate with the date of the observations. 
  The dotted lines at constant wavelength indicate the rest wavelength of each line. 
  The spectra are not adjusted for the systemic velocity of the ER, $\sim 287$ km~s$^{-1}$ \citep{groningsson:2008a}.
  For each line, the wavelength range 
  displayed corresponds to a velocity range of [-7,500, +7,500] km s$^{-1}$. }
  \label{fig:hi_lines}
\end{figure*}

\begin{figure*}[ht] 
\begin{center}
  \includegraphics[width=5.5in]{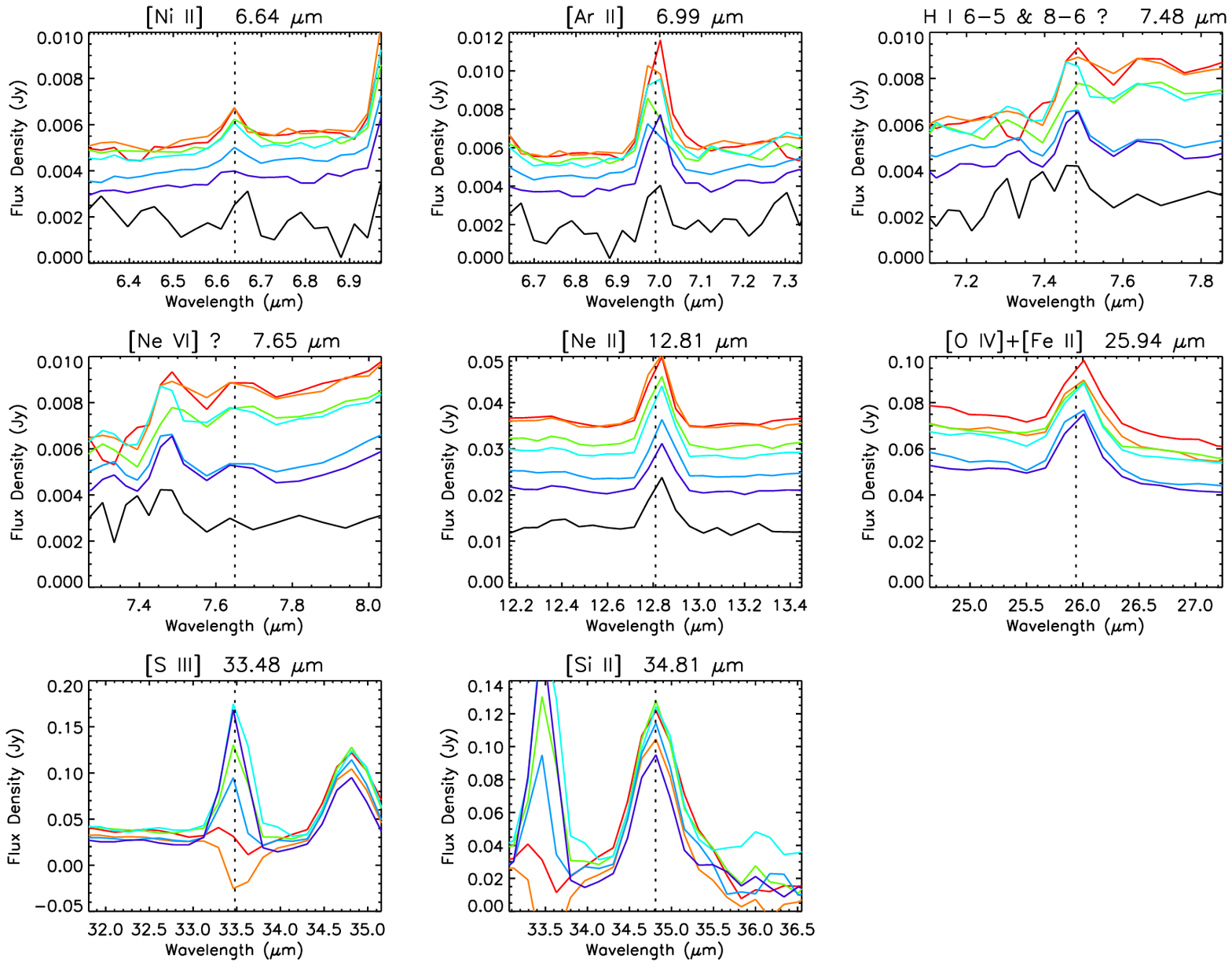}
\end{center}
  \caption{Spectral lines of SN 1987A as seen in the low spectral 
  resolution SL+LL {\it Spitzer} IRS data.
  The rising continuum levels correlate with the date of the observations. 
  The dotted lines at constant wavelength indicate the rest wavelength of each line. 
  The spectra are not adjusted for the systemic velocity of the ER, $\sim 287$ km~s$^{-1}$ \citep{groningsson:2008a}.
  For each line, the wavelength range 
  displayed corresponds to a velocity range of [-15,000, +15,000] km s$^{-1}$. }
  \label{fig:lo_lines}
\end{figure*}

The atomic spectral lines seen in the mid-IR are usually the ground state fine-structure 
transitions of various ionized species. One or more higher-lying H recombination lines 
are seen, and one excited state transition of [\ion{Fe}{2}] is detected. Cutouts of the 
lines in the SH and LH data are shown in Figure \ref{fig:hi_lines}, and lines in 
the SL and LL data are shown in Figure \ref{fig:lo_lines}.
Here we only discuss the lines, as the continuum in the SL and LL spectra 
has been analyzed by \cite{bouchet:2006} and \citep{dwek:2008,dwek:2010}. The continuum
in the SH and LH data adds no new information and is less reliably measured
as the concatenation of many spectral orders.

Nearly all of the observed emission lines can be attributed to the emission ring, 
as the widths of most lines seem to be unresolved at the $R \sim 600$ resolution of
the IRS SH and LH modules. However this spectral resolution is insufficient to 
distinguish between the ionized but unshocked component of the ring and the 
shocked component of the ring, which are clearly seen as narrow (FWHM $\sim$ 10 km s$^{-1}$) 
and intermediate (FWHM $\sim$ 300 km s$^{-1}$) width components 
in high resolution optical spectra \citep[e.g.][]{groningsson:2008a}. 

One apparent exception is the [\ion{Si}{2}] line at 34.81 $\micron$. At all epochs 
(barring the low signal to noise data of the initial observations at day 6,109),
this emission line seems to be flanked by a pair of lines at velocities of 
$\pm$ 1,700 km s$^{-1}$ with respect to the central component. The 
background spectra do not show any indication of the shifted lines.
The velocities of the flanking lines are large compared to the velocities 
associated with shocked emission ring, but fairly small compared to the velocity
of the SN ejecta. The distinctness of the flanking lines implies that they are 
not produced by an expanding spherical distribution (either a shell or a filled sphere) 
of material. Either a bi-polar or an expanding ring-shaped distribution could produce 
the flanking lines. The homunculus in the $\eta$ Car nebula is an 
example of a bipolar structure in the circumstellar medium of a
massive star. However its expansion velocity is lower, at $\sim600-800$~km~s$^{-1}$
\citep[e.g.][]{gehrz:1972,meaburn:1987}.
A ring-shaped distribution would also produce a plateau of emission
between the flanking lines, but the presence of the central line makes it difficult to 
determine whether or not a plateau exists. Furthermore, the strength of the central 
component and any underlying plateau may be affected by imperfect subtraction of 
the background [\ion{Si}{2}] emission which is strong in the vicinity of SN 1987A. 
None of the other lines show this velocity structure, with the 
possible exception of the [\ion{Fe}{2}] 25.99 $\micron$ line.

The high resolution spectra show a complicated blend of lines at 26 $\micron$. The strongest
two components are [\ion{O}{4}] 25.89 $\micron$ and [\ion{Fe}{2}] 25.99 $\micron$. Both of these identifications 
are confirmed by having velocity shifts in agreement with other unblended lines in the spectra.
Additionally, there appear to be wings on both the red and blue sides of the [\ion{O}{4}] and [\ion{Fe}{2}] pair.
The blue wing may be a separate line that is only marginally resolved from the [\ion{O}{4}] line.
Bouchet et al. (2006) had pointed out that this feature could be [\ion{F}{4}] 25.83 $\micron$
(based on the line list used by SMART\footnote{\url{http://www.ipac.caltech.edu/iso/lws/ir\_lines.html}}). 
However, we now believe the [\ion{F}{4}] identification is unlikely because (a) other line 
lists\footnote{\url{http://www.pa.uky.edu/~peter/atomic/} and\\ \url{http://physics.nist.gov/PhysRefData/ASD/index.html}}
cite a wavelength of 25.76-25.773 for this line, making the wavelength agreement worse, (b) no other
fluorine lines are observed ([\ion{F}{1}] 24.75, [\ion{F}{2}] 29.33, nor [\ion{F}{5}] 13.40 $\micron$) despite the fact
that they lie in spectral regions with low noise, and (c) there are many other elements that are 
normally much more abundant than fluorine that are not seen in the spectrum. If not [\ion{F}{4}],
then this blue wing is probably a velocity shifted component of [\ion{O}{4}] or [\ion{Fe}{2}]. If [\ion{O}{4}],
it is a velocity structure not seen in any other lines. If [\ion{Fe}{2}], it may correspond to 
the flanking lines of [\ion{Si}{2}], however there is no corresponding feature on the red side.
The red wing that is present on the 26 $\micron$ complex is much broader than the blue wing.
We presume that it represents a higher velocity distribution of [\ion{Fe}{2}]. 
The 26 $\micron$ complex was fitted iteratively for [\ion{O}{4}], [\ion{Fe}{2}], [\ion{F}{4}] (or blue wing component),
and the broad [\ion{Fe}{2}] component, subtracting each line before fitting the next.) 

The complex of [\ion{Si}{2}] and [\ion{Fe}{2}] lines at 35 $\micron$ was also 
fitted iteratively. The central [\ion{Si}{2}] line was fitted first. 
It usually has the largest width, which is possibly influenced by 
blending with the red and blue 
shifted components. In contrast, a simultaneous fit for all 
three [\ion{Si}{2}] lines often derives a much smaller width and a 
proportionally small flux for the central [\ion{Si}{2}] line. Thus the 
individual [\ion{Si}{2}] line fluxes may be uncertain by nearly a 
factor of 2 due to systematic effects, but the integrated flux of all
three [\ion{Si}{2}] lines is consistent within $5\%$.
The gaussian fits to all lines in the SH and LH spectra at all epochs are listed in Table \ref{tab:lines}.

The IRS SL module provides coverage of the $5 < \lambda < 10\ \micron$ region which is inaccessible 
with the SH module. The strongest line seen here is [\ion{Ar}{2}] 6.99 $\micron$ which is clearly seen at 
all dates. The [\ion{Ni}{2}] 6.64 $\micron$ line (the single ground state fine structure transition of this species) 
is not so clear in the earliest data, but seems to be present in the later observations. 
Additional lines may be present near 7.48 and 7.65 $\micron$. These features lie near the ends 
of the spectral orders of the IRS modules and are thus somewhat less reliable, but they do seem 
be present in multiple orders and at many or all epochs. The shorter wavelength feature
may be a blend of the \ion{H}{1} 6-5 and 8-6 transitions. This identification is 
supported by the weak detection of the \ion{H}{1} 7-6 transition in the SH data. The 7.65 $\micron$ feature
could be [\ion{Ne}{6}], which would make it the most highly ionized species that we observe. 

\subsection{Line Evolution}
\begin{figure*}[ht] 
\begin{center}
  \includegraphics[width=5.5in]{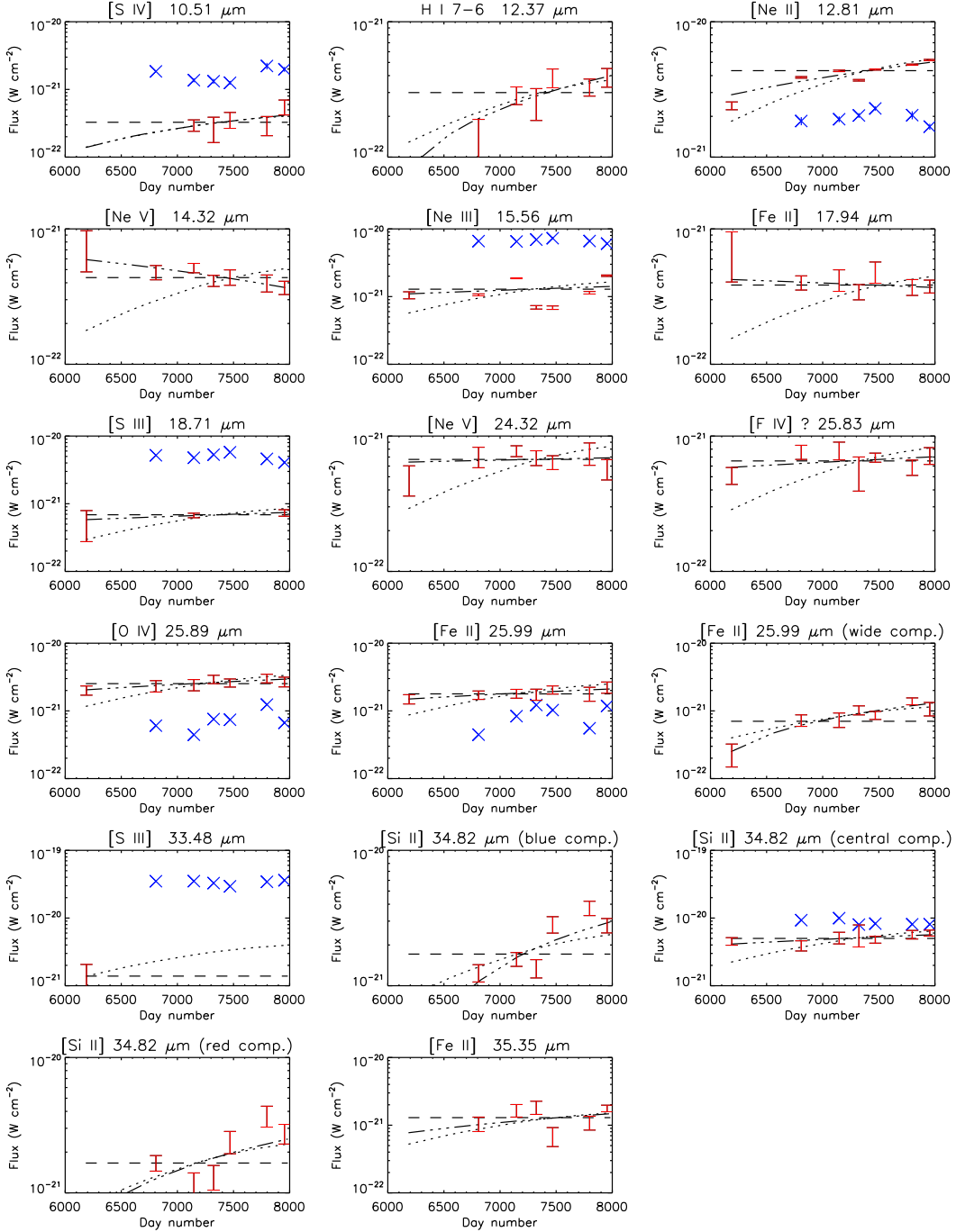}
\end{center}
  \caption{Evolution of spectral line fluxes of SN 1987A as seen in the 
  SH+LH {\it Spitzer} IRS data. Measured fluxes are indicated with 
  1-$\sigma$ error bars. 
  The $\times$ symbols denote flux measurements of the subtracted 
  background line emission, which in some cases is dominant.
  The dotted line shows the evolution of the 24 $\micron$ flux density after being
  normalized to fit the line fluxes. The dashed lines are models of constant line intensity,
  and the dash-triple-dotted lines indicate models where intensity is linear with time.}
  \label{fig:hi_evol}
\end{figure*}

\begin{figure*}[ht] 
\begin{center}
  \includegraphics[width=5.5in]{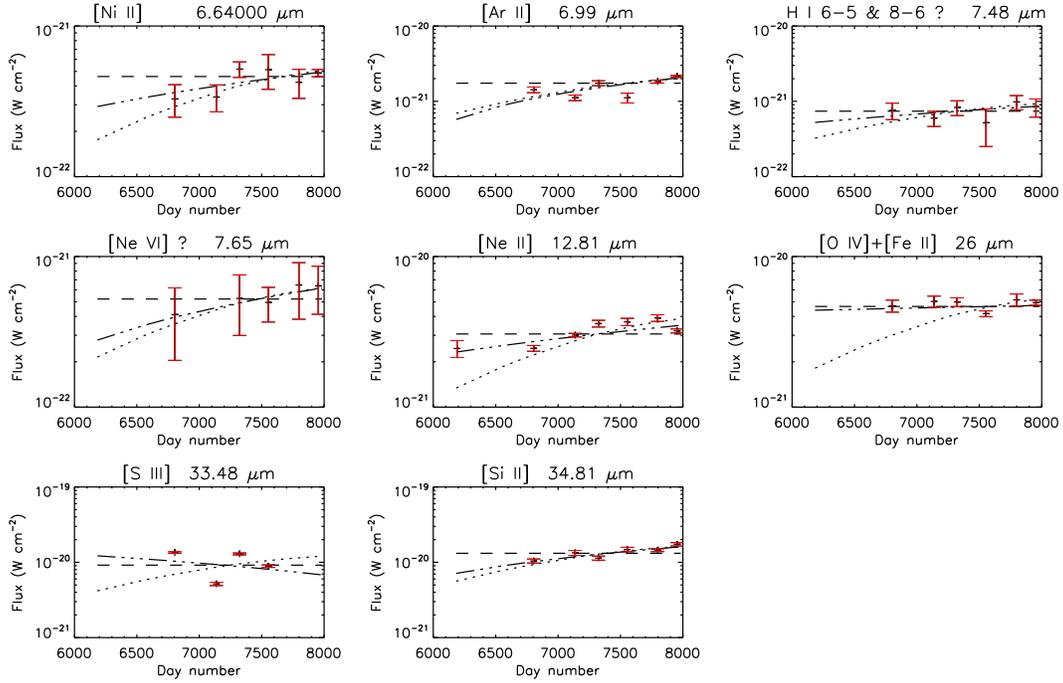}
\end{center}
  \caption{Evolution of spectral line fluxes of SN 1987A as seen in the 
  SL+LL {\it Spitzer} IRS data. Error bars show 1-$\sigma$ uncertainties. 
  The dotted line shows the evolution of the 24 $\micron$ flux density after being
  normalized to fit the line fluxes. The dashed lines are models of constant line intensity,
  and the dash-triple-dotted lines indicate models where intensity is linear with time.}
  \label{fig:lo_evol}
\end{figure*}

Figure \ref{fig:hi_evol} illustrates the changes in the lines fluxes over a 5-year period
for those lines seen in the SH and LH spectra. When present, the intensities 
of lines in the background are also plotted. This provides an indication of the 
potential systematic errors in measurements due to the different slit orientations and background 
fields at each epoch. The comparison with the background also shows that the SN 1987A line 
measurements are suspect or impossible when the background intensity is $\sim$2 or more 
times brighter then the SN (e.g. [\ion{S}{4}], [\ion{Ne}{3}], [\ion{S}{3}], and maybe [\ion{Si}{2}]).
We have fitted these light curves with three models. The first model is that the line intensity is 
proportional to the continuum intensity: $F_{line} = A\ F_{24\micron}$, and is shown as a dotted line
in each panel of the figure. The second model is that the line intensity is 
constant: $F_{line} = B$, as indicated by the dashed lines. The third model uses 2 parameters
to model the evolution as a linear function of time: $F_{line} = \beta + \alpha\ t$ (dash-triple-dot line in the figure).
Table \ref{tab:hi_evol} lists the derived parameters of the fits for each line and each model. 
The probabilities, $P(<\chi^2)$, are the probability of finding a smaller value of $\chi^2$ than the
observed $\chi^2$ under the assumption that the data are a random sampling of the model. Thus 
$P(<\chi^2) \sim 1$ indicates the data are very unlikely to be related to the given model.
Figure \ref{fig:lo_evol} and Table \ref{tab:lo_evol} provide the results for the comparable analysis 
of the low resolution spectral lines.

The formal results of the fitting procedures indicate that the evolution of the line intensities
are usually consistent with a constant value, although in several cases uncertainties are large enough 
that the intensities are also consistent with the models which increase over time.
The only lines to exhibit significantly increasing intensities are the \ion{H}{1} 7--6 and 
the broad [\ion{Fe}{2}] lines.
The rising intensities correlated with the continuum may be associated from the increasing 
volume of the shocked portion of the ER. The line intensities that are flat (or possibly declining) 
may be dominated by emission for unshocked regions of the ER, corresponding to the narrow 
components seen in the optical emission lines \citep{groningsson:2008b}.

\begin{figure*}[ht] 
  \includegraphics[width=2.25in]{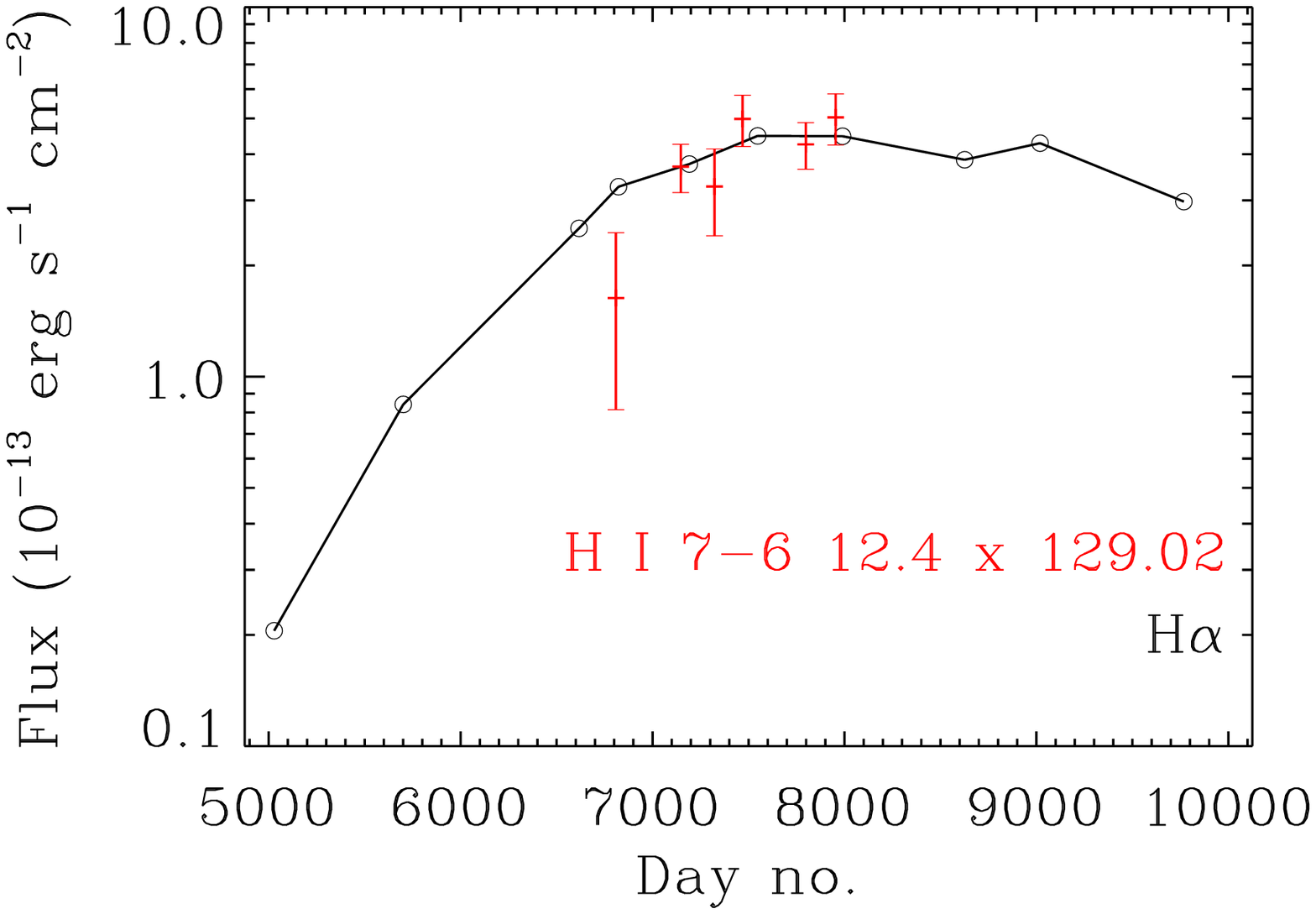}
  \includegraphics[width=2.25in]{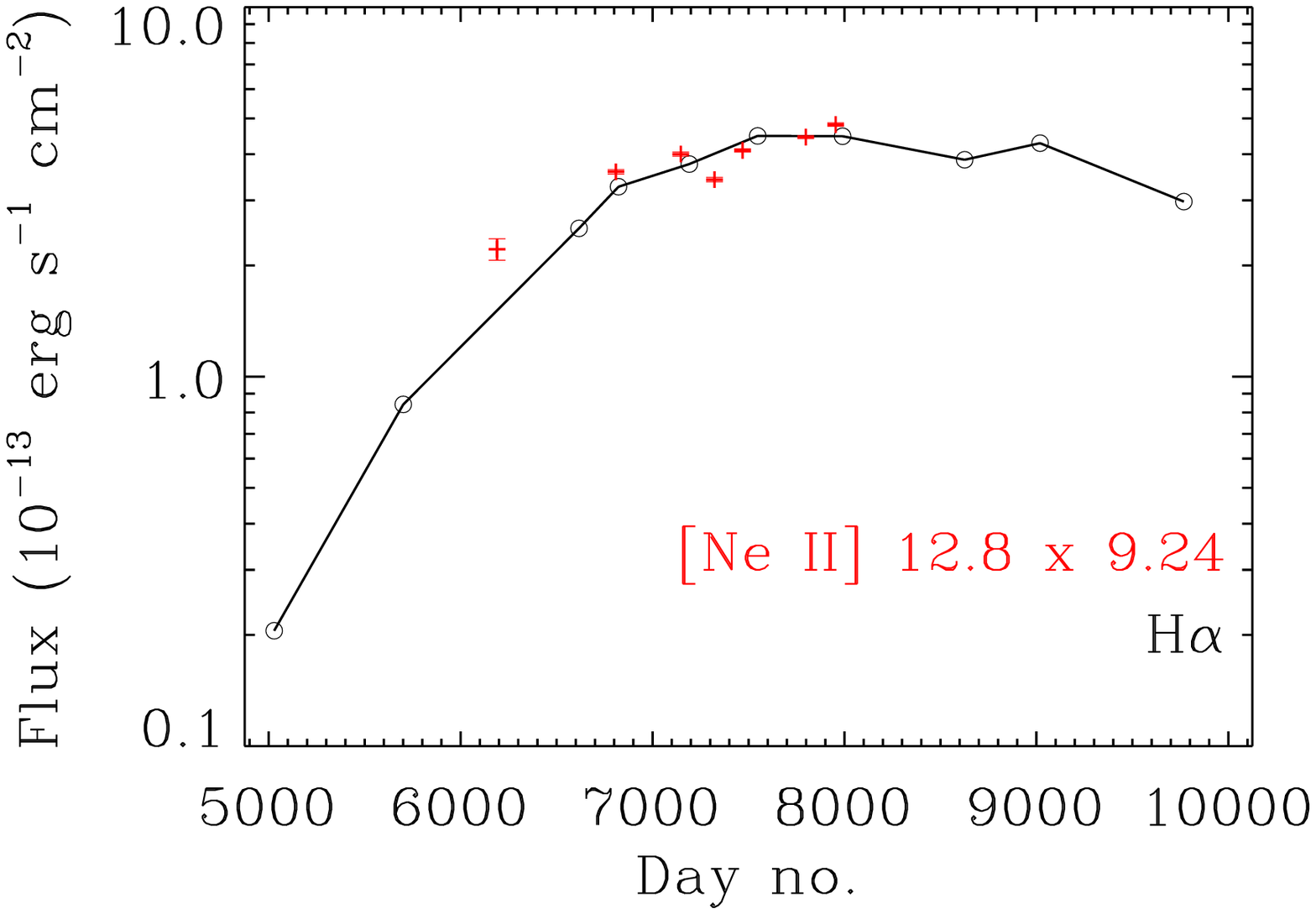}
  \includegraphics[width=2.25in]{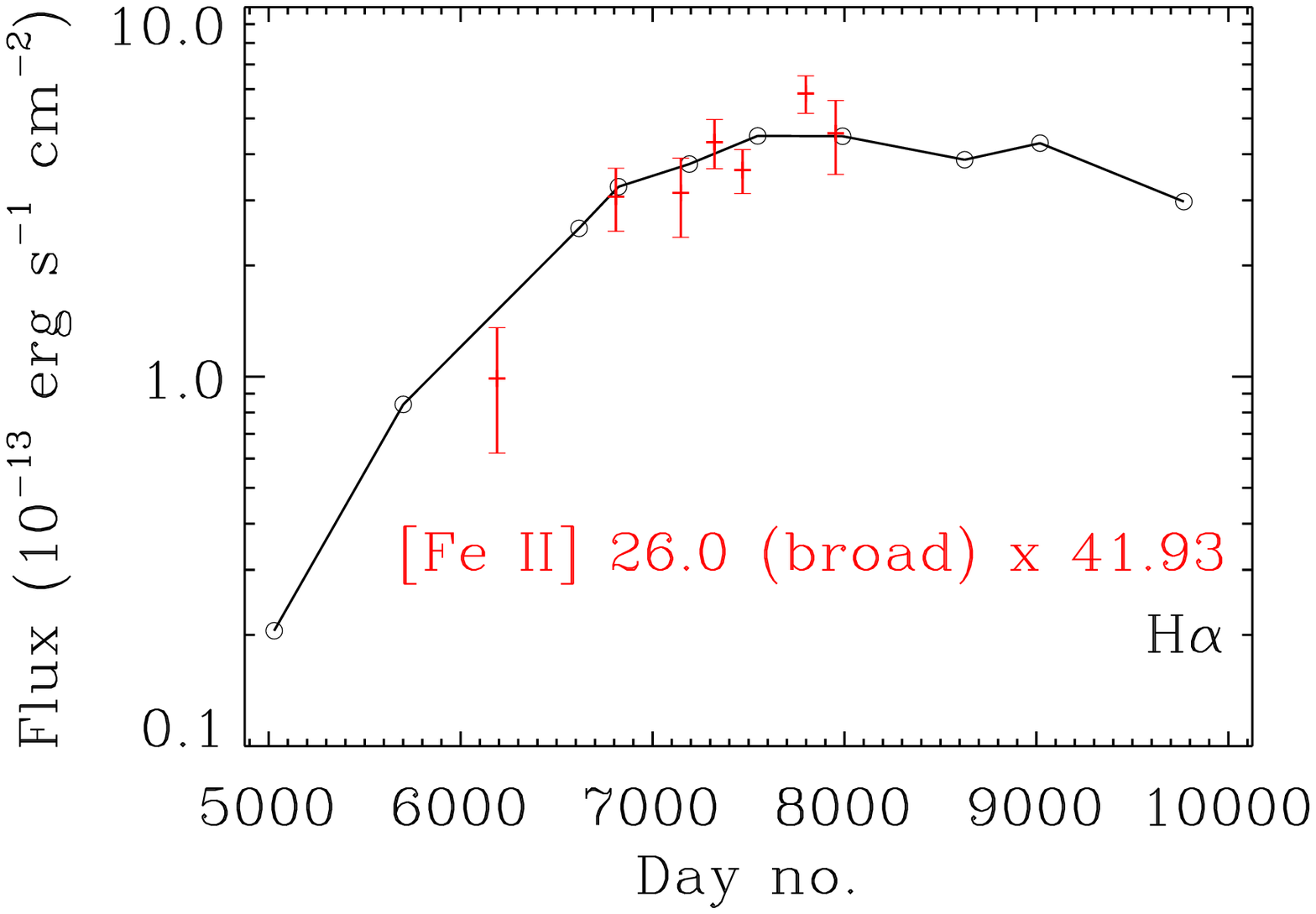}\\
  \includegraphics[width=2.25in]{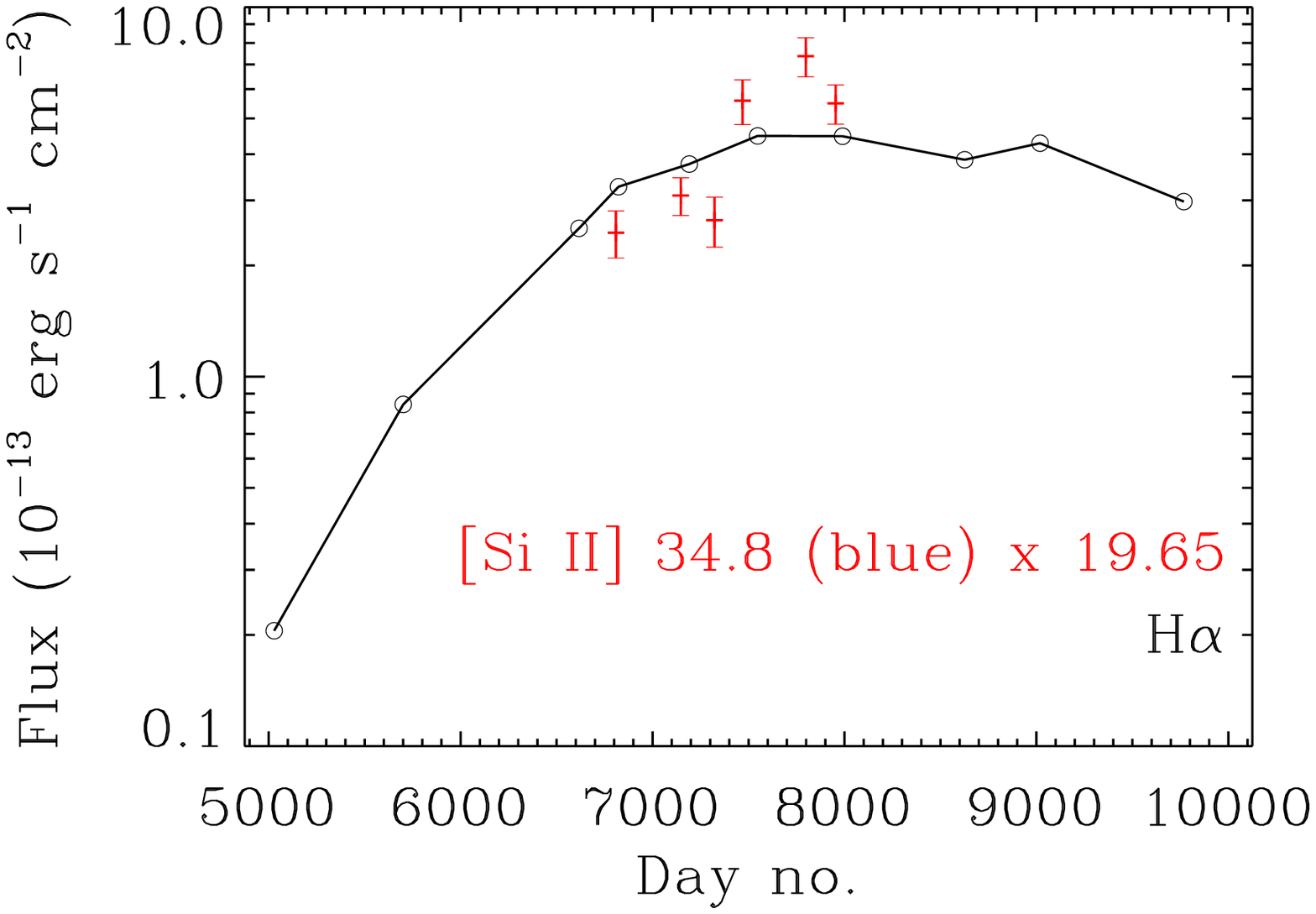}
  \includegraphics[width=2.25in]{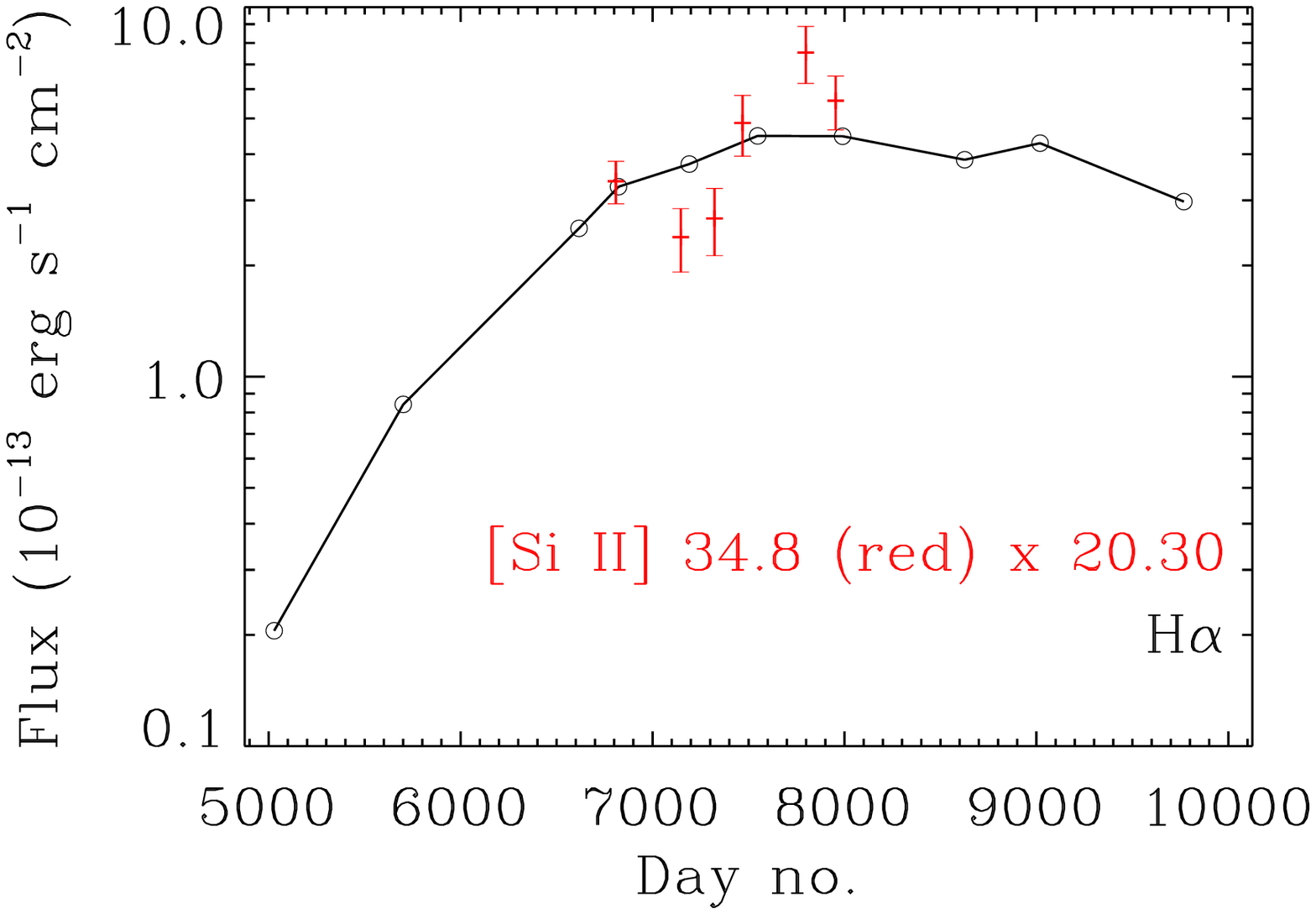}
  \caption{IR emission lines with markedly increasing fluxes 
  (+ symbols with error bars) are reasonably correlated with the 
  evolution of the H$\alpha$ flux (circles) from the {\it shocked} ER material \citep{fransson:2015}. The scale factors
  applied to the IR lines fluxes are given in the legend for each panel.}
  \label{fig:lines_vs_shockedha}
\end{figure*}

\begin{figure*}[ht] 
  \includegraphics[width=2.25in]{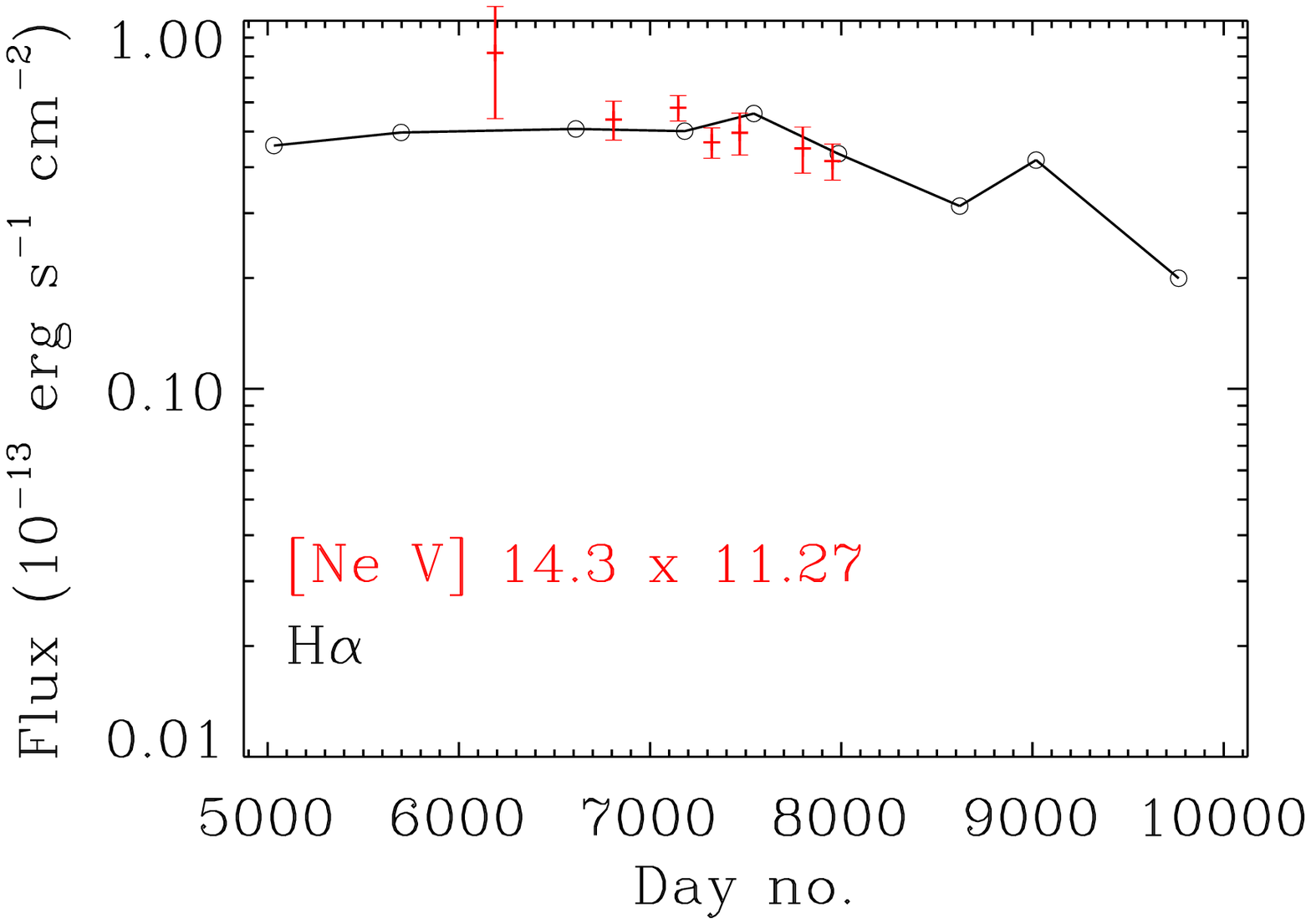}
  \includegraphics[width=2.25in]{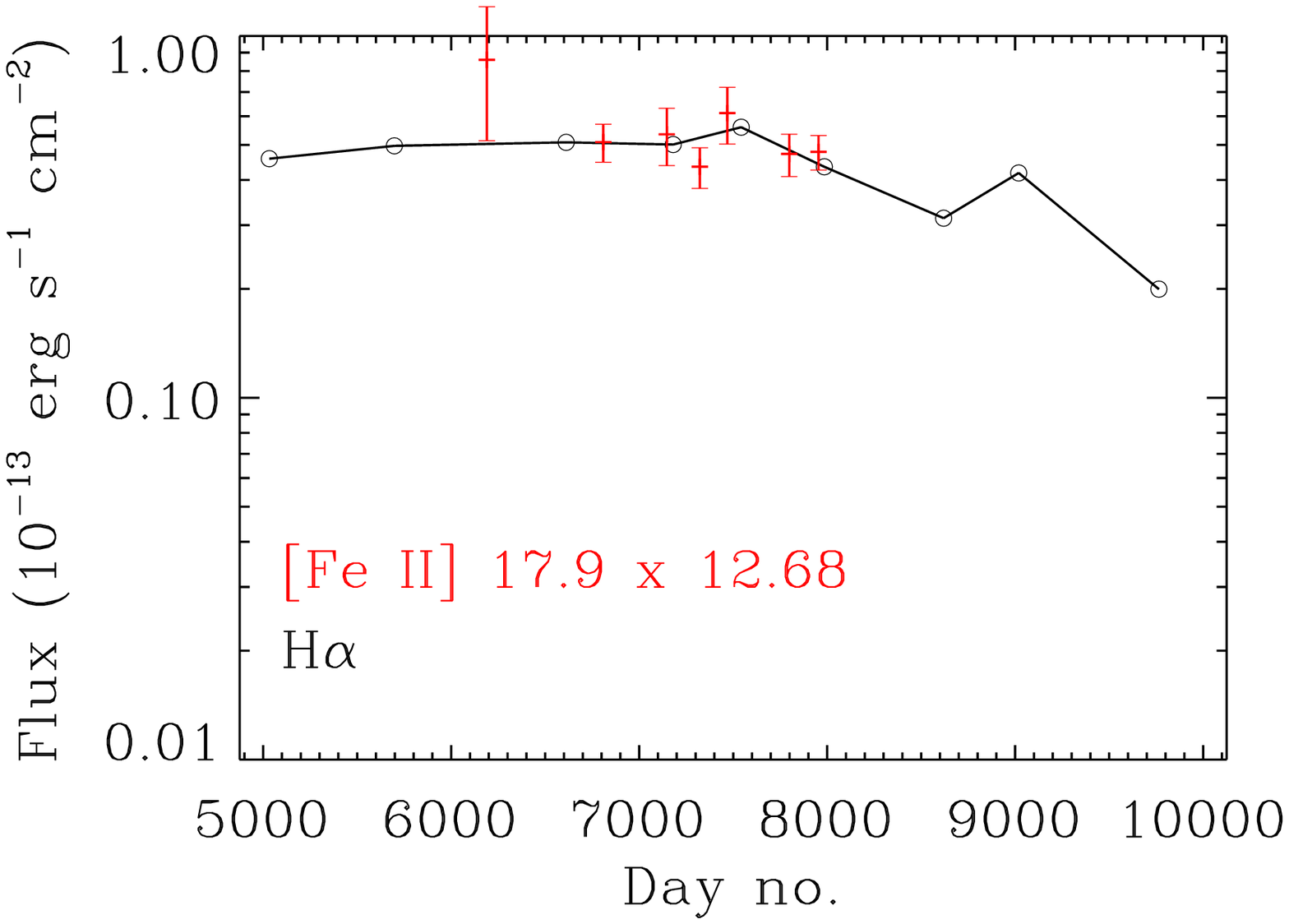}
  \includegraphics[width=2.25in]{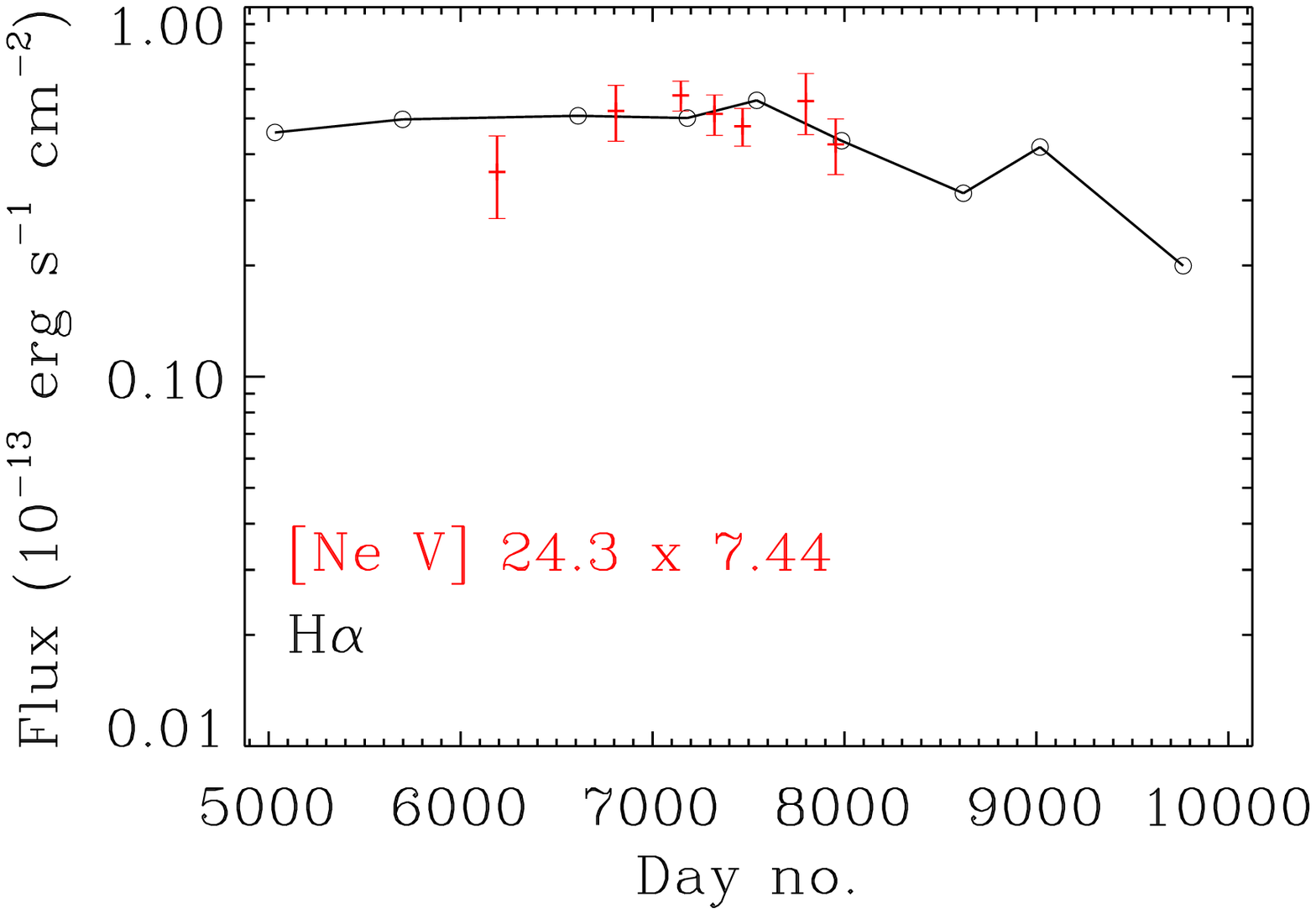}\\
  \includegraphics[width=2.25in]{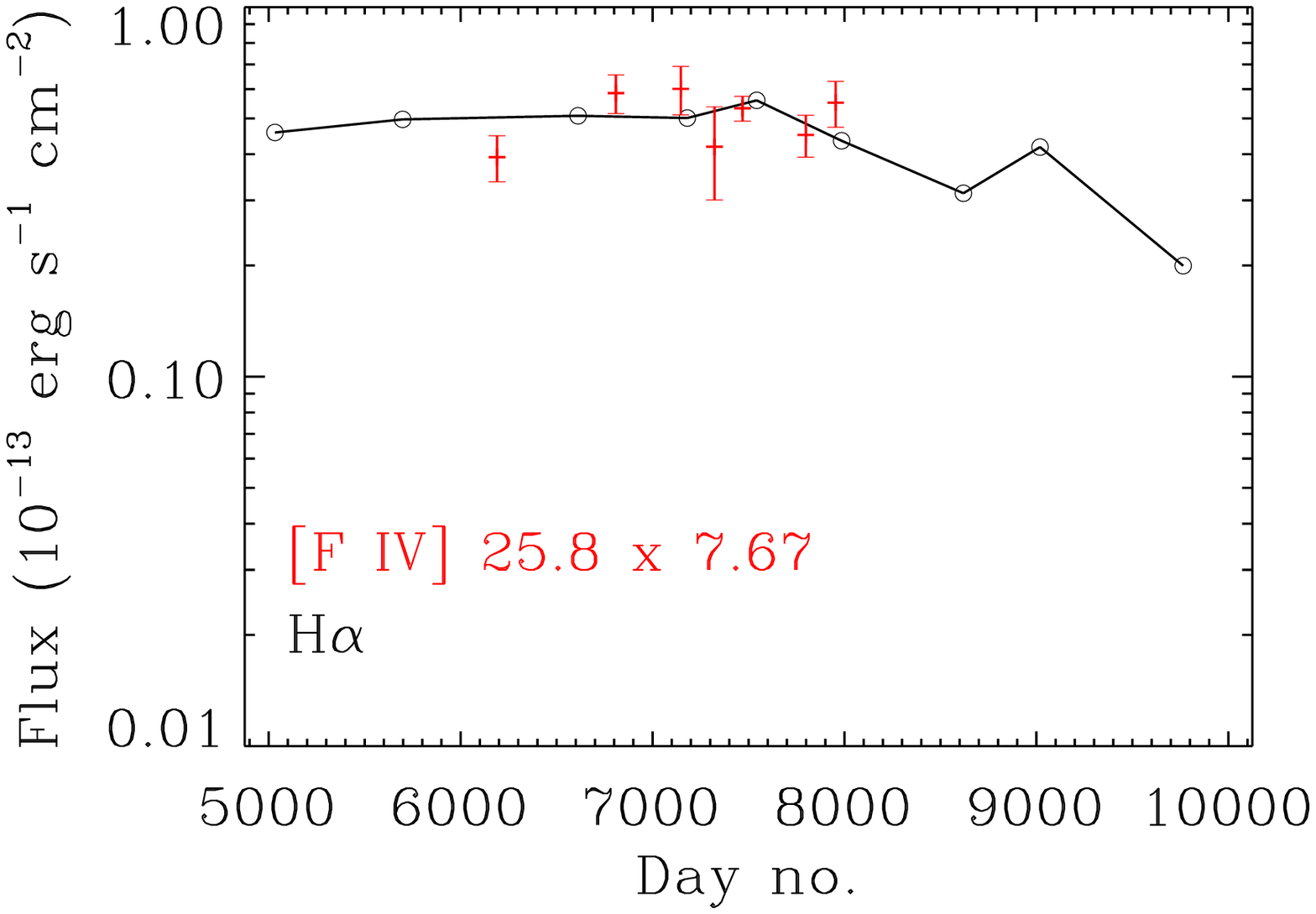}
  \includegraphics[width=2.25in]{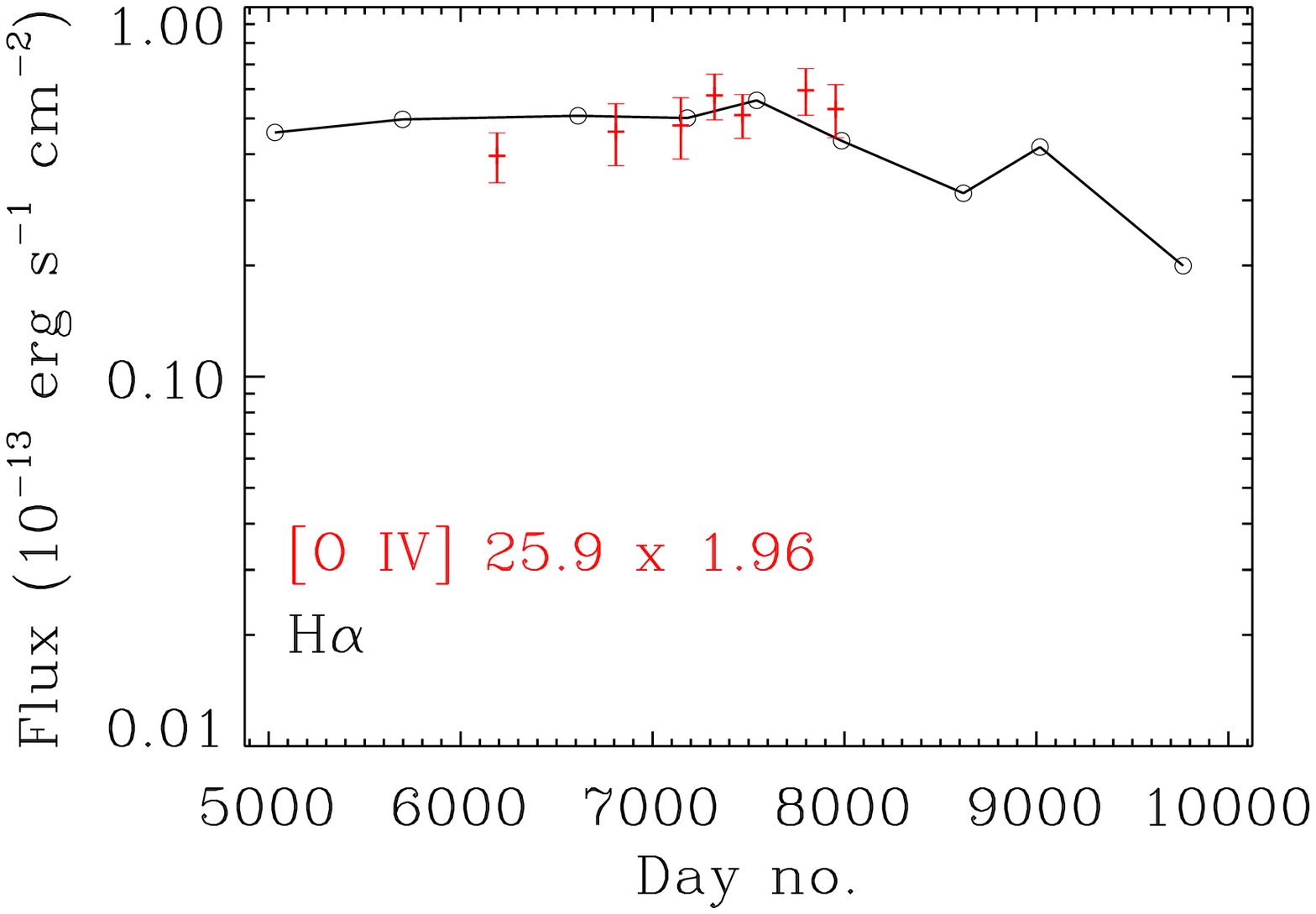}
  \includegraphics[width=2.25in]{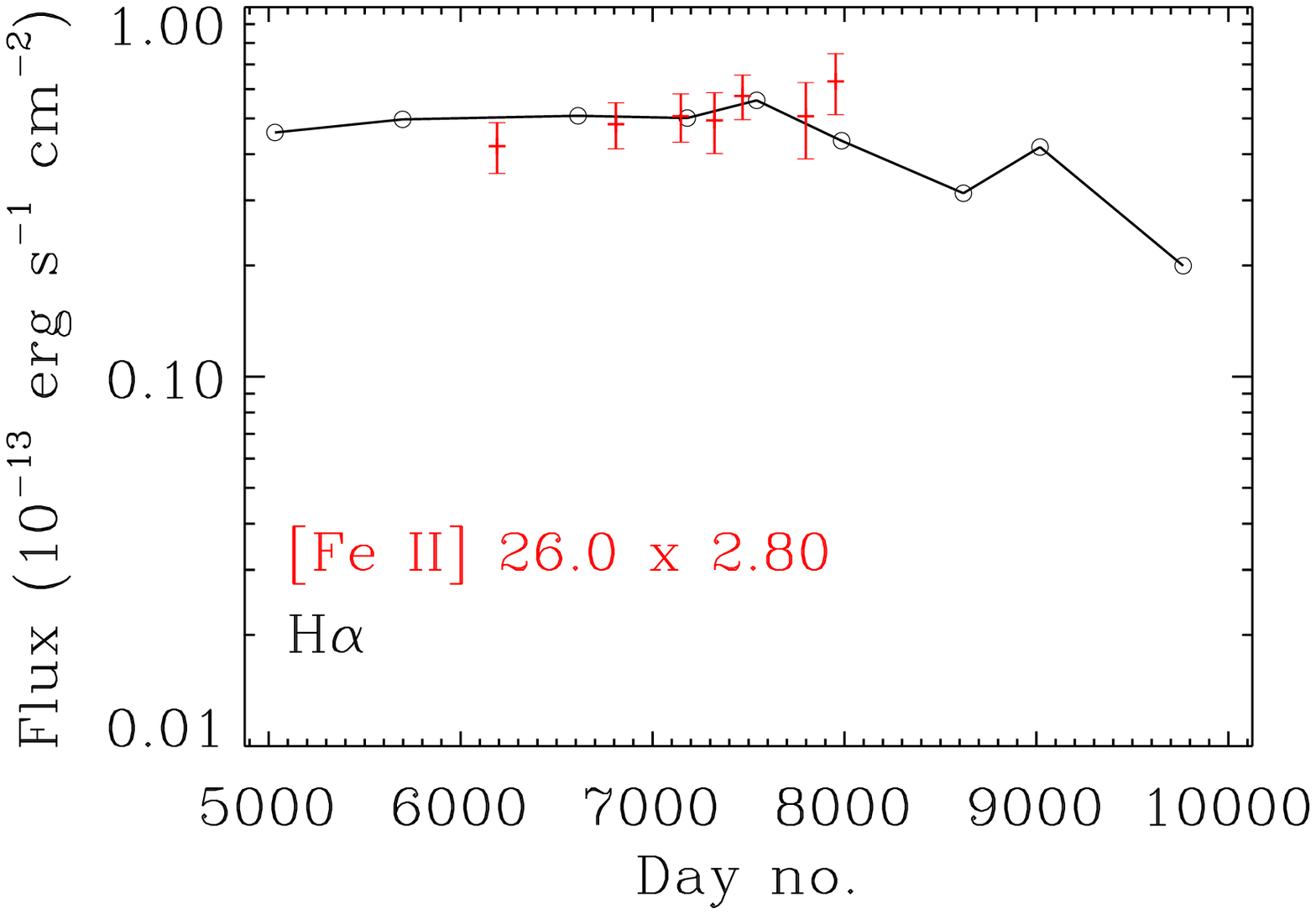}\\
  \includegraphics[width=2.25in]{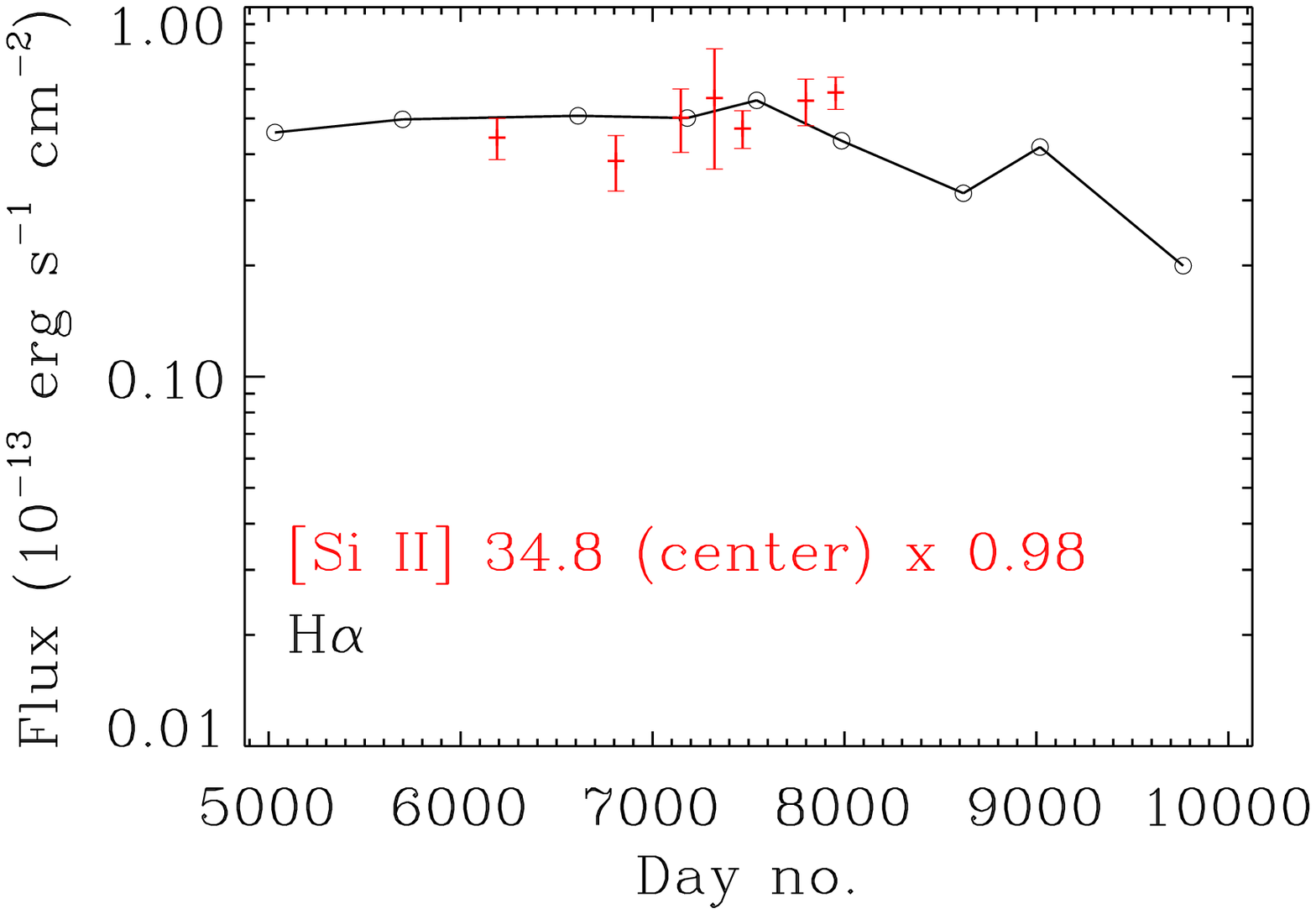}
  \includegraphics[width=2.25in]{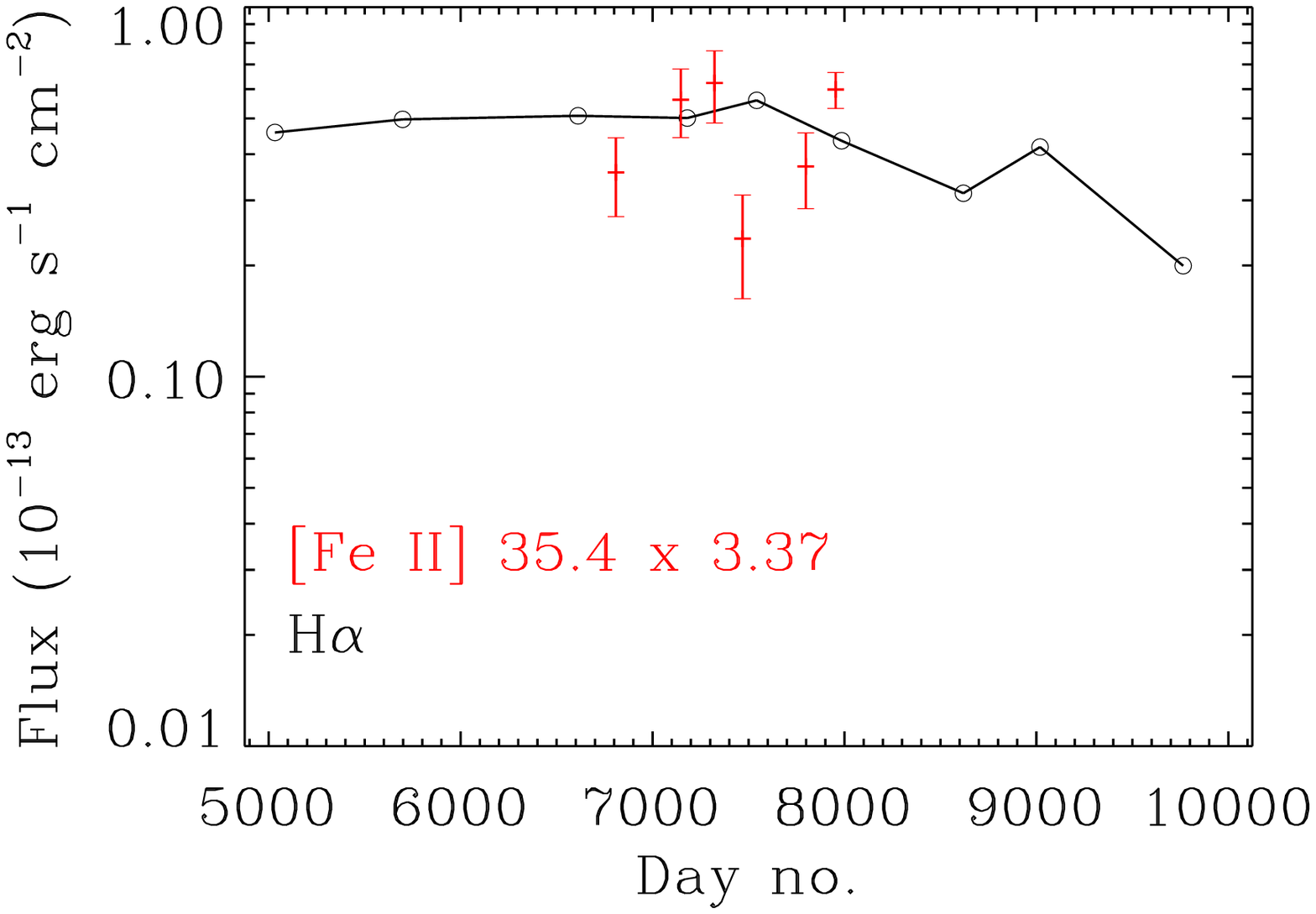}
  \caption{IR emission lines with approximately constant fluxes 
  (+ symbols with error bars) are reasonably correlated with the 
  evolution of the H$\alpha$ flux (circles) from the {\it unshocked} ER material \citep{fransson:2015}. The scale factors
  applied to the IR lines fluxes are given in the legend for each panel.}
  \label{fig:lines_vs_ha}
\end{figure*}

The evolution of the IR emission lines has also been compared to the evolution of the H$\alpha$ 
line components from the shocked and unshocked (pre-shock) portions of the ER \citep{fransson:2015}.
Figure \ref{fig:lines_vs_shockedha} shows the rising, then cresting, H$\alpha$ intensity 
of the shocked material of the ER compared to scaled versions of the IR lines that appear to show similar trends.
The other IR lines seems to show relatively little evolution and appear to be more
consistent with the emission from the unshocked portion of the ER (Figure \ref{fig:lines_vs_ha}).

\subsection{Gas Diagnostics}
\begin{figure}[ht] 
  \includegraphics[width=3.5in]{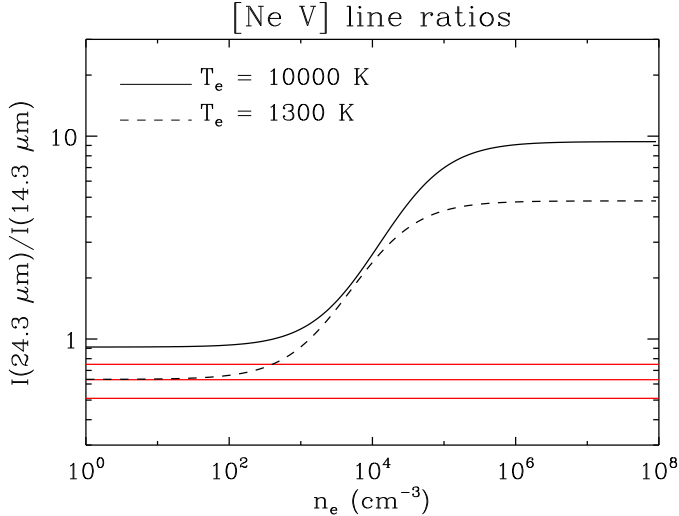}
  \caption{Expected [\ion{Ne}{5}] line ratios as a function of density for $T_e = 10^4$ and 1,300 K. The
  observed line ratios are $\sim 1.5\pm0.6$ at day 6,190 and $\sim0.63\pm0.12$ at later times (red lines).}
  \label{fig:ne_v}
\end{figure}

The ratio of the [\ion{Ne}{5}] lines is the only good density diagnostic these data provide
since other useful line pairs ([\ion{S}{3}], [\ion{Ne}{3}], [\ion{Ar}{3}], [\ion{Ar}{5}]) are either corrupted 
by the bright background or absent from the SN spectrum at this sensitivity. The ratio of the 
[\ion{Ne}{5}] line intensities at the first epoch is $R_{(14.3/24.3)} = 1.5\pm0.6$, 
which implies an electron density of $\sim3\times10^3$ cm$^{-3}$ (for $T_e \sim 10^4$ K) (see Figure \ref{fig:ne_v}). 
However, for all subsequent observations the ratio has a mean
value of $R_{(14.3/24.3)} = 0.63\pm0.05$ which lies below the low density limit ($n_e < 10^3$ cm$^{-3}$) unless
the gas temperature is $T_e < 1,300$ K. 

\begin{figure*}[ht] 
  \includegraphics[width=7in]{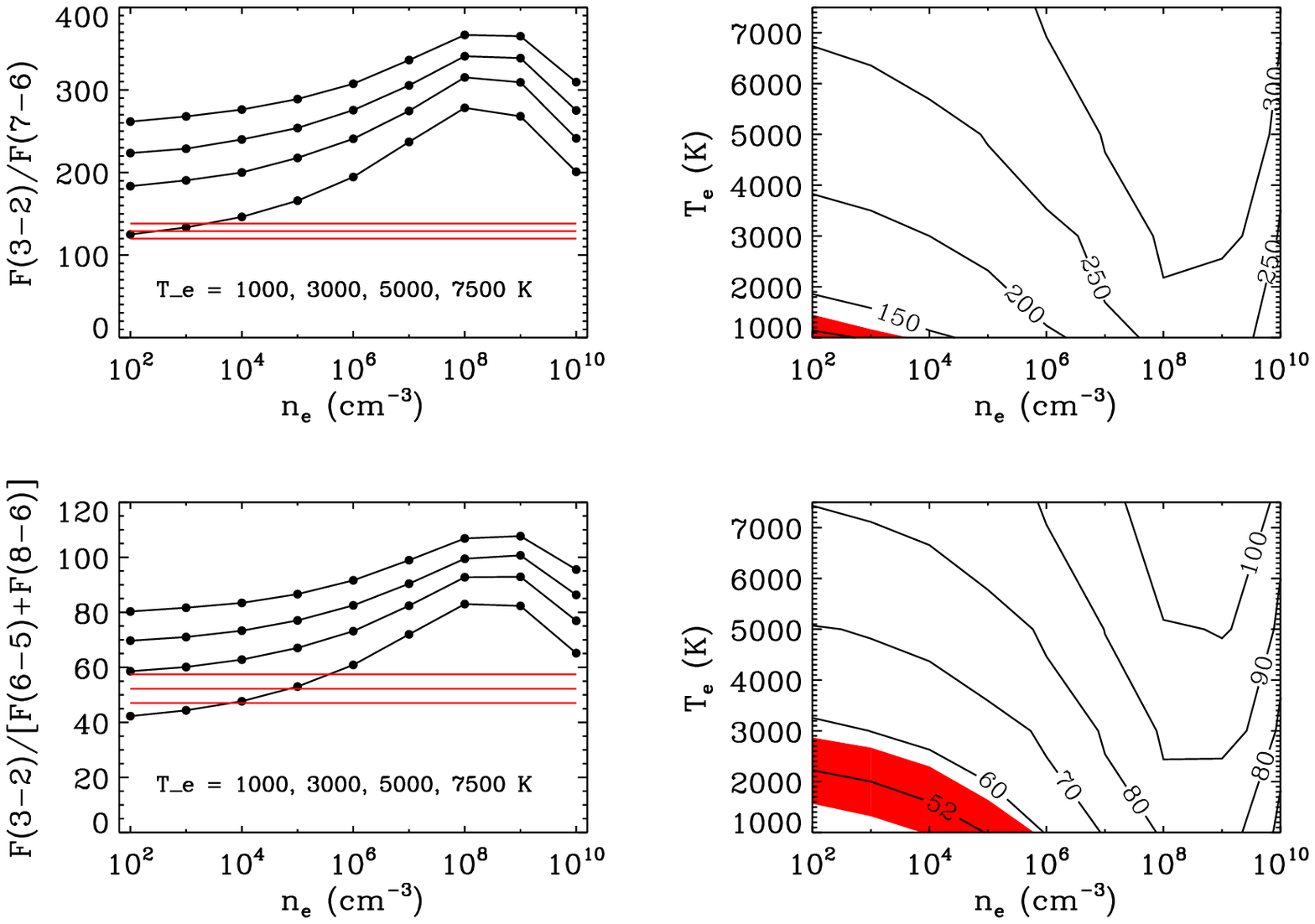}
  \caption{SN 1987A H recombination line ratios compared to 
  Case B recombination \citep{hummer:1987}. (upper left) the Case B ratio of 
  H$\alpha$/H(7-6) is plotted as a function of density for various
  temperatures (1,000 K $< T_e < $ 7,500 K). The observed line ratio and 
  its $\pm 1 \sigma$ uncertainty range are shown by the red lines. (upper right)
  The same H$\alpha$/H(7-6) is plotted as contours as a function of 
  $n_e$ and $T_e$. The $\pm 1 \sigma$ uncertainty range is indicated by 
  the shaded red band at the low temperature and density limits of the 
  calculations. (lower panels) These show similar plots for the case B and 
  observed line ratios of H$\alpha$ / [H(6-5)+H(8-6)].}
  \label{fig:caseb}
\end{figure*}

Similarly low density and temperature are inferred from the ratio of
H$\alpha$/H(7-6) $\approx 129$, assuming case B recombination 
\citep{hummer:1987} (Figure \ref{fig:caseb}). This observed ratio is the scale factor between the 
H(7-6) 12.37 $\mu$m light curve as measured at high resolution with the IRS, 
and the H$\alpha$ light curve of the shocked gas measured with {\it HST}
\citep{fransson:2015}. The possible blend of the H(6-5) and H(8-6) lines
observed near 7.48 $\mu$m in the low resolution IRS spectrum suggests slightly
higher densities and/or temperatures, but this measurement is less reliable.

\cite{dwek:2008} had concluded that the collisionally heated dust must 
reside in gas with $T_e \approx 3.5\times10^6$~K and $n_e \approx (0.3-1)\times 
10^4$ cm$^{-3}$. This is denser, hotter, and at higher pressure than the line 
emitting regions, suggesting that the observed IR line emission does not arise
for the same part of the ER as the dust continuum emission. To reach 
equilibrium temperatures of $T \approx 180$~K for silicate grains, 
collisional heating at the low gas temperatures implied by the emission
line ratios would require densities of $n_e \gtrsim 10^6$~cm$^{-3}$
\citep{bouchet:2006,dwek:2008}.
Such high densities are ruled out by the observed line ratios (Figures 
\ref{fig:ne_v} and \ref{fig:caseb}).


\section{Discussion}
\subsection{Continuum (Dust) Evolution}

With the long interval of observations now available, it has become clear
that the SN emission at 3.6 and 4.5 $\mu$m has peaked and is now in decline.
This suggests, that, as for the optical emission \citep{fransson:2015}, the
SN blast wave has sent shocks through the bulk of the material in the ER, and 
the emission of any remaining swept up material is insufficient to make up 
for the fading of the ER material that has already been shocked.
Despite the likelihood that the emitting dust is collisionally heated 
in the hot gas, it is also clear that 3.6 and 4.5 $\mu$m emission 
is not tracking the X-ray emission. If the X-ray emission is reaching a peak, 
it is $>1,000$~days after the IR peak.
The simplest explanation for this evolving IRX ratio
is that the dust is being destroyed on timescales of a few years, 
$\tau_{\rm sput} \approx 10^6 a (\micron) / n_{\rm H} ({\rm cm}^{-3})$~yr 
\citep{draine:1979} which is 
shorter than the cooling time for the X-ray emitting gas: $\sim5-30$~yr
\citep{dwek:2010}. The apparently short sputtering timescale and high 
dust temperature for the grains the produce the 3.6 and 4.5 $\mu$m emission
are most easily achieved by relatively small grains compared to those that 
produce the 24 $\mu$m emission. However, the exact sizes depend on the
unknown composition of the dust and on more precise estimates of the 
sputtering timescale \citep[see][]{dwek:2010}.
\cite{bocchio:2012} show that very small amorphous hydrocarbons and PAHs
have even shorter lifetimes in a hot gas than classical sputtering 
calculations predict. Although the spectra of SN 1987A show no indications of 
the PAH emission features, the physical mechanisms they consider 
may also act to decrease the lifetimes of small grains of other compositions.

The 5 -- 30 $\mu$m spectrum had been used to estimate the mass of warm dust in the
ER as $1.2\times10^{-6}~M_{\sun}$ at day 7,554 \citep{dwek:2010}. Because there
was little change in the dust temperature, this mass scales linearly with the 
evolving IR flux density. The spectral fits in 
\citet[Figures 4 and 5]{dwek:2010} suggest that the dust responsible for 
the 3.6 and 4.5 $\mu$m emission is $\sim2$ times warmer and $\sim100$ times
fainter (in the Rayleigh-Jeans tail). Therefore the mass of this hot dust is 
only $\sim0.5\%$ of the total dust in the ER, with a large uncertainty depending
on the actual emissivity of these grains of uncertain composition. (If the hot dust 
component is stochastically heated rather than at equilibrium temperatures, 
then its total dust mass would be larger because only a small fraction of the 
dust would be contributing to the short wavelength emission.)
The evident destruction of this hot dust component is not necessarily indicative
of significant reduction in the total dust mass of the ER.

The IRX evolution model of Equation (6), as shown in Figure \ref{fig:t0fits},
indicates a correlation between wavelength and $t_0$. The parameter 
$t_0$ is the {\it extrapolated} time at which the IR emission began
the trend of $S(IR,t') \sim S(X,t')/t'$ where $t' = t-t_0$. It is unclear
if this trend really does extrapolate back to $t_0$, but if so, we can suggest 
two potential reasons for the correlation of wavelength and $t_0$. 

The first possibility is that the UV flash from the SN has preferentially 
evaporated small dust grains, creating a gradient in the size distribution
of dust grains as a function of radius as described by \cite{fischera:2002b}.
In this case, the initial (near day 3,200) IR emission from the interaction 
of the blast wave with the ER should 
be dominated by relatively large grains. These grains do not get heated to 
very high temperatures, and may thus produce strong 24 $\mu$m emission but 
little emission at $<10$ $\mu$m. As the blast wave expands further into the 
ring, near day 4,500 it may begin encountering smaller grains that 
were sufficiently distant or shielded from the SN to have escaped evaporation. 
These smaller grains would be heated to higher temperatures capable of 
producing the observed emission and IRX trends at 3.6 and 4.5 $\mu$m.
If this scenario is valid it would suggest that the small grains emitting
at the shortest wavelengths are {\it not} carbonaceous. \cite{fischera:2002b} 
show that the UV flash is significantly less effective at evaporating 
graphite grains than silicate or iron grains. Therefore, small carbonaceous
grains would be encountered at earlier times than large (or any) silicate 
grains.

The second possibility is that grain-grain collisions in the post-shock
gas are fragmenting larger grains into smaller ones. The result 
$t_0 =$ 3,251 at 24 $\mu$m would be interpreted as there having been sufficient 
time for all of the largest grains to be destroyed in the post-shock region.
The result $t_0 =$ 4,523 at 3.6 $\mu$m would indicate that smaller grains
are present for an additional $\sim$3.5 yr after the destruction of the 
largest grains. However, grain-grain collisions are generally most effective 
in the cases of much slower shocks (e.g. 10s of km~s$^{-1}$) propagating
into media at least as dense as the ER. In these higher velocity shocks, 
thermal and non-thermal sputtering would be the dominant destruction 
mechanism \citep{dwek:2010}, and small grains should be destroyed more
rapidly than the larger grains.

\subsection{Emission Line (Gas) Evolution}

The observed IR emission lines are mostly collisionally excited fine
structure transitions in the ground state of various ions. The lines identified
are typical of those seen in other supernova remnants 
\citep{oliva:1999,arendt:1999,temim:2006,neufeld:2007,smith:2009,andersen:2011,ghavamian:2012,sankrit:2014}. 
The accuracy of the measurement of the lines strengths is limited by the 
modest signal-to-noise ratio of some lines, and the presence of strong 
(sometimes dominant) and structured background emission from the ISM 
in some of the lines. 

The evolution of the IR emission lines is puzzling. Many of the emission
lines seem to show little evolution, consistent with the minimal evolution 
of optical emission lines from the unshocked portion of the ER 
(Figure \ref{fig:lines_vs_ha}). However, these lines include highly ionized 
species such as \ion{Ne}{5} and \ion{O}{4}, which have much higher ionization 
potentials than other lines attributed to the unshocked portion of the ER.

Another puzzle is presented by the higher velocity components
of the \ion{Fe}{2} and \ion{Si}{2} lines. These lines may originate
in gas phase material, but they may also represent material that has 
been eroded from dust grains. The high velocities suggest 
shocked material, and indeed these lines do correlate with the rising
intensity of the shocked H$\alpha$ emission, as well as the increasing
24~$\mu$m (dust) flux density. The velocity structures of 
these two lines is somewhat mysterious. It is not clear why the \ion{Si}{2}
should show {\it distinct} high and low velocity components, whereas the 
\ion{Fe}{2} line seems to show a simple broad component which would be
expected for a spherical distribution of material.
The lines may indicate the presence an expanding fossil remnant from 
the SN precursor as in $\eta$ Carinae. The Fe lines may be from a more 
symmetrical fossil remnant than the Si lines.  $\eta$ Car has undergone 
multiple eruptions (at least 3 in the last $\sim200$ years), not all 
producing the same geometry \citep[see][]{smith:1998,smith:2000}.
Alternatively, it may be that these lines originate in the 
SN ejecta, rather than the ER or other circumstellar structures, 
in which case, the velocity structures of the \ion{Si}{2} 
lines may reflect uneven distribution of Si in the ejecta.
The presence of the redshifted component would indicate that the ejecta 
are not (uniformly) optically thick at $\sim35$ $\mu$m. 

The final puzzle presented by the emission lines is that both the 
H recombination lines and the {\ion{Ne}{5} lines suggest low densities 
($n_e \lesssim 10^3$ cm$^{-3}$) and low temperatures ($T_e \lesssim 2,500$ K).
The densities are much lower than estimated for the ER from pre-interaction 
observations \citep[e.g.][]{plait:1995} or analysis of the earliest hot spot
\citep{michael:2000,pun:2002}. However, \cite{groningsson:2008b} find 
indications of densities nearly this low for the narrow line emission 
of the unshocked material, and \cite{mattila:2010} invoke a $10^2$ 
cm$^{-3}$ component distributed over a wider area to account for the 
evolution of high-ionization lines at late times (days 5,000 -- 7,500).
So the densities suggest that the H and Ne emission lines may 
arise from the regional outside of the ER, 
perhaps associated with multiple ejections from the precursor. 
However implied the low gas temperature remains at odds with prior 
analyses, and seems unlikely since the recombination 
rate for \ion{Ne}{5} is several times higher than the 
cooling rate \cite[assuming solar abundances;][]{osterbrock:1989,draine:2011}.

\section{Summary}

More than ten years of {\it Spitzer} broadband mid-IR photometry of SN 1987A
have now shown that the 3.6 and 4.5 $\mu$m emission associated with the ER 
has peaked and is now in a declining phase. 

The IR emission is not directly proportional to the X-ray emission, which
would be expected if the dust is collisionally heated and the 
dust-to-gas mass ratio remains constant.

The IRX does seem to evolve as $(t-t_0)^{-1}$, which is consistent with
an accumulating mass of shocked gas behind the ER shocks, but a finite
zone of IR emission, limited by the dust destruction timescales 
that are short compared to the gas cooling timescale.

The dust mass itself can grow linearly as $(t-t_0)$ at early times
(the dust temperature remains constant), but must begin decreasing 
after about day 7,500 to match the 3.6 and 4.5 $\mu$m light curves.

The time $t_0$ and its apparent variation with wavelength 
can be interpreted either as a) an indication of 
evaporation of small grains by the UV flash, or b) the 
initial preferential destruction of large grains via grain-grain 
collisions in the post-shock gas.

The IR light curve is also in fair agreement with the latest optical 
light curves (particularly the B band), which have indicated that the 
peak of the interaction with the ER has passed. 

IR emission lines seem to arise from two regions. One set of lines 
is characterized by relatively constant emission, as matched by the 
evolution of narrow (pre-shock) optical emission lines. The other 
set of lines includes odd high-velocity components, and seems 
to match the evolution of the optical lines from the post-shock ER regions.
These lines may indicate possible fossil circumstellar remnants, distinct from the ER, produced by the SN precursor. 

\acknowledgments
This work is based on observations made with the {\it Spitzer Space Telescope}, 
which is operated by the Jet Propulsion Laboratory, California 
Institute of Technology 
under a contract with NASA. Support for this work was provided by NASA.
This research has made use of NASA's Astrophysics Data System Bibliographic Services. ED was supported 
by NASA grants 12-ADAP12-0145 and 13-ADAP13-0094. We thank the referee, A. Jones, for
useful comments which improved this manuscript.

{\it Facilities:} \facility{Spitzer}, \facility{Chandra}

\bibliographystyle{apj}
\bibliography{SN1987Aring-spitzer2015}

\clearpage

\begin{deluxetable*}{rccccc}
\tablewidth{0pt}
\tablecaption{SN 1987A Flux Densities}
\tablehead{
\colhead{Day} &
\multicolumn{4}{c}{IRAC} &
\colhead{MIPS}\\
\cline{2-5}
\colhead{Number} &
\colhead{$S(3.6\ \micron)$} &
\colhead{$S(4.5\ \micron)$} &
\colhead{$S(5.8\ \micron)$} &
\colhead{$S(8\ \micron)$} &
\colhead{$S(24\ \micron)$}
}
\startdata
 6130.09 &  0.99 $\pm$ 0.01 &  1.20 $\pm$ 0.01 &  1.62 $\pm$ 0.02 &  4.69 $\pm$ 0.03 &   \nodata       \\
 6184.08 &  \nodata         &  \nodata         &  \nodata         &  \nodata         &  26.3 $\pm$ 1.8 \\
 6487.93 &  1.10 $\pm$ 0.01 &  1.51 $\pm$ 0.01 &  2.34 $\pm$ 0.04 &  6.98 $\pm$ 0.04 &   \nodata       \\
 6487.94 &  \nodata         &  1.53 $\pm$ 0.02 &  \nodata         &  7.13 $\pm$ 0.07 &   \nodata       \\
 6551.91 &  \nodata         &  \nodata         &  \nodata         &  \nodata         &  36.4 $\pm$ 1.9 \\
 6724.25 &  1.21 $\pm$ 0.01 &  1.67 $\pm$ 0.01 &  2.50 $\pm$ 0.04 &  7.59 $\pm$ 0.05 &   \nodata       \\
 6725.68 &  1.18 $\pm$ 0.02 &  \nodata         &  2.57 $\pm$ 0.05 &  \nodata         &   \nodata       \\
 6734.33 &  \nodata         &  \nodata         &  \nodata         &  \nodata         &  41.7 $\pm$ 1.9 \\
 6823.54 &  1.29 $\pm$ 0.01 &  1.76 $\pm$ 0.01 &  2.68 $\pm$ 0.04 &  8.10 $\pm$ 0.06 &   \nodata       \\
 6824.65 &  1.22 $\pm$ 0.01 &  1.78 $\pm$ 0.01 &  2.71 $\pm$ 0.04 &  8.26 $\pm$ 0.05 &   \nodata       \\
 6828.53 &  \nodata         &  \nodata         &  \nodata         &  \nodata         &  44.4 $\pm$ 1.9 \\
 7156.35 &  1.38 $\pm$ 0.01 &  2.03 $\pm$ 0.01 &  3.23 $\pm$ 0.02 & 10.21 $\pm$ 0.04 &   \nodata       \\
 7158.88 &  \nodata         &  \nodata         &  \nodata         &  \nodata         &  55.3 $\pm$ 1.8 \\
 7298.80 &  1.47 $\pm$ 0.01 &  2.15 $\pm$ 0.01 &  3.39 $\pm$ 0.02 & 11.13 $\pm$ 0.04 &   \nodata       \\
 7309.70 &  \nodata         &  \nodata         &  \nodata         &  \nodata         &  59.8 $\pm$ 1.9 \\
 7489.68 &  \nodata         &  \nodata         &  \nodata         &  \nodata         &  65.2 $\pm$ 1.9 \\
 7502.04 &  1.51 $\pm$ 0.01 &  2.27 $\pm$ 0.01 &  3.75 $\pm$ 0.02 & 12.24 $\pm$ 0.05 &   \nodata       \\
 7687.35 &  1.55 $\pm$ 0.01 &  2.36 $\pm$ 0.01 &  4.02 $\pm$ 0.02 & 13.04 $\pm$ 0.04 &   \nodata       \\
 7689.55 &  \nodata         &  \nodata         &  \nodata         &  \nodata         &  70.3 $\pm$ 1.9 \\
 7974.80 &  1.59 $\pm$ 0.01 &  2.41 $\pm$ 0.01 &  4.08 $\pm$ 0.02 & 13.61 $\pm$ 0.03 &   \nodata       \\
 7983.16 &  \nodata         &  \nodata         &  \nodata         &  \nodata         &  75.7 $\pm$ 1.9 \\
 8576.21 &  \nodata         &  2.51 $\pm$ 0.02 &  \nodata         &  \nodata         &   \nodata       \\
 8585.63 &  1.63 $\pm$ 0.01 &  2.50 $\pm$ 0.01 &  \nodata         &  \nodata         &   \nodata       \\
 8706.09 &  1.60 $\pm$ 0.02 &  \nodata         &  \nodata         &  \nodata         &   \nodata       \\
 8730.61 &  1.63 $\pm$ 0.02 &  \nodata         &  \nodata         &  \nodata         &   \nodata       \\
 8732.20 &  \nodata         &  2.51 $\pm$ 0.02 &  \nodata         &  \nodata         &   \nodata       \\
 8735.25 &  \nodata         &  2.53 $\pm$ 0.02 &  \nodata         &  \nodata         &   \nodata       \\
 8736.63 &  1.57 $\pm$ 0.02 &  \nodata         &  \nodata         &  \nodata         &   \nodata       \\
 8738.06 &  \nodata         &  2.50 $\pm$ 0.01 &  \nodata         &  \nodata         &   \nodata       \\
 8743.47 &  1.68 $\pm$ 0.02 &  2.53 $\pm$ 0.02 &  \nodata         &  \nodata         &   \nodata       \\
 8751.53 &  1.54 $\pm$ 0.02 &  2.42 $\pm$ 0.02 &  \nodata         &  \nodata         &   \nodata       \\
 8757.27 &  1.66 $\pm$ 0.02 &  2.50 $\pm$ 0.01 &  \nodata         &  \nodata         &   \nodata       \\
 8829.32 &  \nodata         &  2.54 $\pm$ 0.02 &  \nodata         &  \nodata         &   \nodata       \\
 8856.29 &  1.62 $\pm$ 0.01 &  2.50 $\pm$ 0.01 &  \nodata         &  \nodata         &   \nodata       \\
 9024.97 &  1.60 $\pm$ 0.01 &  2.52 $\pm$ 0.01 &  \nodata         &  \nodata         &   \nodata       \\
 9232.27 &  1.59 $\pm$ 0.01 &  2.50 $\pm$ 0.01 &  \nodata         &  \nodata         &   \nodata       \\
 9495.25 &  1.58 $\pm$ 0.01 &  2.41 $\pm$ 0.01 &  \nodata         &  \nodata         &   \nodata       \\
 9656.20 &  1.59 $\pm$ 0.01 &  2.39 $\pm$ 0.01 &  \nodata         &  \nodata         &   \nodata       \\
 9810.19 &  1.62 $\pm$ 0.01 &  2.38 $\pm$ 0.01 &  \nodata         &  \nodata         &   \nodata       \\
10034.95 &  1.55 $\pm$ 0.01 &  2.29 $\pm$ 0.01 &  \nodata         &  \nodata         &   \nodata       \\
10244.63 &  1.52 $\pm$ 0.01 &  2.23 $\pm$ 0.01 &  \nodata         &  \nodata         &   \nodata       \\
10377.66 &  1.52 $\pm$ 0.01 &  2.17 $\pm$ 0.01 &  \nodata         &  \nodata         &   \nodata       
\enddata
\tablecomments{units = Jy}
\label{tab:fluxes}
\end{deluxetable*}

\begin{deluxetable*}{ccccccc}
\tabletypesize{\scriptsize}
\tablewidth{0pt}
\tablecaption{High Resolution Lines}
\tablehead{
\colhead{Day number} &
\colhead{Species} &
\colhead{$\lambda_0$ ($\micron$)} &
\colhead{$\lambda$ ($\micron$)} &
\colhead{$v$ (km s$^{-1}$)} &
\colhead{FWHM ($\micron$)} &
\colhead{Flux (10$^{-22}$ W m$^{-2}$)}
}
\startdata
6190 &             [\ion{Ne}{2}] &   12.81355 & $   12.8305\pm    0.0004$ &       397.& $    0.0189\pm    0.0004$ &$      24.0\pm       1.6$ \\
6190 &              [\ion{Ne}{5}] &   14.32170 & $   14.3369\pm    0.0049$ &       318.& $    0.0465\pm    0.0050$ &$       7.3\pm       2.4$ \\
6190 &            [\ion{Ne}{3}] &   15.55510 & $   15.5765\pm    0.0009$ &       413.& $    0.0176\pm    0.0007$ &$      10.5\pm       1.3$ \\
6190 &             [\ion{Fe}{2}] &   17.93595 & $   17.9695\pm    0.0046$ &       561.& $    0.0365\pm    0.0048$ &$       6.8\pm       2.7$ \\
6190 &             [\ion{S}{3}] &   18.71300 & $   18.7411\pm    0.0034$ &       451.& $    0.0251\pm    0.0041$ &$       5.3\pm       2.6$ \\
6190 &              [\ion{Ne}{5}] &   24.31750 & $   24.3404\pm    0.0022$ &       283.& $    0.0304\pm    0.0026$ &$       4.8\pm       1.2$ \\
6190 &              [\ion{F}{4}]? &   25.83000 & $   25.8354\pm    0.0021$ &      63.& $  0.0456\pm   0.0021$ &$      5.1\pm      0.7$ \\
6190 &              [\ion{O}{4}] &   25.89030 & $   25.9069\pm    0.0026$ &     192.& $  0.0538\pm   0.0027$ &$     20.2\pm      3.1$ \\
6190 &             [\ion{Fe}{2}] &   25.98829 & $   26.0065\pm    0.0031$ &     210.& $  0.0616\pm   0.0031$ &$     15.0\pm      2.4$ \\
6190 &     [\ion{Fe}{2}] (broad) &   25.98829 & $   26.0831\pm    0.0051$ &    1094.& $  0.0406\pm   0.0050$ &$      2.4\pm      0.9$ \\
6190 &             [\ion{S}{3}] &   33.48100 & $   33.4806\pm    0.0189$ &        -4.& $    0.1131\pm    0.0201$ &$      13.9\pm       6.8$ \\
6190 &             [\ion{Si}{2}] &   34.81520 & $   34.8546\pm0.0030  $ & 339. & $  0.0713\pm0.0029  $ & $ 45.5\pm 5.8$ \\
\hline
6809 &              [\ion{S}{4}] &   10.51050 & $   10.5222\pm    0.0020$ &       334.& $    0.0147\pm    0.0018$ &$      -1.6\pm       0.6$ \\
6809 &             \ion{H}{1} 7-6 &   12.37190 & $   12.3822\pm    0.0011$ &       250.& $    0.0095\pm    0.0014$ &$       1.3\pm       0.6$ \\
6809 &             [\ion{Ne}{2}] &   12.81355 & $   12.8259\pm    0.0001$ &       289.& $    0.0188\pm    0.0001$ &$      38.8\pm       0.5$ \\
6809 &              [\ion{Ne}{5}] &   14.32170 & $   14.3356\pm    0.0009$ &       291.& $    0.0218\pm    0.0008$ &$       4.8\pm       0.6$ \\
6809 &            [\ion{Ne}{3}] &   15.55510 & $   15.5696\pm    0.0002$ &       280.& $    0.0233\pm    0.0002$ &$      10.6\pm       0.3$ \\
6809 &             [\ion{Fe}{2}] &   17.93595 & $   17.9530\pm    0.0015$ &       285.& $    0.0393\pm    0.0015$ &$       4.0\pm       0.5$ \\
6809 &             [\ion{S}{3}] &   18.71300 & $   18.7153\pm    0.0044$ &        37.& $    0.0134\pm    0.0053$ &$      -1.4\pm       2.1$ \\
6809 &              [\ion{Ne}{5}] &   24.31750 & $   24.3446\pm    0.0013$ &       334.& $    0.0242\pm    0.0014$ &$       7.0\pm       1.2$ \\
6809 &              [\ion{F}{4}]? &   25.83000 & $   25.8495\pm    0.0025$ &     226.& $  0.0613\pm   0.0026$ &$      7.6\pm      0.9$ \\
6809 &              [\ion{O}{4}] &   25.89030 & $   25.9155\pm    0.0029$ &     292.& $  0.0473\pm   0.0029$ &$     23.5\pm      4.5$ \\
6809 &             [\ion{Fe}{2}] &   25.98829 & $   26.0199\pm    0.0033$ &     365.& $  0.0710\pm   0.0033$ &$     17.2\pm      2.5$ \\
6809 &     [\ion{Fe}{2}] (broad) &   25.98829 & $   26.1516\pm    0.0097$ &    1885.& $  0.1615\pm   0.0107$ &$      7.3\pm      1.4$ \\
6809 &             [\ion{S}{3}] &   33.48100 & $   33.5163\pm    0.0050$ &       316.& $    0.0786\pm    0.0120$ &$     -16.1\pm      13.5$ \\
6809 &      [\ion{Si}{2}] (blue) &   34.81520 & $  34.6456\pm0.0040  $ & -1460. & $  0.0851\pm0.0040  $ & $ 12.5\pm 1.8$ \\
6809 &             [\ion{Si}{2}] &   34.81520 & $  34.8538\pm0.0103  $ & 332. & $  0.2449\pm0.0158  $ & $ 39.3\pm 6.7$ \\
6809 &       [\ion{Si}{2}] (red) &   34.81520 & $  35.1001\pm0.0076  $ & 2453. & $  0.1827\pm0.0078  $ & $ 16.7\pm 2.2$ \\
6809 &             [\ion{Fe}{2}] &   35.34865 & $  35.3912\pm0.0134  $ & 361. & $  0.1749\pm0.0135  $ & $ 10.6\pm 2.6$ \\
\hline
7147 &              [\ion{S}{4}] &   10.51050 & $   10.5221\pm    0.0010$ &       331.& $    0.0157\pm    0.0010$ &$       3.0\pm       0.6$ \\
7147 &             \ion{H}{1} 7-6 &   12.37190 & $   12.3857\pm    0.0011$ &       335.& $    0.0229\pm    0.0011$ &$       2.9\pm       0.4$ \\
7147 &             [\ion{Ne}{2}] &   12.81355 & $   12.8269\pm    0.0001$ &       313.& $    0.0191\pm    0.0001$ &$      43.3\pm       0.6$ \\
7147 &              [\ion{Ne}{5}] &   14.32170 & $   14.3364\pm    0.0006$ &       308.& $    0.0234\pm    0.0006$ &$       5.1\pm       0.4$ \\
7147 &            [\ion{Ne}{3}] &   15.55510 & $   15.5703\pm    0.0001$ &       293.& $    0.0253\pm    0.0001$ &$      18.6\pm       0.3$ \\
7147 &             [\ion{Fe}{2}] &   17.93595 & $   17.9529\pm    0.0020$ &       284.& $    0.0353\pm    0.0021$ &$       4.2\pm       0.8$ \\
7147 &             [\ion{S}{3}] &   18.71300 & $   18.7314\pm    0.0007$ &       295.& $    0.0298\pm    0.0008$ &$       6.7\pm       0.5$ \\
7147 &              [\ion{Ne}{5}] &   24.31750 & $   24.3433\pm    0.0010$ &       318.& $    0.0308\pm    0.0009$ &$       7.8\pm       0.7$ \\
7147 &              [\ion{F}{4}]? &   25.83000 & $   25.8516\pm    0.0039$ &     251.& $  0.0693\pm   0.0036$ &$      7.8\pm      1.2$ \\
7147 &              [\ion{O}{4}] &   25.89030 & $   25.9157\pm    0.0029$ &     294.& $  0.0484\pm   0.0030$ &$     24.4\pm      4.6$ \\
7147 &             [\ion{Fe}{2}] &   25.98829 & $   26.0216\pm    0.0036$ &     385.& $  0.0744\pm   0.0036$ &$     18.1\pm      2.7$ \\
7147 &     [\ion{Fe}{2}] (broad) &   25.98829 & $   26.1581\pm    0.0099$ &    1960.& $  0.1293\pm   0.0099$ &$      7.5\pm      1.8$ \\
7147 &      [\ion{Si}{2}] (blue) &   34.81520 & $  34.6283\pm0.0039  $ & -1609. & $  0.1041\pm0.0039  $ & $ 15.7\pm 1.8$ \\
7147 &             [\ion{Si}{2}] &   34.81520 & $  34.8774\pm0.0127  $ & 536. & $  0.3213\pm0.0242  $ & $ 51.5\pm10.1$ \\
7147 &       [\ion{Si}{2}] (red) &   34.81520 & $  35.1013\pm0.0086  $ & 2464. & $  0.1382\pm0.0088  $ & $ 11.7\pm 2.3$ \\
7147 &             [\ion{Fe}{2}] &   35.34865 & $  35.3494\pm0.0100  $ & 6. & $  0.1542\pm0.0107  $ & $ 16.7\pm 3.5$ \\
\hline
7323 &              [\ion{S}{4}] &   10.51050 & $   10.5209\pm    0.0022$ &       297.& $    0.0167\pm    0.0022$ &$       2.8\pm       1.1$ \\
7323 &             \ion{H}{1} 7-6 &   12.37190 & $   12.3869\pm    0.0019$ &       364.& $    0.0236\pm    0.0019$ &$       2.5\pm       0.7$ \\
7323 &             [\ion{Ne}{2}] &   12.81355 & $   12.8287\pm    0.0001$ &       355.& $    0.0167\pm    0.0001$ &$      37.0\pm       0.5$ \\
7323 &              [\ion{Ne}{5}] &   14.32170 & $   14.3363\pm    0.0007$ &       306.& $    0.0212\pm    0.0006$ &$       4.1\pm       0.4$ \\
7323 &            [\ion{Ne}{3}] &   15.55510 & $   15.5749\pm    0.0004$ &       382.& $    0.0227\pm    0.0004$ &$       6.9\pm       0.4$ \\
7323 &             [\ion{Fe}{2}] &   17.93595 & $   17.9571\pm    0.0012$ &       354.& $    0.0273\pm    0.0011$ &$       3.4\pm       0.4$ \\
7323 &             [\ion{S}{3}] &   18.71300 & $   18.7210\pm    0.0014$ &       128.& $    0.0197\pm    0.0009$ &$      -3.1\pm       0.4$ \\
7323 &              [\ion{Ne}{5}] &   24.31750 & $   24.3446\pm    0.0010$ &       334.& $    0.0257\pm    0.0011$ &$       6.9\pm       0.9$ \\
7323 &              [\ion{F}{4}]? &   25.83000 & $   25.8437\pm    0.0042$ &     159.& $  0.0448\pm   0.0041$ &$      5.5\pm      1.5$ \\
7323 &              [\ion{O}{4}] &   25.89030 & $   25.9201\pm    0.0023$ &     345.& $  0.0518\pm   0.0024$ &$     29.4\pm      4.2$ \\
7323 &             [\ion{Fe}{2}] &   25.98829 & $   26.0179\pm    0.0040$ &     342.& $  0.0670\pm   0.0040$ &$     17.6\pm      3.3$ \\
7323 &     [\ion{Fe}{2}] (broad) &   25.98829 & $   26.1416\pm    0.0093$ &    1770.& $  0.1858\pm   0.0093$ &$     10.3\pm      1.6$ \\
7323 &      [\ion{Si}{2}] (blue) &   34.81520 & $   34.6395\pm0.0062  $ & -1513. & $  0.1252\pm0.0062  $ & $ 13.5\pm 2.1$ \\
7323 &             [\ion{Si}{2}] &   34.81520 & $   34.8428\pm0.0246  $ & 238. & $  0.3615\pm0.0529  $ & $ 58.2\pm20.9$ \\
7323 &       [\ion{Si}{2}] (red) &   34.81520 & $   35.1129\pm0.0107  $ & 2563. & $  0.1689\pm0.0115  $ & $ 13.2\pm 2.7$ \\
7323 &             [\ion{Fe}{2}] &   35.34865 & $   35.3835\pm0.0110  $ & 296. & $  0.1577\pm0.0115  $ & $ 18.5\pm 4.1$ 
\enddata
\label{tab:lines}
\end{deluxetable*}

\addtocounter{table}{-1}
\begin{deluxetable*}{ccccccc}
\tabletypesize{\scriptsize}
\tablewidth{0pt}
\tablecaption{-- continued-- High Resolution Lines}
\tablehead{
\colhead{Day number} &
\colhead{Species} &
\colhead{$\lambda_0$ ($\micron$)} &
\colhead{$\lambda$ ($\micron$)} &
\colhead{$v$ (km s$^{-1}$)} &
\colhead{FWHM ($\micron$)} &
\colhead{Flux (10$^{-22}$ W m$^{-2}$)}
}
\startdata
7469 &              [\ion{S}{4}] &   10.51050 & $   10.5171\pm    0.0016$ &       188.& $    0.0186\pm    0.0016$ &$       3.6\pm       1.0$ \\
7469 &             \ion{H}{1} 7-6 &   12.37190 & $   12.3825\pm    0.0012$ &       257.& $    0.0248\pm    0.0013$ &$       3.9\pm       0.6$ \\
7469 &             [\ion{Ne}{2}] &   12.81355 & $   12.8250\pm    0.0001$ &       268.& $    0.0195\pm    0.0001$ &$      44.4\pm       0.4$ \\
7469 &              [\ion{Ne}{5}] &   14.32170 & $   14.3353\pm    0.0009$ &       285.& $    0.0206\pm    0.0008$ &$       4.4\pm       0.6$ \\
7469 &            [\ion{Ne}{3}] &   15.55510 & $   15.5673\pm    0.0004$ &       235.& $    0.0185\pm    0.0004$ &$       6.9\pm       0.4$ \\
7469 &             [\ion{Fe}{2}] &   17.93595 & $   17.9525\pm    0.0027$ &       277.& $    0.0488\pm    0.0028$ &$       4.8\pm       0.9$ \\
7469 &             [\ion{S}{3}] &   18.71300 & $   18.7336\pm    0.0014$ &       330.& $    0.0295\pm    0.0014$ &$      -4.2\pm       0.6$ \\
7469 &              [\ion{Ne}{5}] &   24.31750 & $   24.3447\pm    0.0010$ &       336.& $    0.0275\pm    0.0011$ &$       6.4\pm       0.8$ \\
7469 &              [\ion{F}{4}]? &   25.83000 & $   25.8442\pm    0.0014$ &     165.& $  0.0570\pm   0.0015$ &$      6.9\pm      0.5$ \\
7469 &              [\ion{O}{4}] &   25.89030 & $   25.9168\pm    0.0020$ &     307.& $  0.0468\pm   0.0021$ &$     26.0\pm      3.5$ \\
7469 &             [\ion{Fe}{2}] &   25.98829 & $   26.0166\pm    0.0035$ &     327.& $  0.0792\pm   0.0035$ &$     20.5\pm      2.8$ \\
7469 &     [\ion{Fe}{2}] (broad) &   25.98829 & $   26.1526\pm    0.0071$ &    1897.& $  0.1640\pm   0.0073$ &$      8.6\pm      1.2$ \\
7469 &      [\ion{Si}{2}] (blue) &   34.81520 & $   34.6232\pm0.0071  $ & -1653. & $  0.1609\pm0.0071  $ & $ 28.4\pm 3.9$ \\
7469 &             [\ion{Si}{2}] &   34.81520 & $   34.8516\pm0.0059  $ & 313. & $  0.1826\pm0.0077  $ & $ 48.2\pm 5.6$ \\
7469 &       [\ion{Si}{2}] (red) &   34.81520 & $   35.0706\pm0.0100  $ & 2199. & $  0.1642\pm0.0100  $ & $ 23.9\pm 4.5$ \\
7469 &             [\ion{Fe}{2}] &   35.34865 & $   35.3474\pm0.0090  $ & -11. & $  0.0918\pm0.0093  $ & $  7.0\pm 2.2$ \\
\hline
7798 &              [\ion{S}{4}] &   10.51050 & $   10.5190\pm    0.0019$ &       243.& $    0.0192\pm    0.0019$ &$       3.0\pm       1.0$ \\
7798 &             \ion{H}{1} 7-6 &   12.37190 & $   12.3855\pm    0.0010$ &       330.& $    0.0203\pm    0.0009$ &$       3.3\pm       0.5$ \\
7798 &             [\ion{Ne}{2}] &   12.81355 & $   12.8251\pm    0.0001$ &       270.& $    0.0191\pm    0.0001$ &$      48.2\pm       0.5$ \\
7798 &              [\ion{Ne}{5}] &   14.32170 & $   14.3344\pm    0.0010$ &       266.& $    0.0193\pm    0.0009$ &$       4.0\pm       0.6$ \\
7798 &            [\ion{Ne}{3}] &   15.55510 & $   15.5676\pm    0.0003$ &       241.& $    0.0243\pm    0.0003$ &$      11.4\pm       0.5$ \\
7798 &             [\ion{Fe}{2}] &   17.93595 & $   17.9533\pm    0.0016$ &       290.& $    0.0377\pm    0.0017$ &$       3.7\pm       0.5$ \\
7798 &              [\ion{Ne}{5}] &   24.31750 & $   24.3441\pm    0.0016$ &       328.& $    0.0245\pm    0.0015$ &$       7.5\pm       1.4$ \\
7798 &              [\ion{F}{4}]? &   25.83000 & $   25.8481\pm    0.0019$ &     210.& $  0.0483\pm   0.0022$ &$      5.9\pm      0.8$ \\
7798 &              [\ion{O}{4}] &   25.89030 & $   25.9192\pm    0.0025$ &     335.& $  0.0539\pm   0.0025$ &$     30.4\pm      4.4$ \\
7798 &             [\ion{Fe}{2}] &   25.98829 & $   26.0154\pm    0.0049$ &     313.& $  0.0649\pm   0.0049$ &$     18.1\pm      4.2$ \\
7798 &     [\ion{Fe}{2}] (broad) &   25.98829 & $   26.1326\pm    0.0091$ &    1666.& $  0.2330\pm   0.0091$ &$     13.9\pm      1.6$ \\
7798 &      [\ion{Si}{2}] (blue) &   34.81520 & $   34.6527\pm0.0062  $ & -1399. & $  0.1599\pm0.0063  $ & $ 37.5\pm 4.5$ \\
7798 &             [\ion{Si}{2}] &   34.81520 & $   34.8587\pm0.0058  $ & 375. & $  0.1300\pm0.0062  $ & $ 57.3\pm 8.3$ \\
7798 &       [\ion{Si}{2}] (red) &   34.81520 & $   35.0713\pm0.0098  $ & 2205. & $  0.1740\pm0.0099  $ & $ 37.1\pm 6.5$ \\
7798 &             [\ion{Fe}{2}] &   35.34865 & $   35.3406\pm0.0084  $ & -68. & $  0.1295\pm0.0105  $ & $ 11.0\pm 2.5$ \\

\hline
7954 &              [\ion{S}{4}] &   10.51050 & $   10.5196\pm    0.0019$ &       260.& $    0.0226\pm    0.0019$ &$       5.6\pm       1.4$ \\
7954 &             \ion{H}{1} 7-6 &   12.37190 & $   12.3809\pm    0.0011$ &       218.& $    0.0225\pm    0.0012$ &$       3.9\pm       0.6$ \\
7954 &             [\ion{Ne}{2}] &   12.81355 & $   12.8254\pm    0.0001$ &       277.& $    0.0191\pm    0.0001$ &$      52.1\pm       0.6$ \\
7954 &              [\ion{Ne}{5}] &   14.32170 & $   14.3331\pm    0.0008$ &       239.& $    0.0188\pm    0.0007$ &$       3.7\pm       0.4$ \\
7954 &            [\ion{Ne}{3}] &   15.55510 & $   15.5686\pm    0.0002$ &       260.& $    0.0238\pm    0.0002$ &$      20.4\pm       0.5$ \\
7954 &             [\ion{Fe}{2}] &   17.93595 & $   17.9541\pm    0.0008$ &       304.& $    0.0245\pm    0.0009$ &$       3.8\pm       0.4$ \\
7954 &             [\ion{S}{3}] &   18.71300 & $   18.7289\pm    0.0010$ &       255.& $    0.0305\pm    0.0011$ &$       7.3\pm       0.8$ \\
7954 &              [\ion{Ne}{5}] &   24.31750 & $   24.3436\pm    0.0016$ &       322.& $    0.0241\pm    0.0014$ &$       5.7\pm       1.0$ \\
7954 &              [\ion{F}{4}]? &   25.83000 & $   25.8494\pm    0.0042$ &     225.& $  0.0813\pm   0.0039$ &$      7.2\pm      1.0$ \\
7954 &              [\ion{O}{4}] &   25.89030 & $   25.9167\pm    0.0027$ &     306.& $  0.0509\pm   0.0027$ &$     27.0\pm      4.4$ \\
7954 &             [\ion{Fe}{2}] &   25.98829 & $   26.0270\pm    0.0055$ &     447.& $  0.0906\pm   0.0055$ &$     22.5\pm      4.2$ \\
7954 &     [\ion{Fe}{2}] (broad) &   25.98829 & $   26.1722\pm    0.0143$ &    2123.& $  0.1959\pm   0.0144$ &$     10.9\pm      2.5$ \\
7954 &             [\ion{S}{3}] &   33.48100 & $   33.5225\pm    0.0118$ &       372.& $    0.0935\pm    0.0116$ &$     -15.7\pm       6.0$ \\
7954 &      [\ion{Si}{2}] (blue) &   34.81520 & $   34.6324\pm0.0045  $ & -1574. & $  0.1157\pm0.0045  $ & $ 27.9\pm 3.4$ \\
7954 &             [\ion{Si}{2}] &   34.81520 & $   34.8572\pm0.0058  $ & 362. & $  0.1982\pm0.0068  $ & $ 60.2\pm 6.0$ \\
7954 &       [\ion{Si}{2}] (red) &   34.81520 & $   35.0921\pm0.0092  $ & 2384. & $  0.1721\pm0.0093  $ & $ 27.5\pm 4.6$ \\
7954 &             [\ion{Fe}{2}] &   35.34865 & $   35.3650\pm0.0053  $ & 139. & $  0.1560\pm0.0058  $ & $ 17.8\pm 2.0$

\enddata
\tablecomments{$v_{sys} = 289.2$ km s$^{-1}$ \citep{crotts:1991}.}
\end{deluxetable*}
 
\begin{deluxetable*}{lccccccccccccc}
\tabletypesize{\footnotesize}
\tablewidth{0pt}
\tablecaption{High Resolution Line Evolution}
\tablehead{
\colhead{} &
\colhead{} &
\multicolumn{3}{c}{$F_{line} = A\ F_{24\micron}$} & & 
\multicolumn{3}{c}{$F_{line} = B$} & &
\multicolumn{4}{c}{$F_{line} = \beta + \alpha\ t$} \\
\cline{3-5}\cline{7-9}\cline{11-14}
\colhead{Species} &
\colhead{$N_{data}$} & 
\colhead{$A$} &
\colhead{$\chi^2$} & 
\colhead{$P(<\chi^2)$} & &
\colhead{$B$} &
\colhead{$\chi^2$} & 
\colhead{$P(<\chi^2)$} & &
\colhead{$\beta$} &
\colhead{$\alpha$} &
\colhead{$\chi^2$} & 
\colhead{$P(<\chi^2)$}
}
\startdata
              $[$S IV$]$ &  5 &  5.326e-21 &      2.214 &      0.304 & &  3.270e-22 &      3.418 &      0.510 & & -8.241e-22 &  1.557e-25 &      2.110 &      0.450\\
                 \ion{H}{1} 7-6 &  6 &  4.888e-21 &      4.270 &      0.489 & &  2.997e-22 &     12.553 &      0.972 & & -9.992e-22 &  1.751e-25 &      4.150 &      0.614\\
             $[$Ne II$]$ &  7 &  6.977e-20 &    468.208 &      1.000 & &  4.351e-21 &    732.517 &      1.000 & & -4.534e-21 &  1.200e-24 &    196.869 &      1.000\\
              $[$Ne V$]$ &  7 &  6.731e-21 &     41.675 &      1.000 & &  4.367e-22 &      9.013 &      0.827 & &  1.376e-21 & -1.265e-25 &      2.653 &      0.247\\
            $[$Ne III$]$ &  7 &  2.164e-20 &   1256.515 &      1.000 & &  1.286e-21 &   1123.777 &      1.000 & &  1.703e-23 &  1.744e-25 &   1105.106 &      1.000\\
             $[$Fe II$]$ &  7 &  5.873e-21 &     19.530 &      0.997 & &  3.847e-22 &      3.783 &      0.294 & &  6.075e-22 & -2.985e-26 &      3.410 &      0.363\\
             $[$S III$]$ &  3 &  1.123e-20 &      4.020 &      0.866 & &  6.862e-22 &      0.774 &      0.321 & &  2.720e-23 &  8.967e-26 &      0.047 &      0.172\\
              $[$Ne V$]$ &  7 &  1.106e-20 &     19.626 &      0.997 & &  6.718e-22 &      6.208 &      0.600 & &  4.751e-22 &  2.701e-26 &      6.087 &      0.702\\
              $[$F IV$]$? &  7 &  1.083e-20 &     30.014 &      1.000 & &  6.545e-22 &      8.623 &      0.804 & &  1.909e-22 &  6.409e-26 &      7.182 &      0.793\\
              $[$O IV$]$ &  7 &  4.425e-20 &     11.281 &      0.920 & &  2.527e-21 &      5.414 &      0.508 & & -1.094e-21 &  5.074e-25 &      1.163 &      0.052\\
             $[$Fe II$]$ &  7 &  3.295e-20 &     11.028 &      0.912 & &  1.785e-21 &      3.626 &      0.273 & & -5.983e-22 &  3.387e-25 &      0.760 &      0.020\\
     $[$Fe II$]$ (broad) &  7 &  1.512e-20 &      8.798 &      0.815 & &  7.018e-22 &     55.182 &      1.000 & & -3.373e-21 &  5.859e-25 &      4.875 &      0.569\\
             $[$S III$]$ &  1 &  5.234e-20 &      \nodata &      \nodata & &  1.387e-21 &      \nodata &      \nodata & & \nodata &  \nodata &      \nodata &      \nodata\\
      $[$Si II$]$ (blue) &  6 &  3.135e-20 &     24.539 &      1.000 & &  1.712e-21 &     48.649 &      1.000 & & -9.902e-21 &  1.612e-24 &     17.085 &      0.998\\
             $[$Si II$]$ &  7 &  8.463e-20 &     17.976 &      0.994 & &  4.997e-21 &      7.062 &      0.685 & & -1.202e-21 &  8.615e-25 &      2.852 &      0.277\\
       $[$Si II$]$ (red) &  6 &  3.016e-20 &     17.694 &      0.997 & &  1.661e-21 &     24.299 &      1.000 & & -5.840e-21 &  1.044e-24 &     16.277 &      0.997\\
             $[$Fe II$]$ &  6 &  1.992e-20 &     16.415 &      0.994 & &  1.292e-21 &     17.693 &      0.997 & & -1.644e-21 &  3.909e-25 &     15.376 &      0.996
\enddata
\tablecomments{$A$ in (W m$^{-2}$ / Jy), $B$ in (W m$^{-2}$), 
$\alpha$ in (W m$^{-2}$ / day), $\beta$ in (W m$^{-2}$)}
\label{tab:hi_evol}
\end{deluxetable*}

\begin{deluxetable*}{lccccccccccccc}
\tabletypesize{\footnotesize}
\tablewidth{0pt}
\tablecaption{Low Resolution Line Evolution}
\tablehead{
\colhead{} &
\colhead{} &
\multicolumn{3}{c}{$F_{line} = A\ F_{24\micron}$} & & 
\multicolumn{3}{c}{$F_{line} = B$} & &
\multicolumn{4}{c}{$F_{line} = \beta + \alpha\ t$} \\
\cline{3-5}\cline{7-9}\cline{11-14}
\colhead{Species} &
\colhead{$N_{data}$} & 
\colhead{$A$} &
\colhead{$\chi^2$} & 
\colhead{$P(<\chi^2)$} & &
\colhead{$B$} &
\colhead{$\chi^2$} & 
\colhead{$P(<\chi^2)$} & &
\colhead{$\beta$} &
\colhead{$\alpha$} &
\colhead{$\chi^2$} & 
\colhead{$P(<\chi^2)$}
}
\startdata
             $[$Ni II$]$ &  6 &  6.669e-21 &      4.896 &      0.571 & &  4.619e-22 &      8.121 &      0.850 & & -4.041e-22 &  1.126e-25 &      4.060 &      0.602\\
             $[$Ar II$]$ &  6 &  2.635e-20 &     39.320 &      1.000 & &  1.740e-21 &    109.060 &      1.000 & & -4.537e-21 &  8.266e-25 &     33.909 &      1.000\\
        \ion{H}{1} 6-5 \& 8-6 ? &  6 &  1.229e-20 &      3.427 &      0.366 & &  7.400e-22 &      3.533 &      0.382 & & -6.394e-22 &  1.882e-25 &      2.722 &      0.395\\
           $[$Ne VI$]$ ? &  5 &  8.163e-21 &      0.283 &      0.009 & &  5.220e-22 &      0.818 &      0.064 & & -8.842e-22 &  1.880e-25 &      0.203 &      0.023\\
             $[$Ne II$]$ &  7 &  5.085e-20 &     79.567 &      1.000 & &  3.060e-21 &     73.141 &      1.000 & & -1.700e-21 &  6.501e-25 &     35.054 &      1.000\\
  $[$O IV$]$+$[$Fe II$]$ &  6 &  6.886e-20 &     37.651 &      1.000 & &  4.660e-21 &     11.664 &      0.960 & &  3.246e-21 &  1.862e-25 &     11.360 &      0.977\\
             $[$S III$]$ &  4 &  1.587e-19 &    608.003 &      1.000 & &  9.161e-21 &    472.367 &      1.000 & &  3.072e-20 & -2.996e-24 &    450.078 &      1.000\\
             $[$Si II$]$ &  6 &  2.137e-19 &     14.594 &      0.988 & &  1.317e-20 &     52.898 &      1.000 & & -2.361e-20 &  4.970e-24 &     10.493 &      0.967
\enddata
\label{tab:lo_evol}
\tablecomments{$A$ in (W m$^{-2}$ / Jy), $B$ in (W m$^{-2}$), 
$\alpha$ in (W m$^{-2}$ / day), $\beta$ in (W m$^{-2}$)}
\end{deluxetable*}

\end{document}